\documentclass[trackchanges,twocolumn]{aastex701}

\usepackage{natbib}
\setcitestyle{aysep={}} 
\usepackage{graphicx}	
\usepackage{amsmath}	
\usepackage{amssymb}	
\usepackage{subfigure}
\usepackage{threeparttable}
\usepackage{balance}
\usepackage{indentfirst}
\usepackage{siunitx}
\usepackage{array}
\usepackage{booktabs}
\usepackage{mathrsfs}
\usepackage{soul}         
\usepackage{xeCJK}         
\soulregister{\st}{0}      
\usepackage{textcomp}

\newcommand{\lv}{\ifmmode L_{5100} \else $L_{5100}$\fi}
\newcommand{\kms}{\ifmmode {\rm km\ s}^{-1} \else km s$^{-1}$\fi}
\newcommand{\ergs}{\ifmmode {\rm erg\ s}^{-1} \else erg s$^{-1}$\fi}
\newcommand{\ergc}{\ifmmode {\rm erg\ s^{-1}cm^{-2}} \else $\rm erg\ s^{-1} cm^{-2}$\fi}
\newcommand{\lb}{\ifmmode L_{\rm Bol} \else $L_{\rm Bol}$\fi}
\newcommand{\ledd}{\ifmmode L_{\rm Edd} \else $L_{\rm Edd}$\fi}
\newcommand{\hb}{\ifmmode H\beta \else H$\beta$\fi}
\newcommand{\ha}{\ifmmode H\alpha \else H$\alpha$\fi}
\newcommand{\oiii}{\ifmmode \rm [O\ \textsc{iii}] \else $\rm [O\ \textsc{iii}]$\fi}
\newcommand{\nii}{\ifmmode \rm [N\ \sc{ii}] \else $\rm [N\ \sc{ii}]$ \fi}
\newcommand{\mgii}{\ifmmode \rm Mg\ \sc{ii} \else $\rm Mg\ \sc{ii}$\fi}
\newcommand{\civ}{\ifmmode \rm C\ \sc{iv} \else $\rm C\ \sc{iv}$ \fi}
\newcommand{\mbh}{\ifmmode M_{\rm BH}  \else $M_{\rm BH}$\fi}
\newcommand{\msun}{M_{\odot}}
\newcommand{\rfe}{\ifmmode R_{\rm Fe\ \textsc{ii}} \else $R_{\rm Fe\ \textsc{ii}}$\fi}
\newcommand{\sst}{\ifmmode \sigma_{\rm \ast} \else $\sigma_{\rm \ast}$\fi}
\newcommand{\dhb}{\ifmmode D_{\rm H\beta}  \else $D_{\rm H\beta}$\fi}
\newcommand{\leddR}{\ifmmode L_{\rm Bol} /L_{\rm Edd} \else $L_{\rm Bol} /L_{\rm Edd}$\fi}
\newcommand{\feii}{\ifmmode \rm Fe \textsc{ii} \else $\rm Fe\ \textsc{ii}$ \fi}
\newcommand{\mdot}{\ifmmode \dot{\mathscr{M}}  \else $\dot{\mathscr{M}}$\fi}
\newcommand{\rhb}{\ifmmode R_{\rm BLR}({\rm H\beta})  \else $R_{\rm BLR}({\rm H\beta})$\fi}
\newcommand{\shb}{\ifmmode \sigma_{\rm H\beta} \else $\sigma_{\rm \hb}$\fi}
\newcommand{\RL}{\ifmmode R_{\rm BLR}({\rm H\beta}) - L_{\rm 5100} \else $R_{\rm BLR}({\rm H\beta}) - L_{\rm 5100}$\fi}
\newcommand{\ms}{\ifmmode M_{\rm BH}-\sigma_{\ast} \else $M_{\rm BH}-\sigma_{\ast}$\fi}
\newcommand{\bd}{\ifmmode \rm \ha/\hb\ \else $\rm \ha/\hb$\fi}

\begin{document}

\title{The Dependence of the Mean Spectral Energy Distributions on the Accretion Rate  for Quasars with $z < 0.75$ from the Sloan Digital Sky Survey}

\author[orcid=0009-0008-5295-3977]{Yan-Song Ma(马岩松)}
\affiliation{School of Physics and Technology, Nanjing Normal University, Nanjing 210023, People's Republic of China}
\email{whbian@njnu.edu.cn}
\author{Yu-Meng Guan(官雨蒙)}
\affiliation{School of Physics and Technology, Nanjing Normal University, Nanjing 210023, People's Republic of China}
\email{2304174572@qq.com}
\author{Jian-Xia Jiang(姜建霞)}
\affiliation{School of Physics and Technology, Nanjing Normal University, Nanjing 210023, People's Republic of China}
\email{whbian@njnu.edu.cn}
\author[orcid=0009-0008-3600-670X]{Shao-Jun Li(李少君)}
\affiliation{School of Physics and Technology, Nanjing Normal University, Nanjing 210023, People's Republic of China}
\email{whbian@njnu.edu.cn}
\author[orcid=0009-0004-4431-4535]{Xiang-Wei Ning(宁翔炜)}
\affiliation{School of Physics and Technology, Nanjing Normal University, Nanjing 210023, People's Republic of China}
\email{whbian@njnu.edu.cn}
\author{Yi Tang(唐逸)}
\affiliation{School of Physics and Technology, Nanjing Normal University, Nanjing 210023, People's Republic of China}
\email{whbian@njnu.edu.cn}
\author[orcid=0000-0002-2121-8960]{Wei-Hao Bian (卞维豪)}
\affiliation{School of Physics and Technology, Nanjing Normal University, Nanjing 210023, People's Republic of China}
\email[show]{whbian@njnu.edu.cn}

\begin{abstract}
We construct mean spectral energy distributions (SEDs) for a substantial sample of 56,969 Sloan Digital Sky Survey DR16 quasars with $z < 0.75$, utilizing multiwavelength data from the mid-infrared (MIR) to ultraviolet (UV). These SEDs are built on eigenvector 1  parameters---the relative optical \feii\ strength (\rfe) and the \hb\ line width ($\rm FWHM_{\hb}$)---that capture the principal spectral variance of quasar spectra.
From three \rfe-dependent mean SEDs we find that quasars with a  larger \rfe\ exhibit redder UV and optical and redder MIR and near-infrared (NIR) continua, indicating more dust emission. 
We also split our sample directly into Eddington ratio \leddR\ (or dimensionless accretion rate \mdot) bins to construct different mean SEDs and find that the continua become increasingly red with increasing \leddR\ (or \mdot) in the MIR, NIR, and UV bands. This demonstrates that the shapes of Type 1 AGN SEDs depend on the accretion rate.  However, the optical continuum shows the opposite trend (becoming harder and bluer), indicating the complexity of the optical emission region.
From $\rm FWHM_{\hb}$-dependent mean SEDs we find that quasars with a larger $\rm FWHM_{\hb}$ show redder optical and NIR continua and bluer UV and MIR continua. The bluer MIR continuum suggests that a larger angle between of the line of sight and the torus plane results in weaker torus emission in the MIR.

\end{abstract}


\section{INTRODUCTION}
Active galactic nuclei (AGNs), fueled by the accretion of gas and dust onto their central supermassive black holes (SMBHs), rank among the most powerful cosmic objects \citep[e.g.,][]{Ho2008, Netzer2013}. AGNs exhibit distinctive observational signatures across more than 20 orders of magnitude in wavelength, from radio to $\gamma$ rays. The broadband spectral energy distribution (SED) is a fundamental tool for probing  the underlying  physical processes in AGNs \citep[e.g.,][]{Elvis1994, Richards2006a, Ho2008, Shang2011, Krawczy2013, Cai2023, Azadi2022, Ahmed2025, Chen2025, Garnica2025} . 

Regarding the energy-generation mechanisms in AGNs---driven by gas and dust---at least four primary processes are identified: radio synchrotron emission from jets, thermal reradiation by dust grains, pseudoblackbody emission  in the optical--ultraviolet (UV)--soft X-ray range from the disk accretion, and Compton upscattering in the hot corona, which dominates  in the hard X-ray band \citep[e.g.,][]{ Ho2008, Netzer2013}.  
Optical surveys tend to miss heavily reddened and obscured objects, and even X-ray surveys can fail to detect Compton-thick sources \citep{Richards2006a}. Therefore, the mid-infrared (MIR) and  near-infrared (NIR, about 1-4 $\mu \rm m$)  regions offer an attractive window for probing the black hole accretion history of the Universe. 

In the context of spectroscopic observations, SEDs derived from multiband photometry can be conveniently generated for large samples to investigate several crucial issues related to radiative power across nearly the entire electromagnetic spectrum. These issues include  bolometric correction factors (BCs), the  fraction of obscured AGNs missed by optical surveys, luminosity-dependent mean SEDs, and the unseen extreme-ultraviolet (EUV) continuum  \citep[e.g.,][]{Marconi2004, Richards2006a, Ho2008, Shang2011, Krawczy2013}. 

 \cite{Ho2008} used a sample of 150 nearby AGNs to investigate the variance of SEDs in nearby Type 1 AGNs, spanning $4$ dex in SMBH mass (\mbh; $10^5-10^9 \msun$) and $6.5$ dex in the Eddington ratio ($\leddR$, where  \lb\ is the bolometric luminosity and \ledd\ is the Eddington luminosity; $10^{-6}-10^{0.5}$). He found apparent differences in the  SEDs of radio-quiet AGNs binned by the Eddington ratio. 
Using quasars from the Sloan Digital Sky Survey (SDSS) Data Release (DR7), \cite{Krawczy2013} discovered  that   low-luminosity SEDs, compared to high-luminosity SEDs, exhibit a harder (bluer) far-ultraviolet (FUV) spectral slope and less hot torus dust emission \cite[see also][]{Zhang2014}. Theoretical accretion disk models suggest that the UV-optical continuum slope depends on the Eddington ratio, the SMBH mass, and the SMBH spin, although other components also contribute in complex ways \cite[e.g., ][]{Wang2014, Czerny2019, Hagen2023}.  

It is widely accepted that the diversity of AGN phenomenology can be largely unified by two key factors: orientation and the Eddington ratio \citep{Antonucci1993, Urry1995, Shen2014, Netzer2015, Padovani2017}. Investigating the dependence of SEDs on these factors is essential for  understanding the accretion process and its impact on host galaxies. Theoretical AGN SED templates can be calculated from accretion models parameterized by SMBH mass, accretion rate, spin, viewing angle, and the presence of jets,  torus, inner Comptonization region, or gaps in accretion disk  \citep[e.g.,][]{Shields1978, Czerny1987, Laor1990, Telfer2002, Blaes2004, Shang2005, Davis2007, Davis2011, Wang2014, Siebenmorgen2015, Kubota2018, Czerny2019, Azadi2022, Cai2023,  Hagen2023,  Molina2023, Stolc2023,Ahmed2025, Chen2025, Garnica2025}.  Comparing these theoretical predictions with observed AGN SEDs can help constrain  fundamental AGN parameters.

By adopting different integration limits in the SED, the bolometric luminosity can be estimated---a vital parameter for describing the accretion process. BCs are used to  estimate the bolometric luminosity from  a single-band luminosity. BCs likely depend on luminosity, accretion rate, accretion efficiency, orientation, and possible other factors \citep[e.g.,][]{Marconi2004, Runnoe2012a, Runnoe2012b, Netzer2013, Netzer2019, Pennell2017}. Inferring bolometric luminosity from optical luminosity alone  can introduce errors as large as 50\% for individual quasars \citep{Richards2006a}. Therefore, it is essential to investigate how BCs vary with different quasar SEDs.

Using principal component analysis on a sample of low-$z$ 87 Palomar–Green quasars, \cite{BG92} discovered that principal component 1 (PC1), also known as eigenvector 1 (EV1) in the AGN context,  is associated with the relative strength of optical \feii to broad \hb\ line (\rfe, defined as the ratio of  \feii\ emission within $4435-4685$ \AA\  to the broad \hb), the full width at half-maximum of the \hb\ broad line ($\rm FWHM_{\hb}$) , and \oiii $\lambda 5007$ strength. Principal component 2 (PC2) correlates with optical luminosity  and the optical--X-ray spectral index ($\alpha_{\rm ox}$). Both PC1 and PC2  are linked to the accretion process around SMBHs \citep[e.g.,][]{Boroson2002, Ge2016}. 
With $\rm FWHM_{\hb}$-based \mbh, \cite{Boroson2002} proposed that PC1 is primarily correlated with \leddR, while PC2 shows a strong connection with both \mbh\ and \leddR\ \citep[e.g.,][]{Shen2014, Ge2016}. Investigating the dependence of SEDs on these principal components, as well as on the SMBH accretion process, is essential \citep[e.g.,][]{Gaskell1985, Sulentic2000b, Sulentic2007a, Sulentic2007b, Kuraszkiewicz2009, Zamfir2010, Marziani2015, Panda2019}.

Over the past  2 decades significant progress has been made in measuring SMBH masses in AGNs and in determining their accretion rate or Eddington ratio \citep[e.g.,][]{Bahcall1972, Blandford1982, Peterson1993, Kaspi2000, Vestergaard2002, Greene2005, Vestergaard2006, Bentz2009, Assef2011, Shen2012, KH2013, Netzer2013, Peterson2014, Bentz2015, Shen2015, Dalla2025}. 
For Type 1 AGNs, the virial mass can be calculated as follows  \citep[e.g.,][]{Peterson2004, Netzer2013}:
 \begin{equation}
 \label{eq1}
\mbh=f\times \frac{R_{\rm BLR}~(\Delta V)^2}{G}  .
\end{equation}
where  $R_{\rm BLR}$ is the size of the broad-line region (BLR),  $\Delta V$ is the velocity of the BLR clouds,  $f$ is the  virial factor \citep[e.g.,][]{Yu2020b, Li2025}, and $G$ is the gravitational constant.  
From the standard thin-disk model of \cite{SS73}, the dimensionless accretion rate \mdot\ can be calculated from the 5100 \AA\ continuum luminosity (\lv) and the SMBH mass (\mbh) as follows  \citep[e.g.,][]{Du2016}: 
 \begin{equation}
 \label{eq2}
\mdot \equiv \dot{M}/\dot{M}_{\rm Edd} = 20.1\left(\frac{\it{l}\mathrm{_{44}}}{\cos\,\it{i}}\right)^{3/2} m_{7}^{-2},
\end{equation}
where $\dot{M}$ is the mass accretion rate, $\dot{M}_{\rm Edd}=\ledd/ c^2$ is the Eddington accretion rate, $l_{44}=\lv/10^{44} ~\ergs$,
$m_{7}=M_{\rm BH}/10^7 \msun$, $i$ is the  inclination of the accretion disk relative to the observer, and cos $i$ is assumed to be $0.75$. \lb/\ledd\ and  \mdot\ are two key parameters for describing the SMBH accretion process. \lb\ is usually derived from a single-band luminosity using  BCs  \citep[e.g.,][]{Netzer2013}. Considering $\lb=\eta \dot{M}c^2$, we have $\leddR=\frac{\eta \dot{M}c^2}{\dot{M}_{\rm Edd}c^2}=\eta \mdot$, where $\eta$ is the accretion efficiency \citep[e.g.,][]{Liu2022}. 
Previous studies have revealed a correlation between \rfe\ and \lb/\ledd\ or \mdot\ \citep[e.g.,][]{Du2016, Liu2022}, suggesting that \rfe\ may serve as a tracer of the accretion rate.

To enhance our understanding of how the mean SED depends on the accretion process, we have constructed mean SEDs for quasars from the SDSS DR16 catalog  ($z<0.75$) compiled by \cite{Wu2022}. These mean SEDs are categorized by optical \rfe\ bins, as well as by \leddR\ and \mdot, and cover a wavelength range from the MIR ($\sim 20~ \rm \mu m$ in the rest frame) to the FUV (900 \AA\  in the rest frame). The data used in this analysis are presented in Section \ref{s2}.  Our results and discussions are presented in Section \ref{s3}, and our conclusions are summarized in Section \ref{s4}. Throughout this paper, we assume a cosmology with $\Omega_{\Lambda}=0.7$, $\Omega_{M}=0.3$, and $H_{0}=70~ \kms~ {\rm Mpc}^{-1}$. 

\section{Sample} \label{s2}
Our parent sample consists of 750,414 Type 1 quasars from the SDSS DR16 catalog compiled by \cite{Wu2022}, all brighter than $M_i = -22.0$  and featuring at least one broad emission line ($\rm FWHM > \rm 1000~ km~ s^{-1}$) or interesting and/or complex absorption features.
For all quasars, we utilized multibands photometric data from DR16Q \citep{Lyke2020}.  DR16Q includes force photometry or crossmatched data from the following surveys and catalogs:  the Galaxy Evolution Explorer \citep[GALEX;][]{Martin2005}, the UKIRT Infrared Deep Sky Survey \citep[UKIDSS;][]{Lawrence2007}, the Wide-Field Infrared Survey Explorer \cite[WISE;][] {Wright2010}, the FIRST radio survey \citep{Becker1995}, the Two Micron All Sky Survey \citep[2MASS;][] {Skrutskie2006}, the Second ROSAT All-Sky Survey \citep[][]{Boller2016}, the Third XMM-Newton Serendipitous Source Catalog \citep{Rosen2016}, and Gaia Data Release 2 \citep{Gaia2018}. For a detailed discussion of the multiband photometric data, please refer to \cite{Lyke2020} . 

Here we briefly summarize the photometric data from SDSS, GALEX, 2MASS, and WISE.
In the optical bands, each quasar has photometric data in the five SDSS optical bandpasses $ugriz$, with central wavelengths of $u$: 3551 \AA, $g$: 4686 \AA, $r$: 6165 \AA, $i$: 7481 \AA, and $z$: 8931 \AA\  \citep{Fukugita1996}. 
For UV data from two GALEX bands,  NUV covers 1750–2750 \AA\  and FUV  covers 1350–1750 \AA.  
DR16Q includes data from the 2MASS All-Sky Point Source Catalogue
in three bands with central wavelengths:
$J$ (1.25 $\mu \rm m$ ), $H$ (1.65 $\mu \rm m$) and $K$ (2.16 $\mu \rm m$). Catalog objects were matched to 2MASS sources within $2''.0$. 
The WISE mission provides data in four bands, referred to as $W1$ through $W4$, with effective observed frame wavelengths of 3.36, 4.61, 11.82, and 22.13 $\mu \rm m$, respectively. 
Due to differences in depth between $W1, ~W2, ~W3$, and $W4$, \cite{Lang2014} used WISE data to create coadded “unWISE” images, which were subjected to force photometry at the locations of known SDSS sources by \cite{Lang2016}. Many DR16 quasar targets had fluxes below the WISE detection limit used for the AllWISE Data Release, so forced photometry was applied to custom “unWISE” stacks for $W1$ and $W2$ \citep{Lyke2020}. For $W3$ and $W4$, we used data from the shallower $W3$ and $W4$ photometry provided in the ALLWISE release, as included in SDSS DR16.

For the 14 bands, we plot the magnitude (mag) versus the error ($\sigma_{mag}$), or  the flux ($f$) versus the error ($\sigma_f$) for the two GALEX UV bands. For all bands except for $W1$ and $W2$, we adopt a criterion of $\sigma_{mag} \le  0.5~\rm mag$ or $\sigma_f/f \le 0.5$, equivalent to a signal-to-noise ratio (SNR) of two. For $W1$ and $W2$, we adopt a stricter criterion of $\sigma_{mag} \le  0.1~\rm  mag$, consistent with  \cite{Krawczy2013}. 
Considering these criteria, the numbers of selected quasars in different bands are as follows: 464,271 in $W1$, 363,991 in $W2$, 67,940 in $W3$, 40,842 in $W4$, 17,577 in 2MASS $J$, 16,192 in 2MASS $H$, 16,484 in 2MASS $K$  \footnote{For our sample of DR16 quasars with $z<0.75$, the amount of data from 2MASS is larger than from UKIDSS, therefore, we use 2MASS NIR data instead of UKIDSS data.}, 705,794 in SDSS $u$, 748,212 in SDSS $g$, 747,905 in SDSS $r$, 747,047 in SDSS $i$, 721,311 in SDSS $z$, 360,382 in GALEX NUV, and 175,336 in GALEX FUV. For 750,414 DR16 quasars, we select 749,436 quasars (99.87\% of the parent sample) with photometric data available in at least one of 14 bands. The method of “gap repair” is used for the missing data in other bands.
For SDSS $ugriz$, the zero-points used to obtain AB magnitudes are -0.04, 0, 0, 0, 0.02 mag, respectively \footnote{\href{ "https://www.sdss4.org/dr12/algorithms/fluxcal/\#SDSStoAB"}  {https://www.sdss4.org/dr12/algorithms/fluxcal/\#SDSStoAB}  }. 
For WISE $W1, W2, W3, W4$, the zeropoints are 2.699, 3.339, 5.174, 6.620 mag, respectively 
\footnote{
\href{"https://wise2.ipac.caltech.edu/docs/release/allsky/expsup/sec4\_4h.html\#conv2ab"}
{https://wise2.ipac.caltech.edu/docs/release/allsky/ \\
expsup/sec4\_4h.html\#conv2ab} 
}. 
For 2MASS $JHK$, the zeropoints are 0.91, 1.39, and 1.85 mag, respectively \footnote{\href{"https://www.astronomy.ohio-state.edu/martini.10/usefuldata.html"}{https://www.astronomy.ohio-state.edu/martini.10/usefuldata.html}}.
These zero-points are used to obtain AB magnitudes from Vega magnitudes and then to derive fluxes using the zero-point flux density of 3631 Jy, where $1 \rm Jy =10^{-23} erg~ s^{-1}~ Hz^{-1}~cm^{-2}$. Using a subsample with an SNR larger than two, this will result in a mean SED of brighter sources.

To investigate the  dependence of SEDs on the accretion process, we select quasars based on reliable measurements of \hb\ and \rfe, with the following criteria: $z<0.75$ , \hb\ $\rm SNR > 2$, broad \hb\ equivalent width $ >0$,  $0\le \rfe\le 3$ and the $\rm SNR > 2$ limit for the 14 bands. In addition, to select a cleaned sample of measurements, the following quality cuts are used \mbox{\citep{Wu2022}}: $f_{\rm \hb}/\sigma_{\rm \hb}>2$, $38 <  \log(L_{\rm \hb}/\ergs) < 48$, and $N_{\rm pix, \hb, complex} > 0.5×N_{\rm max}$. The number of selected SDSS DR16 quasars is  $56,969$.  The low-$z$  limit ensures coverage of the  \hb\ emission line, while the SNR limit is crucial for the spectral decompositions of \hb\ and \feii. These criteria enable us to construct a sample of quasars with reliable measurements of  \rfe\ and \mbh, which are used to generate $\rfe$-dependent mean SEDs, as well as mean SEDs dependent on \leddR, \mdot, and $\rm FWHM_{\hb}$.

\section{RESULTS AND DISCUSSIONS} \label{s3}
\subsection{The flux corrections}
When studying the mean SEDs of quasars, several potential contaminations need to be considered. 
These include Galactic extinction, emission lines, absorption by intergalactic hydrogen clouds, host-galaxy contamination, and beaming effects \citep[e.g.,][]{Krawczy2013}. To account for Galactic extinction, we used the Milky Way extinction $E(B - V )$ maps from \cite{Schlegel1998}, and corrected all magnitudes in the  SDSS $ugriz$, 2MASS  $JHK$, and WISE $W1$ bands according to \cite{Fitzpatrick1999}. For GALEX photometric data (NUV and  FUV),  we applied corrections for Galactic extinction using  $A_{\rm NUV}/E(B - V)= 8.741$ and  $A_{\rm FUV}/E(B -  V)=8.376$ \citep{Wyder2005}.
The  $K$-correction, $K(z)$, which adjusts observed magnitudes to intrinsic magnitudes, is defined as $m_{\rm intr}= m_{\rm obs} - K(z)$ \citep{Oke1968}. This correction includes contributions from  the continuum ($K_{\rm cont}$), emission lines ($K_{\rm em}$), and intergalactic medium (IGM) attenuation for higher-$z$ quasars \citep[$K_{\rm IGM}$; redshift-dependent effective optical depth;][]{Meiksin2006,Krawczy2013}.  
For a power-law continuum, $ K_{\rm cont} =-2.5(1+\alpha_{\nu})\log_{10}(1+z)$, where the first term corrects the effective narrowing of the filter width with redshift $z$, and the second term accounts for the slope effect $f_{\nu} \propto \nu^{\alpha_{\nu}}$ (with a canonical $\alpha_{\nu}=-0.5$ for the optical - UV spectrum). 
For $K_{\rm em}$ we adopt identical methodology as presented in \cite{Krawczy2013}. The emission-line template adopts the 13 strongest spectral lines \citep{Vanden2001,Krawczy2013} and a double power-law  continuum ($\alpha_{\nu}=-0.46\ \text{for }\lambda_{\text{rest}}\le4600\,\text{\AA}\quad\text{and}\quad\alpha_{\nu}=-1.58\ \text{for }\lambda_{\text{rest}}>4600\,\text{\AA}$) without correcting for the small blue bump, though its residuals appear in the mean SEDs. The correction is computed as $K_{\rm em} = -2.5 \times {\rm log}(\frac{\int \lambda S_{\lambda}F_{\lambda}(c\And l)d\lambda}{\int \lambda S_{\lambda}F_{\lambda}(c)d\lambda})$, where $S_{\lambda}$ is the transmission filter curve, $F_{\lambda}(c)$ is the continuum, and $F_{\lambda}(c\And l)$ is the continuum with spectral lines. This value is calculated independently for each of the filters. 

To address host contamination, we combine two models: one for higher-luminosity quasars from \cite{Shen2011} and another for lower-luminosity quasars from \cite{Richards2006a}. For high-luminosity quasars, the luminosity fraction of the host galaxy is given by \citep{Shen2011, Ge2016} :
 \begin{equation}
 \label{eq1}
\frac{L_{\rm 5100,host}}{L_{\rm 5100, QSO}}=0.8052-1.5502x+0.9121x^2-0.1577x^3
\end{equation}
where $x=\log (L_{\rm 5100, total}/\ergs)-44$ and $0 \le x< 1.053$.
For low-luminosity quasars, the host fraction is  described by \citep{Richards2006b} :
 \begin{equation}
 \label{eq2}
\log L^{\nu}_{\rm 6156, host}=0.87\log L^{\nu}_{\rm 6156, QSO}+0.2887-\log(L_{\rm Bol}/L_{\rm Edd})
\end{equation}
where  $L^{\nu}_{\rm 6156}$ is in units of $\rm \ergs Hz^{-1}$. Following \cite{Richards2006a}, we assume $L_{\rm Bol}/L_{\rm Edd}$ to be unity, which provides the minimum correction needed. This sets the relative scaling at $6156$ \AA.  We then extend this to all wavelengths using the elliptical galaxy template from \cite{Fioc1997}. To ensure a smooth transition between these two methods, we chose the crossover luminosity where  both methods agree, which is at  $\log (L_{\rm 5100,total}/\ergs)=44.75$. 

\subsection{Method to generate the mean Spectral Energy Distributions}

\begin{figure*}
\centering
\includegraphics[angle=0,width=6.5in, clip]{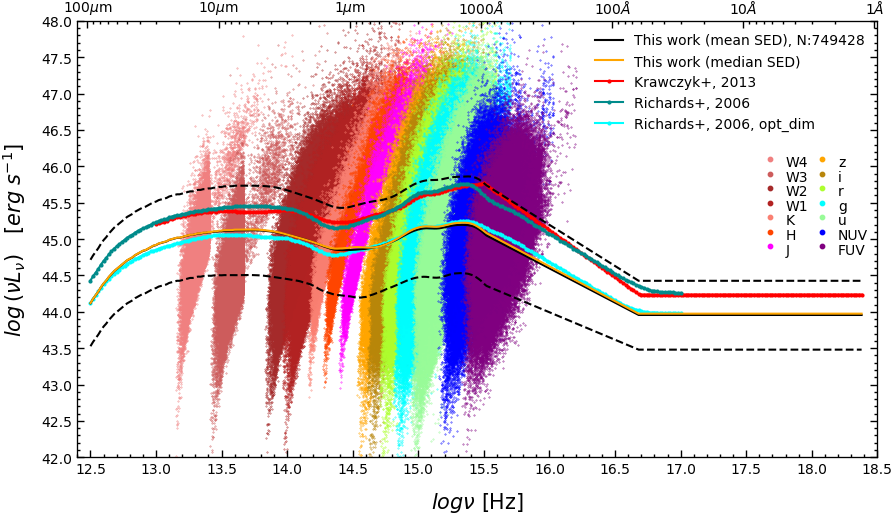}
\caption{
Mean SED (black) and median SED (orange)---both constructed with gap-repaired photometry---together with the underlying data points (without gap repair) from 749,436 SDSS DR16 quasars used to derive them. The dotted lines are 1 $\sigma$ levels. The colors from left to right denote data from WISE $W4,W3,W2$, and $W1$, 2MASS $KHJ$, SDSS $zirgu$, and GALEX NUV and FUV. The mean SEDs from \cite{Richards2006a} (teal represents the full quasar sample; cyan denotes the optical dim subset) and \cite{Krawczy2013} (red) are included for comparison.  
}
\label{fig1}
\end{figure*}

To determine the mean quasar SEDs, we first convert all the flux densities for each quasar to luminosity densities and shift each broadband observation to the rest frame. 
We employ a gap repair method, similar to that described in \cite{Richards2006a}, to derive missing data points. 
Specifically, we replace missing values with those obtained by normalizing the previously constructed SED from \cite{Richards2006a} in the next nearest bandpass for which we have data, the errors are estimated from the relation between the magnitude errors versus magnitudes for each of the filters \citep{Krawczy2013}. 
SED examples for two  individual sources are presented in Appendix \ref{fig_a1}.
For the long-wavelength end, we extrapolate using the mean SED from \cite{Richards2006a}. There is a well-established relation between the monochromatic UV luminosity at 2500 \AA\ ($L_{2500}$)  and the monochromatic X-ray luminosities at 2 keV ($L_{\rm 2~keV}$), $\log (L_{\rm 2 ~keV}) =(0.73\pm 0.05) \log(L_{\rm 2500})+(4.3\pm 1.4)$\citep{Tananbaum1979, Steffen2006, Vasudevan2009, Marchese2012, Lusso2017, Liu2021}. For the X-ray portion of the SED,  we use this relationship and linearly interpolate between the X-ray data point and the bluest data point after gap repair.  The errors in this region are determined from the uncertainties in the relationship. In the X-ray region, the mean SED appears to be flat ($\alpha_{\nu} \sim -1$)  before cutting off at $\sim 500$ keV  \citep{George2000, Zdziarski1995}. The flux at 2 keV is calculated assuming $\alpha_{\nu} = -1$. 
The diversity of the X-ray spectra was also discussed  by others \citep[e.g., ][]{Jin2012, Chen2022}. 
However, previous studies have shown $\alpha_{\nu} $ at 2 keV can vary along the EV1 sequence \citep{Wang1996, Lu1999, Bian2003, Grupe2010}. A softer $\alpha_{\nu} $ (indicating a steeper spectrum) would lead to a smaller bolometric luminosity in narrow-line Seyfert 1 galaxies (NLS1s).
Finally, the data are placed onto a grid with points separated by 0.02 in $\log \nu$. The values on the grid points are derived from the linear interpolation between the 14 rest-frame effective frequencies populated by our gap-repaired photometric data \citep{Richards2006a}. The alternative method of  kriging has been suggested to align the broadband luminosities to the grid points \citep{Krawczy2013}.

At each frequency grid point, we calculate the mean luminosity $\log (\nu L_{\nu})$ and the corresponding 1 $\sigma$ uncertainties (corresponding to the 16th and 84th percentiles of the distribution).
Figure \ref{fig1} shows the raw data points (without gap repair) and the mean SED  (black line, with gap repair) for all SDSS DR16 quasars. The dotted lines indicate the 1 $\sigma$ uncertainties at each grid point.  On the red end of the mean SED, we use the extrapolated mean SED from \cite{Richards2006a}, which shows a drop due to synchrotron  self-absorption with a slope of $\alpha_\nu=2.5$.
On the blue end, shortward of the Ly$\alpha$ line, there is a clear drop due to intergalactic extinction \citep[also see][]{Cai2023}. 
We also compare our mean SED with those from other studies \citep{Richards2006a, Krawczy2013}. The slight differences are attributed to variations in the multibands photometric data used in each sample. 
Mean SED from DR16 quasars includes more low-luminosity sources compared to DR7 quasars \citep{Richards2006a} .
In Figure \ref{fig1} , we also show the mean SED (cyan) of  the optical dim subset from \cite{Richards2006a}, which  is defined as quasars with optical luminosity $\log(L /\ergs)<  46.02$. 
Overall, these mean SEDs are very similar. 
The median SED is also shown in Figure \ref{fig1} (orange line) for comparison.  It  is similar to the mean SED, although with some minor differences, confirming that the results are resistant to outliers.
In this work, we construct  \rfe-dependent SEDs to investigate differences in the mean SEDs  for $z<0.75$.

\subsection{The \rfe\ Dependence of the mean Spectral Energy Distributions}
\begin{figure*}
\centering
\includegraphics[angle=0,width=6in]{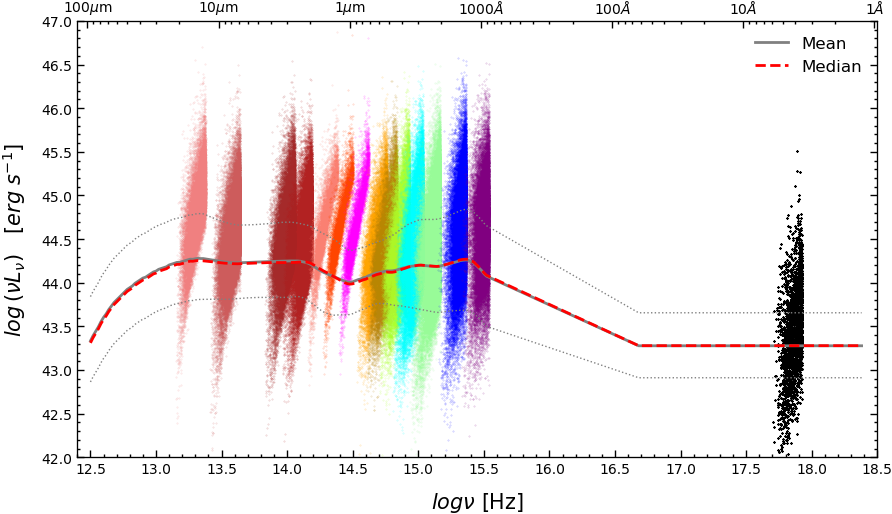}
\includegraphics[angle=0,width=6in]{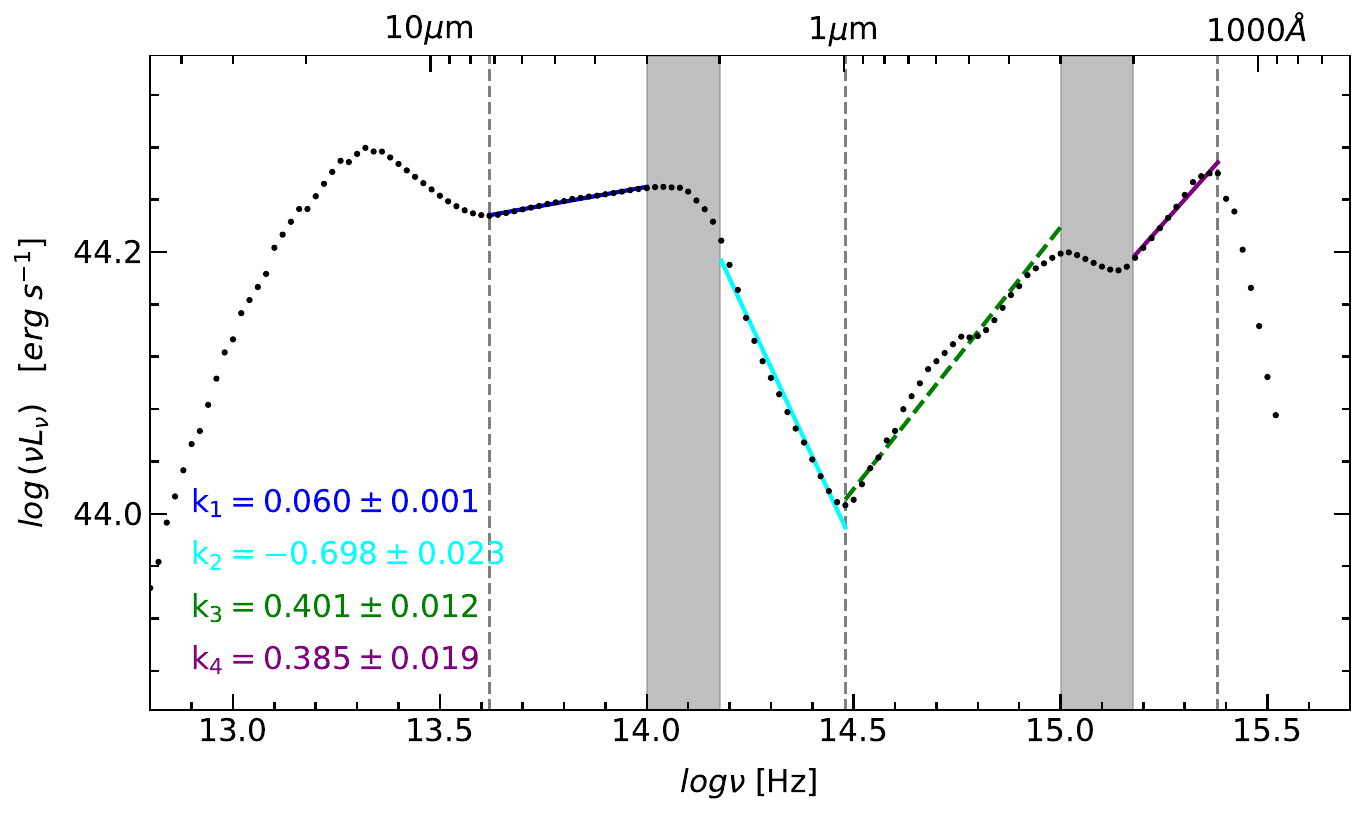}
\caption{Top: mean corrected SED (gray) , median SED (red) and data points  (without gap repair) for  56,969 quasars at $z< 0.75$ and with $0\le \rfe\le 3$. The colors  are the same as Figure \ref{fig1}.  The X-ray luminosities at 2 keV are also shown using a flat spectrum between 2-12 keV from \protect \cite{Lyke2020}. Bottom: zoom-in view of the mean SED about four slopes.  From left to right, the three dotted--dashed vertical lines and two shaded bands denote a $\lambda$ of $8.1,~ 3, ~2$, and $1~\rm \mu m$, as well as $3000,~ 2000$, and $1250~ \text{\AA}$ for the four different regimes. We also show the spectral slopes in MIR, NIR, optical, and UV regimes (blue, cyan, green, and purple, respectively) from the top to bottom in the left corner.}
\label{fig2}
\end{figure*}
Figure \ref{fig2} shows the mean SED for 56,969 quasars with  $z< 0.75$ and $0\le \rfe\le 3$.  
For our sample of quasars with $z<0.75$, the mean SED, which peaks at $\log \nu L_{\nu}= 44.28\ {\rm erg\ s^{-1}}$, is lower than that of the entire SDSS DR16 sample, which peaks at $\log \nu L_{\nu}= 45.20\ {\rm erg\ s^{-1}}$ (Figure \ref{fig1}). 
Two small bumps are  observed in the UV range of the mean SED at approximately  3000 \AA\ and 5000 \AA, likely originating from emissions of the Balmer continuum and \feii, respectively.  
We  fit the SED using  a power-law function  $\nu L_{\nu} =b \nu^{k}$  in four different regimes: MIR  ($3-8.1 ~\rm \mu m$),  NIR ($1-2~\rm  \mu m$),  optical ($3000 ~\text{\AA} -1~\rm \mu m$), and UV ($1250- 2000 ~\text{\AA}$),  obtaining  slopes $k_1,~ k_2,~ k_3$, and $k_4$, respectively.  
We use the standard error of the mean in the fit for four different regimes, defined as $1\sigma/\sqrt{N}$, where $N$ is the number of the sources and $\sigma$ is the standard deviation.
Similar NIR and MIR slope differences have been noted in SEDs derived from Spitzer Space Telescope spectra by other researchers  \citep{Shang2011, Netzer2013, Shi2014}. These studies also identified a dip around  $8 ~\mu m$ and two silicate emission features around 10 and 18 $\rm \mu m$.  These slopes indicate the blueness or hardness of the continuum spectrum.  For the mean SED shown in Figure \ref{fig2}, these four slopes are $k_1=0.060\pm 0.001,~ k_2=-0.698\pm 0.023,~ k_3=0.401\pm 0.012$, and $ k_4=0.385\pm 0.019$.  \cite{Vanden2001} reported an optical slope of $0.56$  for their UV--optical composite spectrum, which is steeper than our value for the low-$z$ sample. This  softer optical spectrum for quasars of lower luminosity is consistent with findings by  others \citep[Figure 12  in][]{Krawczy2013}.

\begin{figure*}
\centering
\includegraphics[angle=0,width=6.5in]{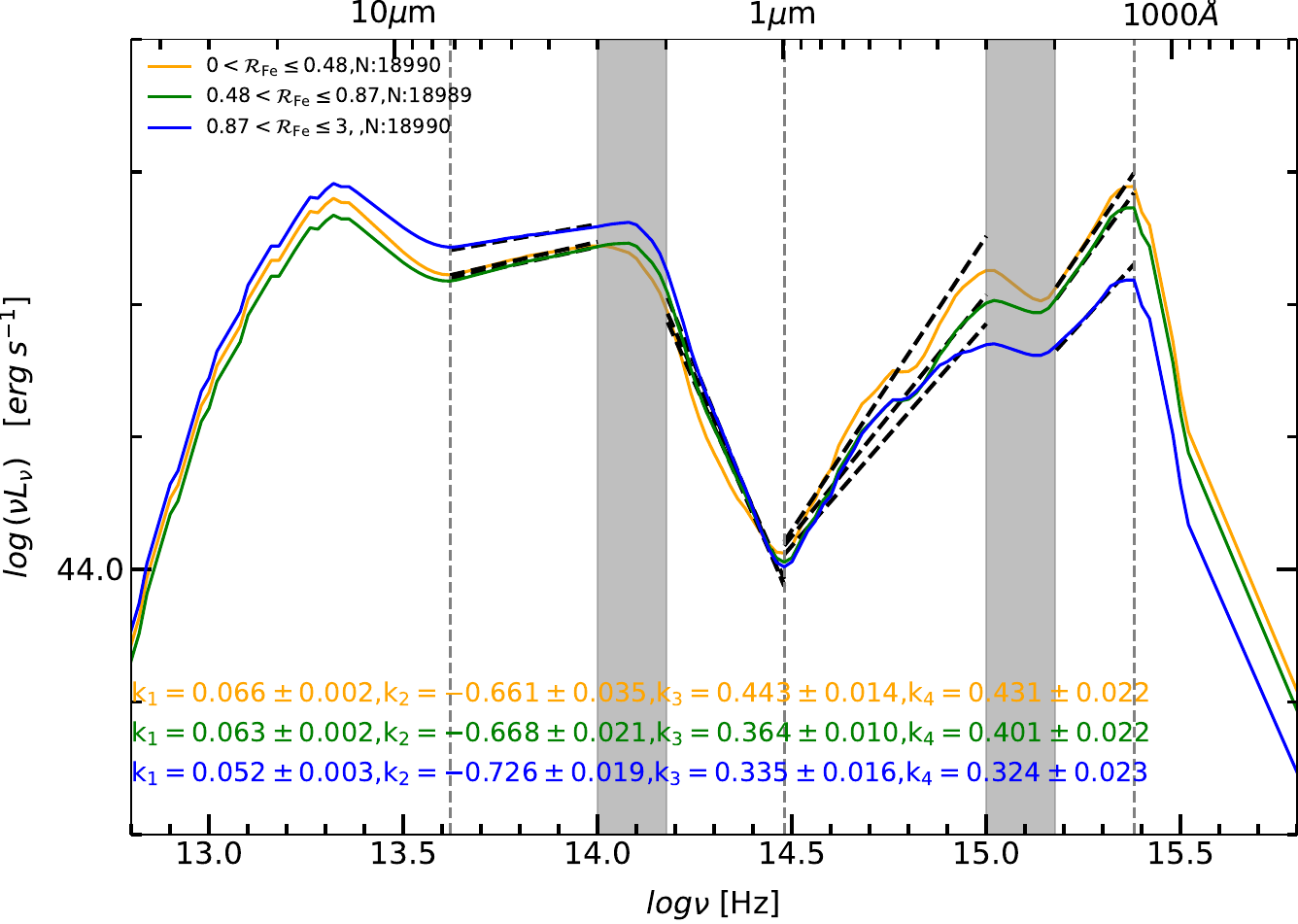}
\caption{The $\rfe$ dependence of the mean SEDs for three \rfe\ bins, i.e.,  [0, 0.48], (0.48, 0.87], (0.87, 3] with the same sample sizes of about 18,990 (orange, green, and blue lines, respectively). At the bottom of the figure, the four slopes in the MIR, NIR, optical and UV are shown for the three  \rfe\ bins. The mean SED with the largest \rfe\ bin shows the reddest UV, NIR, MIR, and optical continuum. Black dash lines show the fit result. Other symbols are the same as Figure \ref{fig2}.}
\label{fig3}
\end{figure*}

Given the strong correlation between \rfe\ and the accretion process \citep{BG92, Boroson2002}, we investigate the SED for different \rfe\ bins.  Based on \rfe, we divide our sample ($z<0.75 $, $0\le \rfe\le 3$) into three subsamples with nearly equal  numbers: $0 \le \rfe\ \le 0.48$, $0.48\le \rfe\ < 0.87$, $0.87\le \rfe\ \le 3$. The SEDs are shown in Figure \ref{fig3}. 
The SEDs with a larger \rfe\ exhibit more radiation at longer wavelengths ($\lambda > 1~\mu m$).  
The UV--optical spectrum is softer  for SEDs with a larger \rfe. 


For the three \rfe\ bins with increasing \rfe\ ($0 \le \rfe\ \le 0.48$, $0.48\le \rfe\ < 0.87$, and $0.87\le \rfe\ \le 3$), the UV slope changes from 0.43 to 0.4 to 0.32. The UV slope is flatter than others for the strongest \rfe\ bin (see Figure \ref{fig3}). For the optical spectrum, the slope changes from 0.44 to 0.36 to 0.33, indicating a reddest optical continuum for the strongest \rfe\ bin. For the NIR spectrum, the slope changes from -0.66 to -0.67 to -0.73, showing a softer NIR continuum with increasing \rfe. Relative to the accretion disk emission,  a  softer NIR continuum indicates stronger NIR emission associated with a more intense accretion process \citep[e.g.,][]{Wang2012}. A similar trend is observed for the MIR continuum along the \rfe.

From \rfe-binned SEDs, we observe that both the UV--optical and NIR--MIR spectra become redder with increasing \rfe. 
Fitting the four spectral slopes for individual-sources in each \rfe\ bin yields results similar to those from the mean SED (see Appendix \ref{app_D}). However, the intrinsic scatter of individual source slope distributions is not considered when analyzing the mean SED slopes.

This trend is consistent with previous suggestions that the NIR continuum slope is a reliable indicator of the relative amount of hot dust emission compared to accretion disk emission \citep{Wang2012, Zhang2014}.  Our findings reveal that quasars with a larger \rfe\ exhibit more pronounced dust emission, implying that the NIR--MIR radiation is stronger for higher accretion rates as traced by \rfe.

\subsection{ The \leddR\ Dependence and \mdot\ Dependence of the mean Spectral Energy Distributions }
\begin{figure*}
\centering
\includegraphics[angle=0,width=5in]{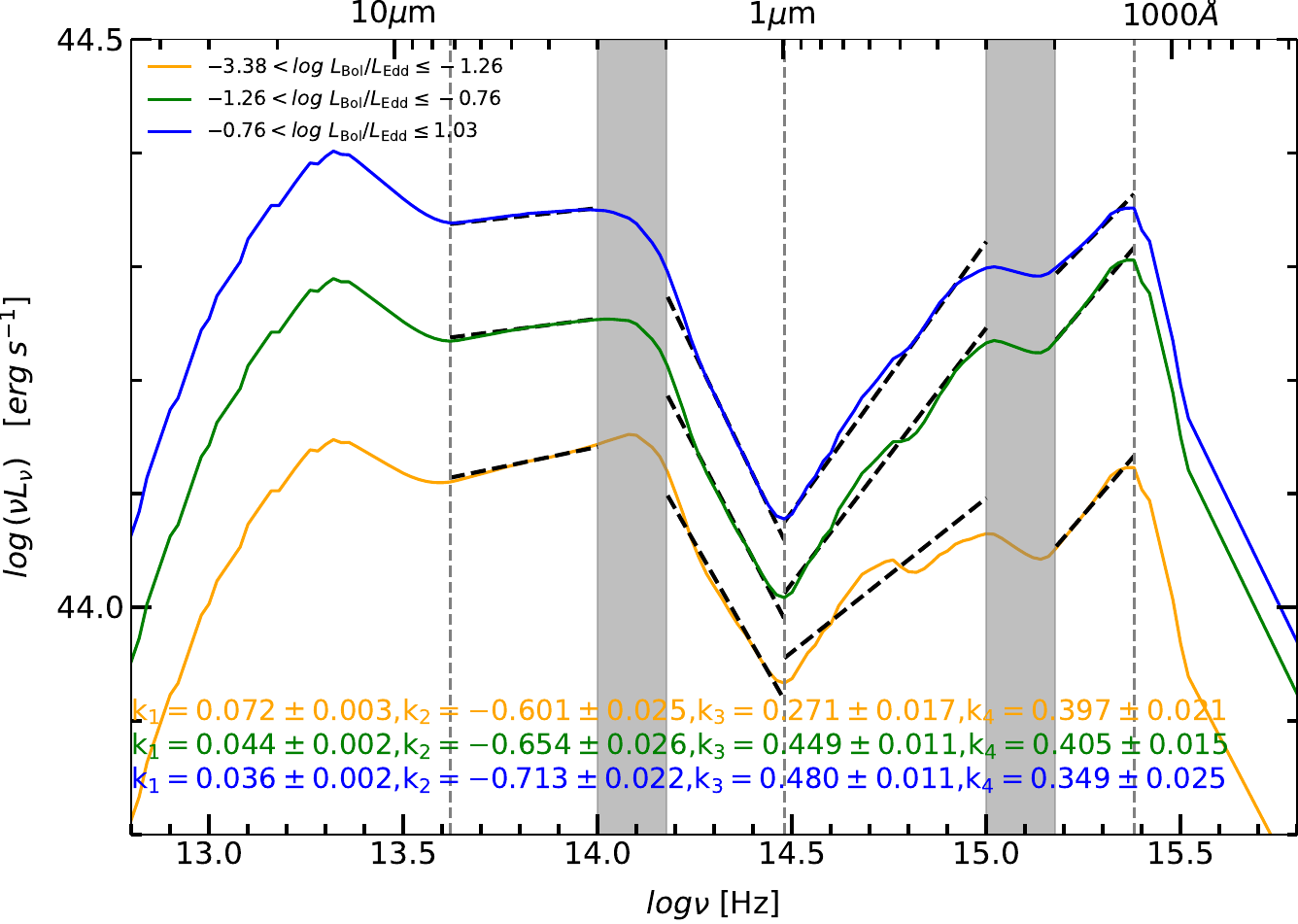}
\includegraphics[angle=0,width=5in]{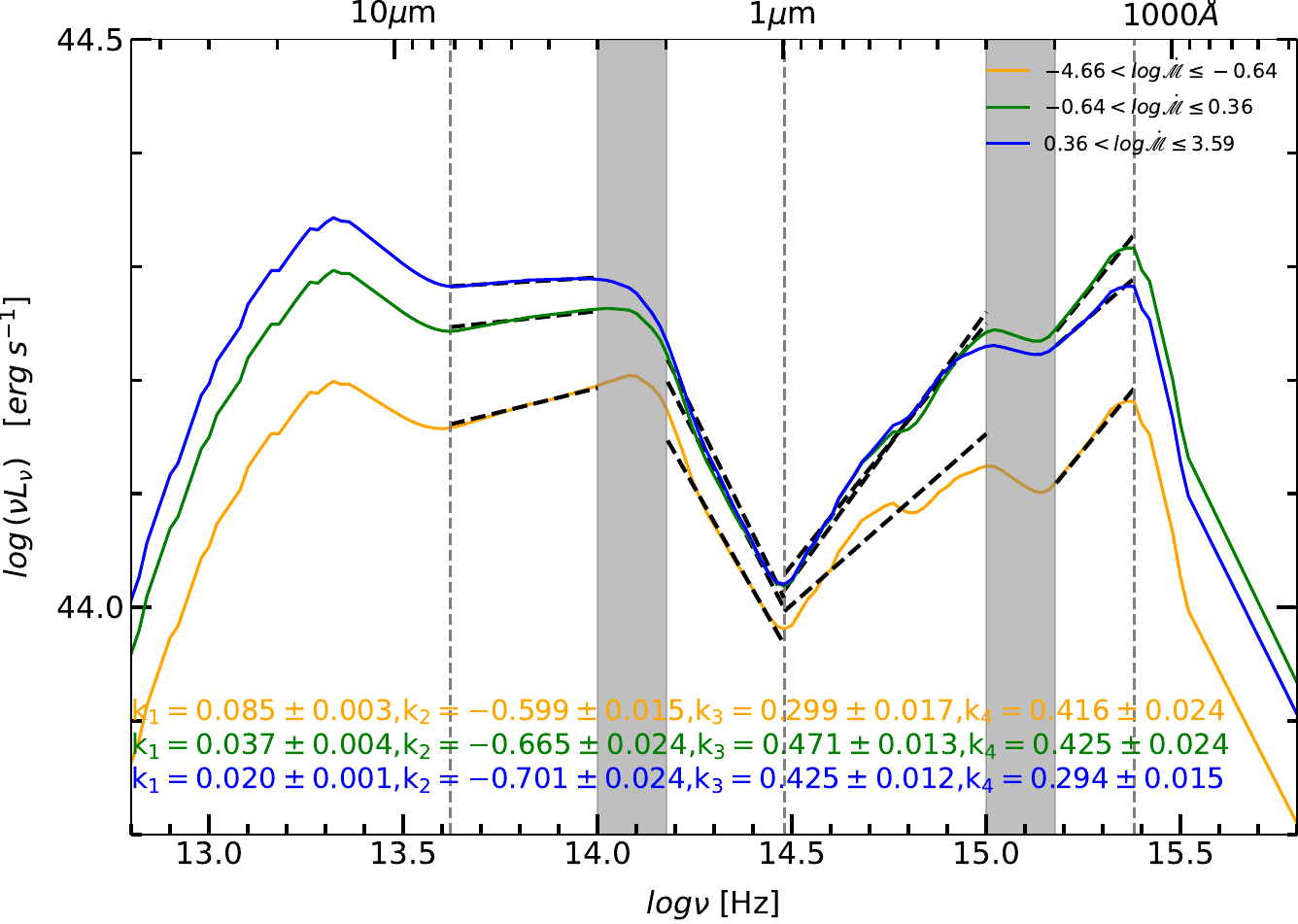}
\caption{Top: the $\leddR$ dependence of the mean SEDs for three \leddR\ bins, i.e.,  (-3.38,-1.26], (-1.26,-0.76], and (-0.76, 1.03] with the same sample sizes of 18,990 (orange, green, and blue lines, respectively). In the bottom, the four slopes in the MIR, NIR, optical, and UV are shown for the three  \leddR\ bins. The SED of the bin with the largest \leddR\ shows the reddest UV, NIR, and MIR  continua and  the bluest optical continuum. Bottom: the $\mdot$ dependence of the mean SEDs for three \mdot\ bins, i.e.,  (-4.66,-0.64], (-0.64, 0.36], and (0.36, 3.59].  The SED of the bin with the largest \mdot\ shows reddest UV, NIR, and MIR continua and  the bluest optical continuum.  Symbols are the same to Figure \ref{fig3}. }
\label{fig4}
\end{figure*}

We not only generate mean SEDs using \rfe\ bins but also calculate \mbh\ and directly determine the accretion rates through two parameters (\leddR\ and \mdot) to generate their mean SEDs based on their respective bins.
Considering the extended empirical \RL\  relation, which includes \rfe\ \citep{Du2019, Yu2020a} and $f=1$, \mbh\ is calculated as follows: $ \log \frac{\mbh}{\msun}=6.96+2\log\frac{\rm FWHM_{\hb}}{1000~\rm km/s}+0.48\log l_{44}-0.38\rfe$ \citep{Liu2022},  where $l_{44}$, $\rm FWHM_{\hb}$ and optical \rfe\  are measured from the single-epoch spectra \citep{Wu2022}. The host contribution at 5100 \AA\ is corrected \citep{Ge2016, Chen2022}. 
We use a correction factor to calculate \lb\ from \lv, $\lb= BC_{\rm 5100} \lv$, where  $BC_{\rm 5100}=53-\log (\lv/\ergs)$ \citep{Marconi2004, Netzer2013}.  The Eddington luminosity is given by $\ledd=1.5\times 10^{38} (\mbh/\msun)~ \ergs$ for a fully ionized solar composition gas with  a mean molecular weight of 1.17 \citep{ Netzer2013}.

Based on \leddR, we divide our sample into three subsamples with equal numbers of samples (18,990): 
$-3.38 < \log \leddR \le -1.26$, $ -1.26 <  \log \leddR \le -0.76$, and $-0.76 < \log  \leddR \le 1.03$. The mean SEDs  for these subsamples are shown in top panel of Figure \ref{fig4}.
We observe that the mean SEDs have higher luminosities with increasing \leddR. Specifically, the UV, NIR, and MIR continua are redder  with increasing  \leddR, while the optical continuum becomes bluer. These findings are consistent with the trends observed in \rfe-binned SEDs except for the optical continuum. The optical slope varies by 0.21 across these subsamples.
 
Based on \mdot, we also divide our sample into three subsamples with nearly equal numbers  of 18,990: $-4.66< \log \mdot\le -0.64, -0.64 < \log \mdot \le 0.36$, and $0.36 < \log \mdot \le 3.59$. A criterion of $ \mdot > 3$  has been suggested  to select super-Eddington accreting massive black holes \citep[SEAMBHs, e.g.,][]{ Du2016}. The third \mdot\ bin here roughly corresponds to quasars with SEAMBHs.  The mean SEDs  for these subsamples  are shown in bottom panel of Figure \ref{fig4}. 
Similar to the \mbox{\leddR} subsamples, the mean SEDs exhibit higher luminosities with increasing \mbox{\mdot}. However, the luminosity difference inverts in the UV band.
The spectral changes with increasing \mdot, as indicated by the changes in the three slopes in the UV,  NIR, and MIR continua, are also consistent with the trends observed in the \rfe-binned SEDs. The differences observed in the optical band may stem from the complexity of the optical wavelength region.

From the theoretical  optical--UV spectrum of the standard thin disk \citep{Wang2014},  the optical--UV spectrum depends on parameters such as \mbh, \leddR, and SMBH spin. Spin significantly impacts the high-energy SED, extending its influence into the NIR--optical for high-mass ($\log (\mbh/\msun)$ $\approx$ 10) and low-accretion-rate (\mdot $\approx$ 0.1) systems. Note that our definition of \mdot\ is ten times that used in \cite{Wang2014}. The flux peak shifts to larger wavelengths with larger \mbh, smaller \mdot, and smaller spin. 
In lower-luminosity quasars, the host galaxy's contribution can significantly contaminate the optical spectrum, complicating comparisons to accretion disk models. For the NIR below 3 $\rm \mu m$ and the MIR around 
$10 ~\rm \mu m$,  contributions come from host stars (peaked in the NIR), AGNs torus (with a polar wind component, peaked in NIR and MIR), very hot dust (peaked in the MIR), and host cold dust \citep[peaked in the far-infrared (FIR);  e.g.,][]{Zhuang2018, Molina2023}. 
The AGN-heated dust emission from the torus contributes a significant fraction ($\sim 70\%$) of the total infrared ($1-1000 ~\mu m$) luminosity. The torus luminosity fraction decreases with larger \leddR, and  the torus opening angle declines with increasing accretion rate until \leddR\ reaches $\sim 0.5$ \citep{Zhuang2018}. 
For UV--soft X-ray regions,  the accretion disk, soft X-ray excess, and tail of the hot corona  make these region complex \citep[e.g., ][]{Jin2012, Hagen2023}.  Considering the SED dependence on the accretion rate, mean SEDs with larger accretion rates (traced by \rfe, \leddR, and \mdot) have redder UV,  NIR and MIR continua. A bluer UV continuum with larger accretion rates can be interpreted by the accretion model,  though it is mixed by the distributions of \mbh\ and spin \citep[e.g.,][]{Wang2014}. 
The bluer optical continuum is likely due to the small blue bump from the Balmer continuum and the optical \mbox{\feii} multiplets \mbox{\citep[e.g.,][]{Vanden2001}}. 
Considering that the host stellar light contribution is corrected, the dip around 1$~\mu \rm m$ would have an effect on the optical-- NIR continua. 
A redder NIR--MIR continuum indicates that  quasars with larger accretion rates have more hot dust emission relative to  the accretion disk. This may be because the torus opening angle declines and the dusty outflow gets stronger with larger \leddR,  leading to more hot dust emission \citep[e.g.,][]{Zhuang2018, Montalvo2026}.
In the future, we need to investigate the mean SEDs considering the plane of \mbh\ versus \mdot, as well as spin, and compare them with the accretion disk model using both photometric and spectroscopic data. Particularly, this should be tested for  samples  with large $\log (\mbh/\msun)$, such as about 9-10 \citep{Wang2014}. 

\subsection{The FWHM Dependence of the mean Spectral Energy Distributions}
\begin{figure*}
\centering
\includegraphics[angle=0,width=6.5in]{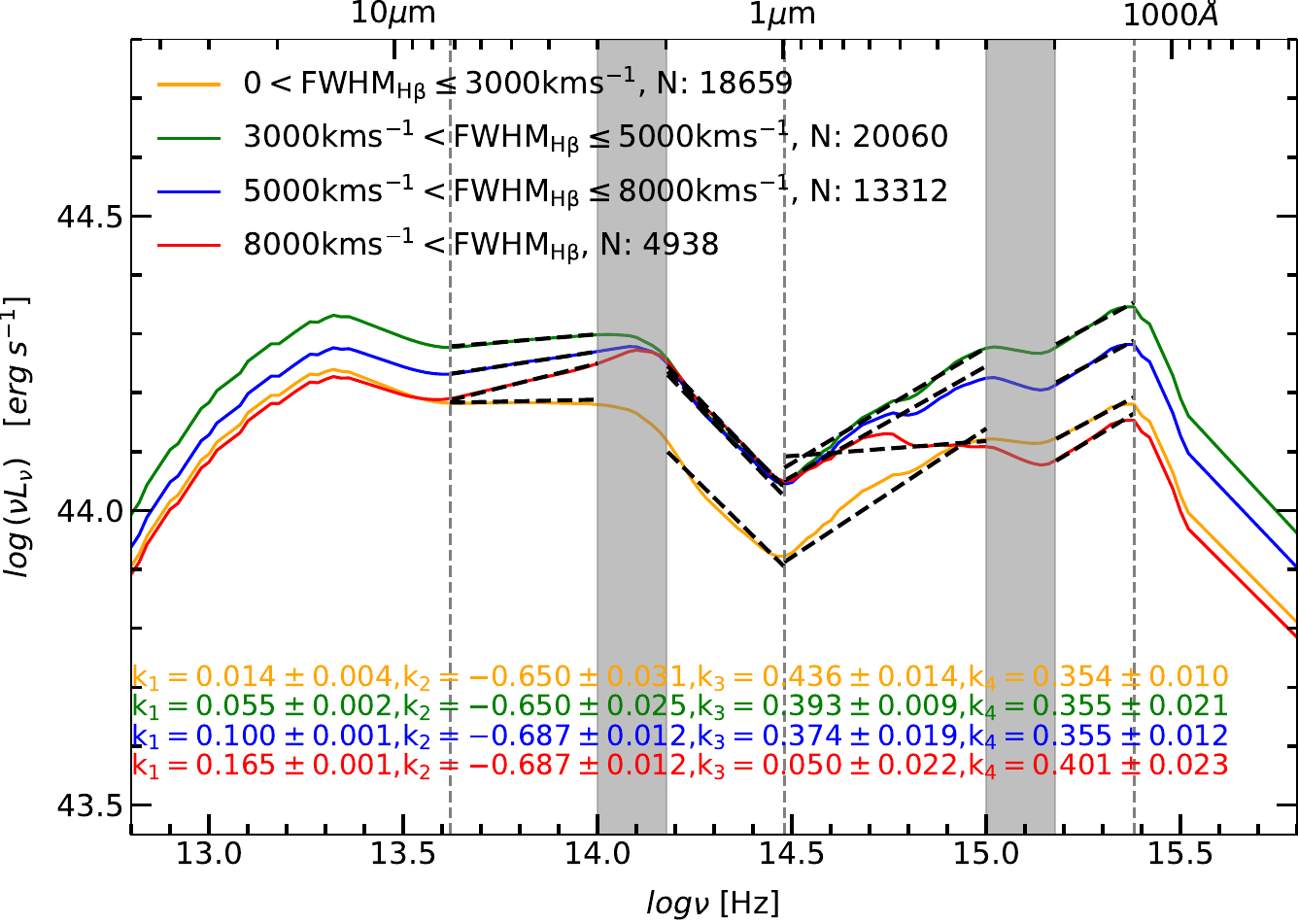}
\caption{ The FWHM dependence of the mean SEDs for four $\rm FWHM_{\hb}$ bins, i.e., (0, 3000], (3000,5000], (5000, 8000], and [8000-)  (km/s) with sample sizes of 18,659, 20,060, 13,312, 4,938 (orange, green, blue,  and red lines, respectively). In the bottom, the four slopes in the MIR, NIR, optical and UV are shown for the four  $\rm FWHM_{\hb}$  bins. The SED of the bin with the  smaller $\rm FWHM_{\hb}$ shows bluer optical and NIR continua and redder UV and MIR continua. Symbols are the same as Figure \ref{fig3}.}
\label{fig5}
\end{figure*}

\cite{Shen2014} proposed that the sequence in \rfe\ is driven by \mbh; but the dispersion in $\rm FWHM_{\hb}$ at fixed \rfe\ is largely due to an orientation effect, as expected in a flattened broad-line region geometry \citep[but also see][]{Panda2018}.  
Following the EV1 plane ($\rm FWHM_{\hb}$ versus \rfe) shown in Figure 1 in \cite{Shen2014}, we divide our sample into four bins based on $\rm FWHM_{\rm \hb}$:  $\rm FWHM_{\rm \hb}< 3000$, $ \rm 3000 \le FWHM_{\rm \hb} < 5000$, $\rm 5000 \le FWHM_{\rm \hb} < 8000$, and $\rm 8000 \le FWHM_{\rm \hb} $ (in units of \kms).
The  first FWHM-bin ($\rm FWHM_{\rm \hb} \le 3000$ \kms) is similar to that  for NLS1s or Population A  AGNs \citep{Bian2003, Sulentic2000, Garnica2025}. The SEDs for four $\rm FWHM_{\hb}$ bins are shown in Figure \ref{fig5}. 
We find that the SED in the max FWHM bin (\mbox{$\rm  FWHM  \ge 8000$ \kms} ) shows a lower luminosity around the UV band. The optical--NIR continuum in this bin is the softest and reddest compared to other FWHM-binned SEDs, while its UV and MIR continuum are conversely the hardest and bluest, consistent with our findings from the \rfe\ bins. With increasing $\rm FWHM_{\rm \hb}$, the UV slope increases  from 0.35 to 0.4, the optical slope decreases from 0.43 to 0.05,  the NIR slope decreases from -0.65 to -0.69, and the MIR slope increases from 0.01 to 0.16.
In a study with  a sample of 51 unobscured Type 1 AGNs,  it was found that the mean SEDs are connected with $\rm FWHM_{\rm \hb}$, specifically, the big blue bump weakens as $\rm FWHM_{\rm \hb}$ increases \citep{Jin2012}.  
The very flat optical--UV spectrum observed in red quasars is similar to that found in quasars with hot dust distributions in quasar winds \citep[e.g., ][]{Rivera2021}.  Red QSOs exhibit significantly brighter radio emission and steeper radio spectral slopes compared to  blue QSOs \citep[e.g., ][]{Glikman2022}. 
According to  the theoretical UV--optical spectrum model \citep{Wang2014}, the  flat optical--UV spectrum may be associated with quasars that have large \mbh\ and small \mdot.

Considering the anti-correlation between $\rm FWHM_{\rm \hb}$ and \rfe\ in the DR16 sample ($r_{\rm s}=-0.33$, $p_{\rm null}<10^{-6}$), there are a few quasars with larger \rfe\ and larger $\rm FWHM_{\rm \hb}$ (only 8.9\% (5,055/56,969) of quasars have high \rfe\ $\geq$ 1.5, and merely 0.9\% (542/56,969) exhibit the combination of broad $\rm FWHM_{H\beta} > 5000$ km s$^{-1}$ and high \rfe\ $\geq$ 1.5). Comparing sources between $\rm FWHM_{H\beta} > 8000$ km s$^{-1}$ and those with $\rm FWHM_{H\beta} < 5000$ km s$^{-1}$, quasars with larger $\rm FWHM_{\rm \hb}$ tend to have smaller \rfe\ and have bluer MIR/UV continua and redder optical/NIR continua. 
The four spectral slopes for individual sources exhibit weak but significant correlations with $\rm FWHM_{\rm \hb}$ (The $p_{\rm null}$ for all bands are less than $10^{−6}$). Despite the anti-correlation between $\rm FWHM_{\rm \hb}$ and \rfe, these relations remain significant when controlling for \rfe\ in a partial correlation analysis.
​

In EV1 plane,  \cite{Shen2014} suggested that the differences of the \oiii\ profiles in the same column cells are due to the velocity scatter derived from  $\rm FWHM_{\hb}$, which is assumed to have an  orientation effect. A larger angle of  the line of sight relative to the BLR plane results in a steeper (harder, bluer) MIR slope, indicating  weaker cold dust emission relative to the accretion disk emission. If the torus is coplanar with the accretion disk, a large inclination angle  would lead  weak torus emission, as suggested by \cite{Zhuang2018}, although the polar wind may have a complex effect on the NIR/MIR emission. Considering the anti-correlation between $\rm FWHM_{\hb}$ and \rfe, it is possible that the hot dust emission is primarily driven by \rfe\ for smaller $\rm FWHM_{\hb}$ values. This suggests that hot dust emission increases with \rfe, as previously suggested by \cite{Shen2014}.

\subsection{Bolometric Corrections}
\begin{figure}
\centering
\includegraphics[angle=0,width=3in]{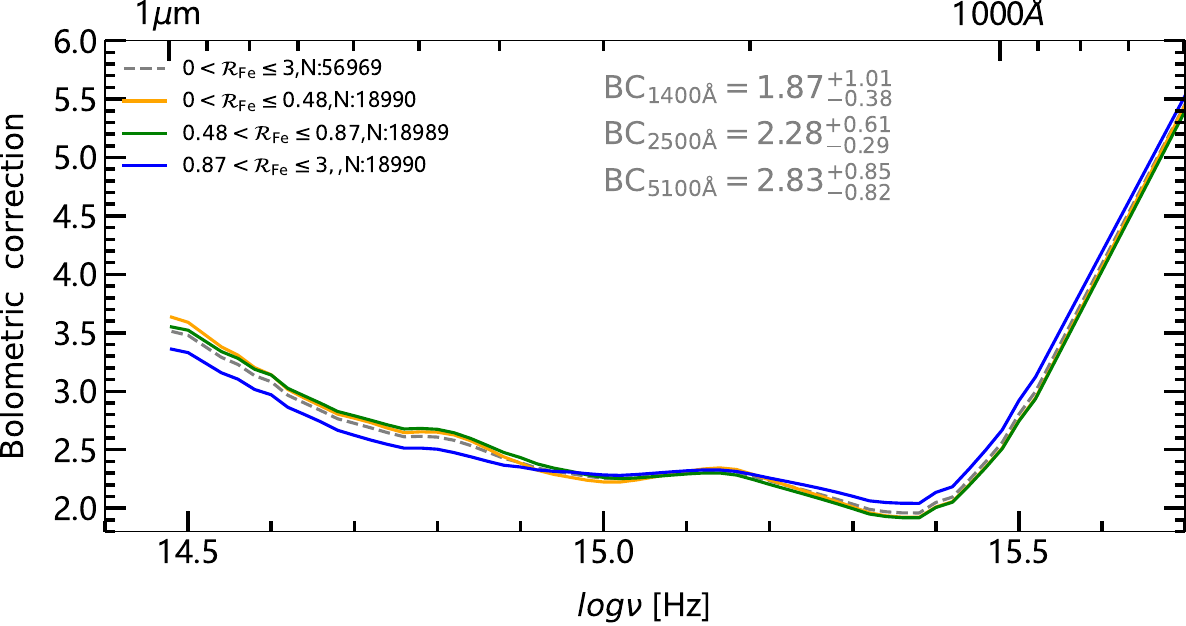}
\includegraphics[angle=0,width=3in]{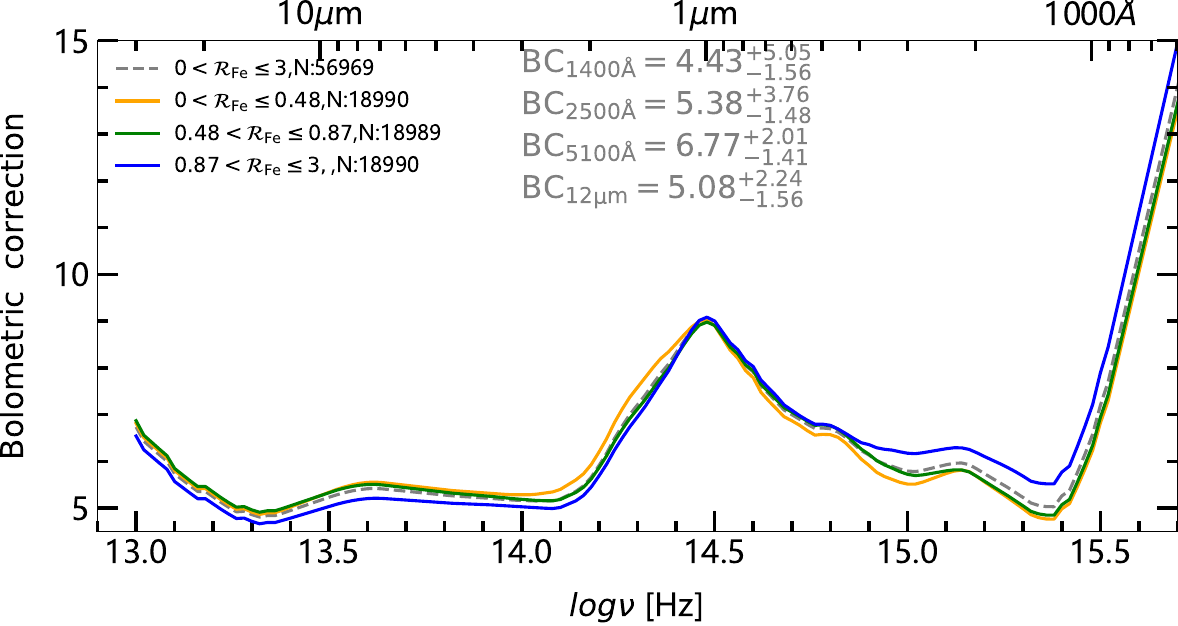}
\caption{Top: BCs from the integration limits of $\rm 1~ \mu m - 2~ keV$ as a function of frequency for the \rfe-binned SEDs in Figure \ref{fig3}. The color coding is the same as for Figure \ref{fig3} , where the gray dashed line shows SED for the total sample of  56,969 quasars in Figure \ref{fig2} . Bottom: BCs from the integration limits of   $\rm 30~ \mu m - 10~keV$. }
\label{fig6}
\end{figure}

\begin{figure}
\centering
\includegraphics[angle=0,width=3in]{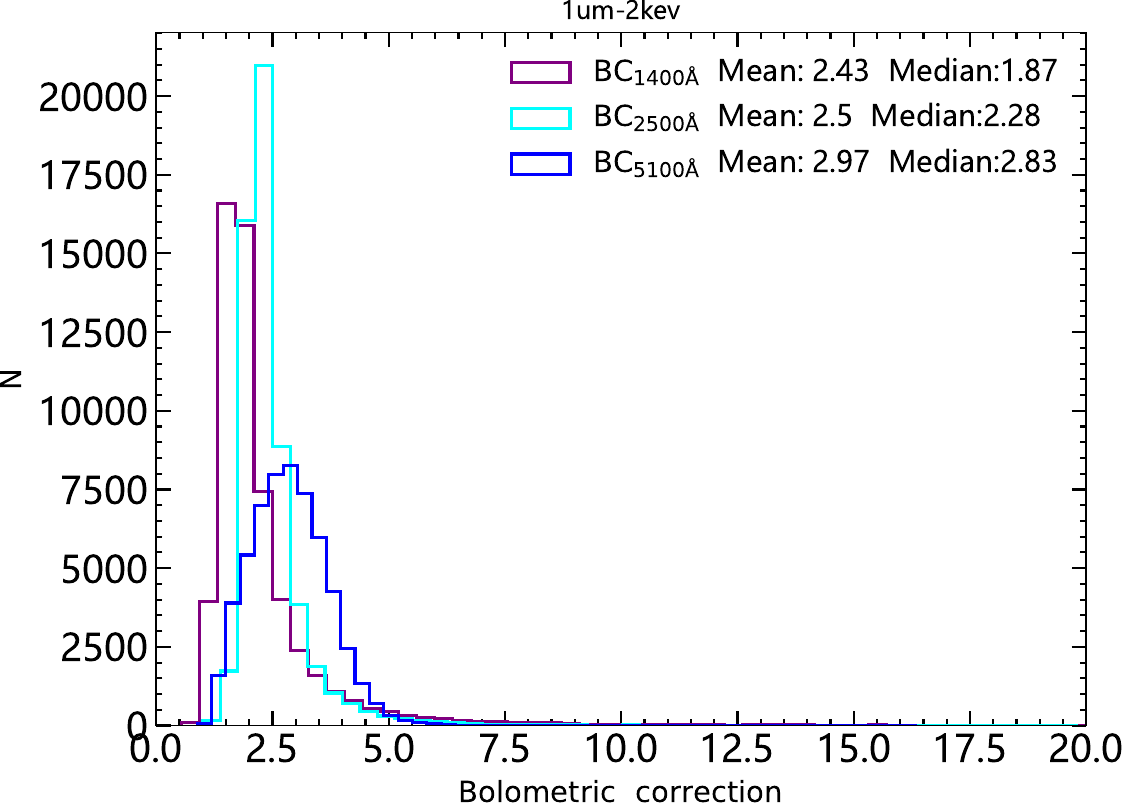}
\includegraphics[angle=0,width=3in]{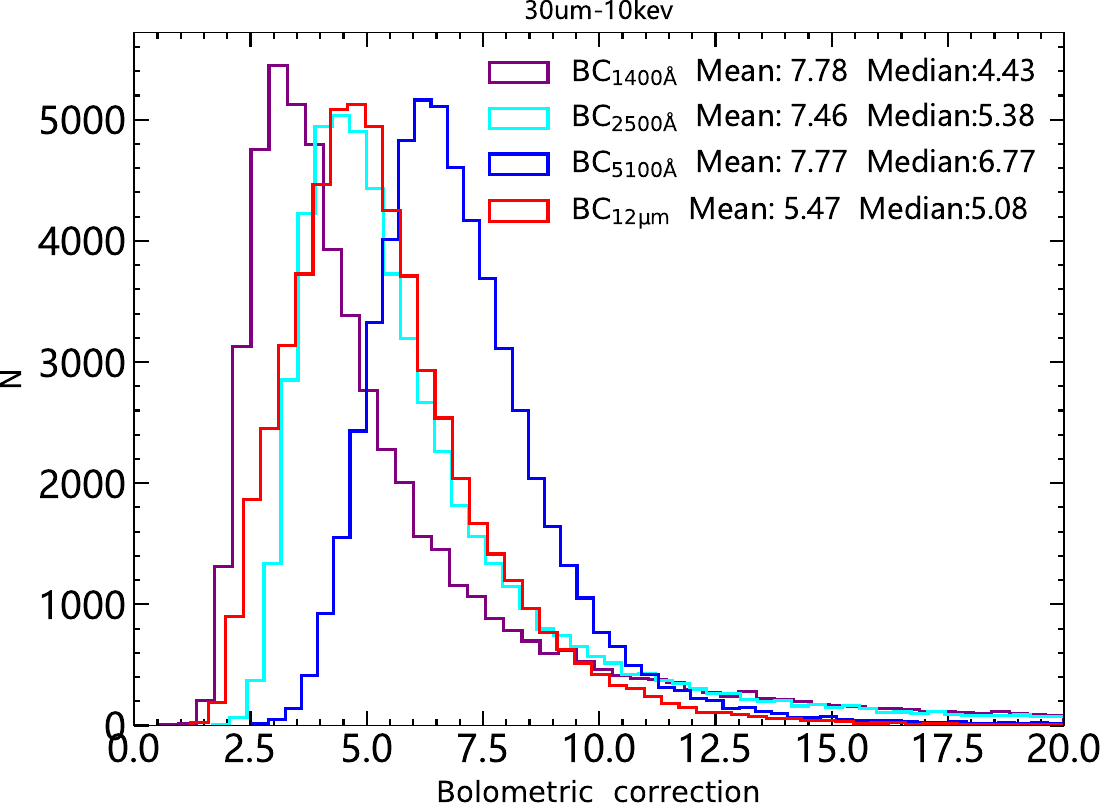}
\caption{Top: histograms of BCs for individual quasars from the integration limits of $\rm 1~ \mu m- 2 ~keV$ at 1400 \AA, 2500 \AA, and 5100 \AA\ (purple, cyan, and blue, respectively).  Bottom: histogram of BCs for individual quasars from the integration limits of $\rm 30~ \mu m- 10~keV$ at 1400 \AA, 2500 \AA, 5100 \AA\ , and $12~\rm  \mu m$ (purple, cyan, blue, and red, respectively).}
\label{fig7}
\end{figure}

\begin{deluxetable}{cc|ccc|cccc}

\tablecolumns{9}
\renewcommand\arraystretch{1.5}
\tabletypesize{\scriptsize}
\setlength{\tabcolsep}{3pt}
\tablewidth{2pt}
\tablecaption{The median and 1 $\sigma$ uncertainties of the Bolometric Corrections Distribution for individual quasars in Different \rfe\ Bins  from the integration limits of $\rm 1 ~\mu m- 2~keV$ and $\rm 30~ \mu m- 10 ~keV$.
\label{tab:1}}
\tablehead{
\multicolumn{2}{c}{} & \multicolumn{3}{c}{$\rm 1~ \mu m- 2~ keV$}  &\multicolumn{4}{c}{$\rm 30~ \mu m- 10 ~keV$} \\
\hline
\rfe-binned range     &$N$ & $\rm BC_{5100}$  & $\rm BC_{2500}$ & $\rm BC_{1400}$ &  $\rm BC_{12 \mu m}$ & $\rm BC_{5100}$ & $\rm BC_{2500}$ & $\rm BC_{1400}$
}

\startdata
$[0, 3]$ & 56,969 & $2.83^{+0.58}_{-0.82}$ & $2.28^{+0.61}_{-0.29}$ &    $1.87^{+1.01}_{-0.38}$ & $5.08^{+2.24}_{-1.56}$ & $6.77^{+2.01}_{-1.41}$ &  $5.38^{+3.76}_{-1.48}$ &  $4.43^{+5.05}_{-1.56}$  \\
\hline
$[0, 0.48]$ & 18,990 & $2.87^{+0.9}_{-0.85}$ & $2.29^{+0.6}_{-0.3}$ &    $1.84^{+1.01}_{-0.4}$ &  $5.19^{+2.47}_{-1.66}$ &  $6.63^{+2.01}_{-1.37}$ &  $5.21^{+3.82}_{-1.46}$ &  $4.22^{+5.13}_{-1.53}$\\
$(0.48, 0.87]$ &18,989 & $2.89^{+0.86}_{-0.82}$ & $2.27^{+0.59}_{-0.28}$ &    $1.84^{+0.92}_{-0.36}$ & $5.18^{+2.3}_{-1.58}$ &  $6.86^{+1.95}_{-1.4}$ &  $5.27^{+3.49}_{-1.39}$ &  $4.29^{+4.48}_{-1.45}$       \\
$(0.87, 3]$ &18,990 & $2.72^{+0.78}_{-0.77}$ & $2.29^{+0.62}_{-0.28}$ &    $1.93^{+1.11}_{-0.38}$ & $4.89^{+2.03}_{-1.46}$ &  $6.82^{+2.04}_{-1.46}$ &  $5.68^{+3.91}_{-1.56}$ &  $4.78^{+5.51}_{-1.64}$       \\
\enddata
\end{deluxetable}

The BC is defined as ${\rm BC}_{\lambda}=L_{\rm Bol}/(\lambda L_\lambda)$, where $\lambda L_{\lambda}$ is the monochromatic luminosity at wavelength $\lambda$. 
There are different integration wavelength limits to derive \lb, such as $\rm 1~\mu m-500~ keV$ \citep{Marconi2004}, $\rm 100~\mu  m - 10~ keV$ \citep{Richards2006a}, $\rm 1~ \mu  m-2 ~keV$ \citep{Krawczy2013}.  The choice of integration limits depends on whether to double count the  torus and/or corona  emission, which arises from the reprocessing of accretion disk photons \citep{Marconi2004, Krawczy2013}. In this section, we calculate BCs using two different integration limits to investigate their dependence on \rfe.

Using the integration limits of $\rm 1~ \mu m-2~keV$ (which avoids double counting torus emission),  we show the BC versus the frequency $\nu$ in top panel in Figure \ref{fig6}. We investigate  the dependence of  $\rm BC_{\lambda}$ on the SEDs for different \rfe\ bins in the optical--UV bands. 
It is found that, for the largest \rfe\ -bin ($0.87 < \rfe\ \le 3$), $\rm BC_{5100}$  is the smallest at 2.72, while $\rm BC_{1400}$ in the UV is the largest at 1.93.
For each quasar in our sample, we calculate BCs using the same integration limits and present  their distributions in the top panel of Figure \ref{fig7}. 
No clear trend between BCs and \rfe\ can be derived from the values  considering their 1 $\sigma$ uncertainties.
These results are summarized in Table \ref{tab:1}.  For the DR16 sample of 56,969 quasars, we find that  $\rm BC_{5100}=2.83^{+0.58}_{-0.82}$, $\rm BC_{2500}=2.28^{+0.61}_{-0.29}$, $\rm BC_{1400}=1.87^{+1.01}_{-0.38}$. With the same integration limits, \cite{Krawczy2013} found that  $\rm BC_{5100}=4.33\pm 1.29$ and $\rm BC_{2500}=2.75\pm 0.40$. Our results are consistent with theirs within the uncertainties.  

Using the integration limits of $\rm 30~ \mu m-10~keV$, we show the BC versus the frequency $\nu$ in the bottom panel in Figure \ref{fig6}.  
For each quasar in our sample, we calculate the BCs using the same integration limits and present  their distributions in the bottom panel of Figure \ref{fig7}. 
We find $\rm BC_{5100}=6.77^{+2.01}_{-1.41}$, $\rm BC_{2500}=5.38^{+3.76}_{-1.48}$, $\rm BC_{1400}=4.43^{+5.05}_{-1.56}$, and $\rm BC_{12~\mu m}=5.08^{+2.24}_{-1.56}$. The ratio $\rm BC_{12 \mu m} / BC_{5100}=0.75$ is consistent with the value of 0.8 suggested by \cite{Netzer2013}.  
Given the large scatter in BC distributions ($\sim$ 0.7 for $\rm 1 ~\mu m -2 ~keV$ and $\sim$ 1.7 for  $\rm 30 ~\mu m-10~ keV$), adopting a single mean quasar SED to compute the bolometric luminosity from an optical luminosity can introduce errors of 25\% in $\rm BC_{5100}$. 
Using  integration limits of $\rm 100 ~ \mu m-10~ keV$, \cite{Elvis1994} and \cite{Richards2006a}  found $\rm BC_{5100}$ to be $11.8^{+12.9}_{-6.3}$ and $10.3 \pm 2.1$, respectively. These larger values are attributed to the inclusion of more radiation within the broader wavelength coverage.  

\begin{figure}
\centering
\includegraphics[angle=0,width=3in]{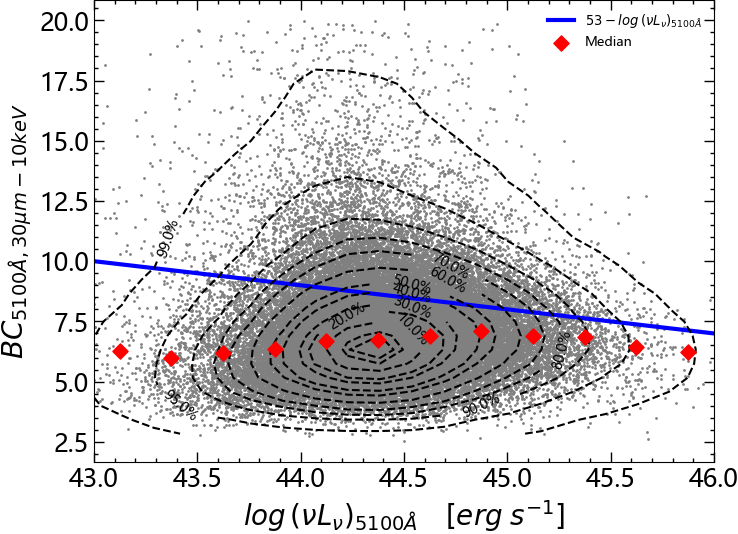}
\caption{ Relation between $\rm BC_{5100}$ and \lv. The blue line shows the luminosity-independent BC that assuming a linear dependence $\mathrm{BC}_{5100} = 53 - \log(L_{5100})$ \citep{Netzer2013}. Grey points show every source in DR16, black dashed curves are density contours, and red diamonds mark the medians.}
\label{fig_c1}
\end{figure}

\begin{figure}
\centering
\includegraphics[angle=0,width=3in]{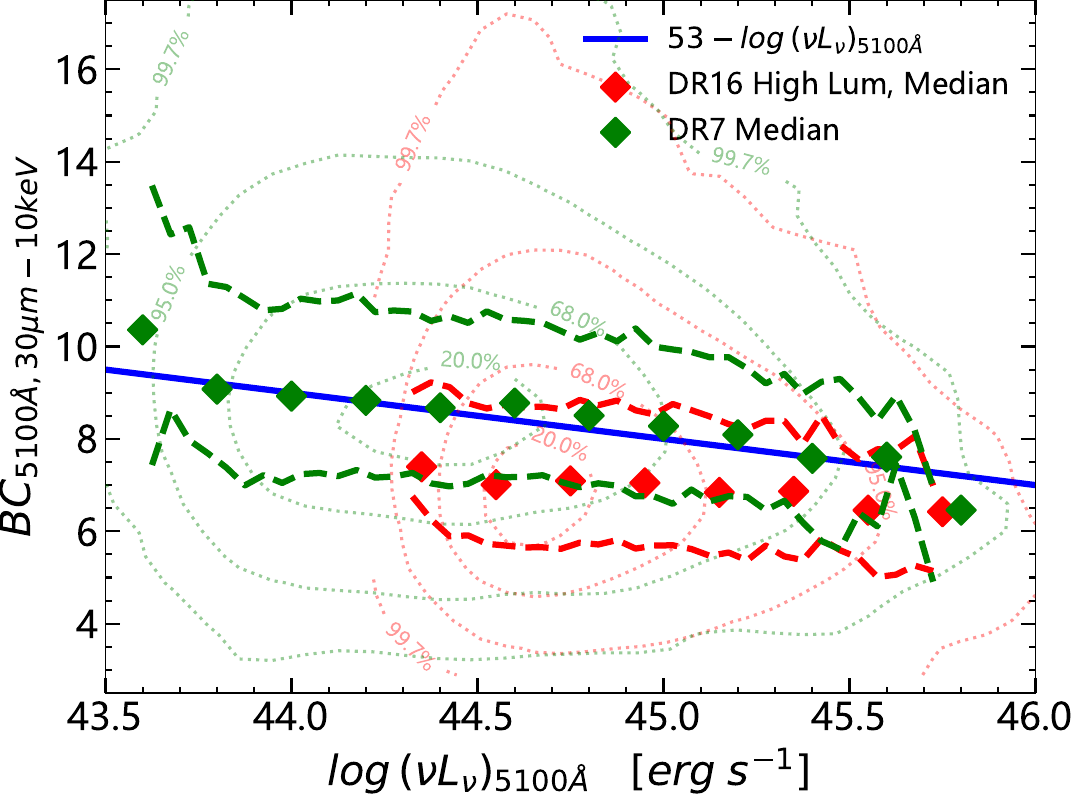}
\caption{ Comparison of the $\rm BC_{5100}$ and \lv\ relation between DR7 and DR16 high-luminosity samples. The blue line shows $\mathrm{BC}_{5100} = 53 - \log(L_{5100})$ \citep{Netzer2013}. Red diamonds indicate the median values for the DR16 high-luminosity sample, while green diamonds show the median values for the DR7 sample. The contour lines represent the source density.}
\label{fig_d1}
\end{figure}

For the 56,969 DR16 sample ($z<0.75$, $0<\rfe\ \le 3$), our present $\rm BC_{5100}$ and \lv\ relation in Figure \ref{fig_c1} with a Spearman correlation coefficient $r_s=0.08$ and a null hypothesis probability $p_{\rm null}=1.07\times 10^{-89}$. It is different from previous work by \cite{Netzer2013}.
We split the sample into three equally populated luminosity bins according $\log(\nu L_{\nu})$ at 2500 \AA\  (see Figure \ref{fig_b1} in Appendix \ref{app_B} ). 
For the high-luminosity subsample in Figure \ref{fig_d1}, we find a relation between $\rm BC_{5100}$ and \lv, with $r_s=-0.06$ and  $p_{\rm null}=4.08\times 10^{-16}$. The median value is consistent with the result reported by \cite{Netzer2013}. For a detailed discussion on the luminosity dependence of the mean SEDs and the relation between $\rm BC_{5100}$ and \lv\ in different luminosity bins, please refer to Appendix \ref{app_B} and Appendix \ref{app_C}. 
SDSS DR16 quasar catalog contains 7 times as many quasars as DR7, while covering broad ranges in luminosity ($44< \rm{log}(L_{\rm bol}/erg\ s^{-1})<48$), and probes lower luminosities than SDSS DR7 quasars ($45< \log (L_{\rm bol}/\rm erg\ s^{-1})<48$) \citep{Wu2022}. 
We also compare the result with DR7 in Figure \ref{fig_d1}. The DR16 median lies at the 1 $\sigma$ lower bound of DR7, whereas the DR7 median sits at the 1 $\sigma$ upper bound of DR16; the DR7 median itself agrees excellently with the published empirical relation \citep{Netzer2013} .
The DR16 median values lie systematically below the DR7 medians and the empirical relation, especially at higher luminosities. 
For the relation between $\rm BC_{5100}$ and \lv\ for DR7 sample, $r_s=-0.13$ and  $p_{\rm null}=6.4\times 10^{-68}$). 
The decrease in the correlation coefficient for the 56,969 DR16 sample may be related to the particular properties of faint sources ($\log(L_{\rm bol}/\rm erg\ s^{-1})<45.41$, \cite{Krawczy2013}), as the DR16 sample includes a larger number of such sources. For DR16 sample, there are weak but significant correlations between BCs and \rfe, \leddR, \mdot\ with $r_s=0.04, ~0.09$, and $0.07$, respectively, and with $p_{\rm null} < 10^{-6}$.


\section{Conclusions} \label{s4}
For a large sample of 56,969 SDSS DR16 quasars ($z<0.75,~ \rfe \le 3$), multiwavelength data spanning from the MIR to the UV are used to construct mean SEDs based on the eigenvector 1 plane, with rest-frame coverage from $\rm \sim 20~ \mu m$ to $\rm 10~ keV$.
We use \rfe, \leddR, and \mdot\ to trace the accretion rate and  the \hb\ line width to  trace the orientation effect. The main conclusions are summarized as follows:   
\begin{itemize}
\item For our sample of  56,969 SDSS DR16 quasars ($z<0.75$) with available \hb\ and \rfe\ measurements,  the mean SED has a redder optical continuum compared to \cite{Vanden2001}. This is consistent with previous findings that low-luminosity SEDs tend to have redder optical continua.

\item  For three \rfe-binned SEDs, it is found that the UV slope depends on \rfe,  with quasars having larger \rfe\ exhibiting redder UV continua.  Additionally,  high-\mbox{\rfe} quasars show signs of redder optical continua and more hot dust emission compared to low-\mbox{\rfe} quasars. 
We also split our sample directly into  \leddR\ (or  \mdot) bins to construct different mean SEDs and find that the continua become increasingly red with increasing \leddR\ (or \mdot) in the MIR, NIR, and UV bands. It demonstrates that the shape of Type 1 AGN SEDs depends on the accretion rate.  However, the optical continuum shows the opposite trend (becoming harder and bluer), indicating the complexity of the optical emission region.

\item  For the four $\rm FWHM_{\hb}$-binned SEDs, the UV and MIR slopes become larger (bluer continua), while the NIR and optical slopes become smaller (redder continua) with increasing  $\rm FWHM_{\hb}$.  A larger angle of  the line of sight relative to the torus plane results in a steeper (harder and bluer) MIR slope, indicating weaker cold dust emission relative to the accretion disk emission.

\item 
For different integration wavelength limits,  the BCs at several wavelengths are calculation from the mean SEDs. 
We also calculate the BCs for each quasar to investigate the BC distributions and find a large scatter in the BCs. 
Using integration limits of 1 $\mu$m--2 keV, $\rm BC_{5100}$ is $2.83^{+0.58}_{-0.82}$ , and it is  $6.77^{+2.01}_{-1.41}$ when using the integration limits of 30 $\mu$m--10 keV.  Considering the large BCs scatter, the dependence of the BCs at several wavelengths on \rfe\ is not obvious. For the DR16 sample of 56,969 quasars ($z<0.75$), the relation between $\rm BC_{5100}$ and \lv\ becomes weaker compared to the DR7 sample. There are weak but significant correlations between BCs and \rfe, \leddR, and \mdot\ with $r_s=0.04, ~0.09, ~0.07$ and corresponding $p_{\rm null}$ values of $7.3\times 10^{-23}, ~3.0\times 10^{-119}$, and $9.3\times 10^{-79}$, respectively.

\end{itemize}


\section*{Data Availability}
The \cite{Wu2022} data underlying this article, which include the DR16Q data from \cite{Lyke2020}, are available on the website of Y. Shen: \url{http://quasar.astro.illinois. edu/paper_data/DR16Q/}. We used the 2024 May 1 version of this data. Following \cite{Krawczy2013}, we crossmatched the $W3$ and $W4$ bands with the ALLWISE database (available on the SDSS DR16 website) within a $2''.0$ matching radius.


\begin{acknowledgments}
We are grateful to the anonymous referee for their useful comments and suggestions, which helped us to improve the paper. We also thank professor Avery Meiksin for providing helpful advice on the $K$-correction on the IGM. This work has been supported by the National Science Foundation of China (No. 12573008).
\end{acknowledgments}

\bibliographystyle{aasjournal}
\bibliography{ref}

@ARTICLE{Montalvo2026,
       author = {{Montalvo Hernandez}, Miguel A. and {Goulding}, Andy D. and {Greene}, Jenny E.},
        title = "{Obscured at the Core: Evidence for Nuclear Dust in Reddened Type-1 AGN}",
      journal = {arXiv e-prints},
         year = 2026,
        month = apr,
          eid = {arXiv:2604.13031},
        pages = {arXiv:2604.13031},
          doi = {10.48550/arXiv.2604.13031},
archivePrefix = {arXiv},
       eprint = {2604.13031},
 primaryClass = {astro-ph.GA},
       adsurl = {https://ui.adsabs.harvard.edu/abs/2026arXiv260413031M}
}

@ARTICLE{Li2025,
       author = {{Li}, Shao-Jun and {Ning}, Xiang-Wei and {Ma}, Yan-Song and {Tang}, Yi and {Bian}, Wei-Hao},
        title = "{The Virial Factor f of the H{\ensuremath{\beta}} Broad Line for NGC 5548 and NGC 4151}",
      journal = {\apj},
         year = 2025,
        month = aug,
       volume = {988},
       number = {2},
          eid = {273},
        pages = {273},
          doi = {10.3847/1538-4357/adea4b},
archivePrefix = {arXiv},
       eprint = {2507.03948},
 primaryClass = {astro-ph.GA},
       adsurl = {https://ui.adsabs.harvard.edu/abs/2025ApJ...988..273L}
}

@ARTICLE{Yu2020b,
       author = {{Yu}, Li-Ming and {Bian}, Wei-Hao and {Zhang}, Xue-Guang and {Zhao}, Bi-Xuan and {Wang}, Chan and {Ge}, Xue and {Zhu}, Bing-Qian and {Chen}, Yu-Qin},
        title = "{The Supermassive Black Hole Masses of Reverberation-mapped Active Galactic Nuclei}",
      journal = {\apj},
         year = 2020,
        month = oct,
       volume = {901},
       number = {2},
          eid = {133},
        pages = {133},
          doi = {10.3847/1538-4357/abb01e},
archivePrefix = {arXiv},
       eprint = {2008.06623},
 primaryClass = {astro-ph.GA},
       adsurl = {https://ui.adsabs.harvard.edu/abs/2020ApJ...901..133Y}
}

@ARTICLE{Antonucci1993,
       author = {{Antonucci}, Robert},
        title = "{Unified models for active galactic nuclei and quasars.}",
      journal = {\araa},
         year = 1993,
        month = jan,
       volume = {31},
        pages = {473-521},
          doi = {10.1146/annurev.aa.31.090193.002353},
       adsurl = {https://ui.adsabs.harvard.edu/abs/1993ARA&A..31..473A}
}

@ARTICLE{Ahmed2025,
       author = {{Ahmed}, Harum and {Gallagher}, Sarah C. and {Shemmer}, Ohad and {Brotherton}, Michael S. and {Dix}, Cooper and {Parrott}, Leigh and {Richards}, Gordon T.},
        title = "{Exploring the Spectral Energy Distributions of Luminous Broad Absorption Line Quasars at High Redshift}",
      journal = {\apj},
         year = 2025,
        month = jun,
       volume = {985},
       number = {2},
          eid = {207},
        pages = {207},
          doi = {10.3847/1538-4357/adca38},
archivePrefix = {arXiv},
       eprint = {2504.03959},
 primaryClass = {astro-ph.GA},
       adsurl = {https://ui.adsabs.harvard.edu/abs/2025ApJ...985..207A}
}

@ARTICLE{Cai2023,
       author = {{Cai}, Zhen-Yi and {Wang}, Jun-Xian},
        title = "{A universal average spectral energy distribution for quasars from the optical to the extreme ultraviolet}",
      journal = {Nature Astronomy},
         year = 2023,
        month = dec,
       volume = {7},
        pages = {1506-1516},
          doi = {10.1038/s41550-023-02088-5},
archivePrefix = {arXiv},
       eprint = {2309.01541},
 primaryClass = {astro-ph.GA},
       adsurl = {https://ui.adsabs.harvard.edu/abs/2023NatAs...7.1506C}
}

@ARTICLE{Chen2025,
       author = {{Chen}, Jie and {Jiang}, Linhua and {Sun}, Shengxiu and {Zhang}, Zijian and {Sun}, Mouyuan},
        title = "{Estimating Bolometric Luminosities of Type 1 Quasars with Self-organizing Maps}",
      journal = {\apj},
         year = 2025,
        month = aug,
       volume = {988},
       number = {2},
          eid = {204},
        pages = {204},
          doi = {10.3847/1538-4357/ade307},
archivePrefix = {arXiv},
       eprint = {2506.04329},
 primaryClass = {astro-ph.CO},
       adsurl = {https://ui.adsabs.harvard.edu/abs/2025ApJ...988..204C}
}

@ARTICLE{Garnica2025,
       author = {{Garnica}, Karla and {Dultzin}, Deborah and {Marziani}, Paola and {Panda}, Swayamtrupta},
        title = "{The spectral energy distribution of extreme population A quasars}",
      journal = {\mnras},
         year = 2025,
        month = jul,
       volume = {540},
       number = {4},
        pages = {3289-3310},
          doi = {10.1093/mnras/staf862},
archivePrefix = {arXiv},
       eprint = {2505.22912},
 primaryClass = {astro-ph.GA},
       adsurl = {https://ui.adsabs.harvard.edu/abs/2025MNRAS.540.3289G}
}

@ARTICLE{Azadi2022,
       author = {{Azadi}, Mojegan and {Wilkes}, Belinda and {Kuraszkiewicz}, Joanna and {Ashby}, Matthew and {Willner}, S.~P.},
        title = "{The Bolometric Luminosity Correction of Radio-Quiet and Radio-Loud Quasars at 1<z<2}",
      journal = {arXiv e-prints},
         year = 2022,
        month = apr,
          eid = {arXiv:2204.12697},
        pages = {arXiv:2204.12697},
          doi = {10.48550/arXiv.2204.12697},
archivePrefix = {arXiv},
       eprint = {2204.12697},
 primaryClass = {astro-ph.GA},
       adsurl = {https://ui.adsabs.harvard.edu/abs/2022arXiv220412697A}
}

@ARTICLE{Becker1995,
       author = {{Becker}, Robert H. and {White}, Richard L. and {Helfand}, David J.},
        title = "{The FIRST Survey: Faint Images of the Radio Sky at Twenty Centimeters}",
      journal = {\apj},
         year = 1995,
        month = sep,
       volume = {450},
        pages = {559},
          doi = {10.1086/176166},
       adsurl = {https://ui.adsabs.harvard.edu/abs/1995ApJ...450..559B}
}

@ARTICLE{Bian2003,
       author = {{Bian}, W. and {Zhao}, Y.},
        title = "{On X-ray variability in narrow-line and broad-line active galactic nuclei}",
      journal = {\mnras},
         year = 2003,
        month = jul,
       volume = {343},
       number = {1},
        pages = {164-168},
          doi = {10.1046/j.1365-8711.2003.06650.x},
archivePrefix = {arXiv},
       eprint = {astro-ph/0303546},
 primaryClass = {astro-ph},
       adsurl = {https://ui.adsabs.harvard.edu/abs/2003MNRAS.343..164B}
}

@ARTICLE{Boller2016,
       author = {{Boller}, Th. and {Freyberg}, M.~J. and {Tr{\"u}mper}, J. and {Haberl}, F. and {Voges}, W. and {Nandra}, K.},
        title = "{Second ROSAT all-sky survey (2RXS) source catalogue}",
      journal = {\aap},
         year = 2016,
        month = apr,
       volume = {588},
          eid = {A103},
        pages = {A103},
          doi = {10.1051/0004-6361/201525648},
archivePrefix = {arXiv},
       eprint = {1609.09244},
 primaryClass = {astro-ph.HE},
       adsurl = {https://ui.adsabs.harvard.edu/abs/2016A&A...588A.103B}
}

@ARTICLE{BG92,
       author = {{Boroson}, Todd A. and {Green}, Richard F.},
        title = "{The Emission-Line Properties of Low-Redshift Quasi-stellar Objects}",
      journal = {\apjs},
         year = 1992,
        month = may,
       volume = {80},
        pages = {109},
          doi = {10.1086/191661},
       adsurl = {https://ui.adsabs.harvard.edu/abs/1992ApJS...80..109B}
}

@ARTICLE{Boroson2002,
       author = {{Boroson}, Todd A.},
        title = "{Black Hole Mass and Eddington Ratio as Drivers for the Observable Properties of Radio-loud and Radio-quiet QSOs}",
      journal = {\apj},
         year = 2002,
        month = jan,
       volume = {565},
       number = {1},
        pages = {78-85},
          doi = {10.1086/324486},
archivePrefix = {arXiv},
       eprint = {astro-ph/0109317},
 primaryClass = {astro-ph},
       adsurl = {https://ui.adsabs.harvard.edu/abs/2002ApJ...565...78B}
}

@ARTICLE{Chen2022,
       author = {{Chen}, Yu-Qin and {Liu}, Yan-Sheng and {Bian}, Wei-Hao},
        title = "{The {\ensuremath{\sigma}}$_{H{\ensuremath{\beta}} }$-based Dimensionless Accretion Rate and Its Connection with the Corona for AGNs}",
      journal = {\apj},
         year = 2022,
        month = nov,
       volume = {940},
       number = {1},
          eid = {50},
        pages = {50},
          doi = {10.3847/1538-4357/ac947e},
archivePrefix = {arXiv},
       eprint = {2210.01316},
 primaryClass = {astro-ph.GA},
       adsurl = {https://ui.adsabs.harvard.edu/abs/2022ApJ...940...50C}
}

@ARTICLE{Czerny2019,
       author = {{Czerny}, Bo{\.z}ena and {Wang}, Jian-Min and {Du}, Pu and {Hryniewicz}, Krzysztof and {Karas}, Vladimir and {Li}, Yan-Rong and {Panda}, Swayamtrupta and {Sniegowska}, Marzena and {Wildy}, Conor and {Yuan}, Ye-Fei},
        title = "{Interpretation of Departure from the Broad-line Region Scaling in Active Galactic Nuclei}",
      journal = {\apj},
         year = 2019,
        month = jan,
       volume = {870},
       number = {2},
          eid = {84},
        pages = {84},
          doi = {10.3847/1538-4357/aaf396},
archivePrefix = {arXiv},
       eprint = {1811.09559},
 primaryClass = {astro-ph.GA},
       adsurl = {https://ui.adsabs.harvard.edu/abs/2019ApJ...870...84C}
}

@ARTICLE{DU2016,
       author = {{Du}, Pu and {Lu}, Kai-Xing and {Zhang}, Zhi-Xiang and {Huang}, Ying-Ke and {Wang}, Kai and {Hu}, Chen and {Qiu}, Jie and {Li}, Yan-Rong and {Fan}, Xu-Liang and {Fang}, Xiang-Er and {Bai}, Jin-Ming and {Bian}, Wei-Hao and {Yuan}, Ye-Fei and {Ho}, Luis C. and {Wang}, Jian-Min and {SEAMBH Collaboration}},
        title = "{Supermassive Black Holes with High Accretion Rates in Active Galactic Nuclei. V. A New Size-Luminosity Scaling Relation for the Broad-line Region}",
      journal = {\apj},
         year = 2016,
        month = jul,
       volume = {825},
       number = {2},
          eid = {126},
        pages = {126},
          doi = {10.3847/0004-637X/825/2/126},
archivePrefix = {arXiv},
       eprint = {1604.06218},
 primaryClass = {astro-ph.GA},
       adsurl = {https://ui.adsabs.harvard.edu/abs/2016ApJ...825..126D}
}

@ARTICLE{Du2019,
       author = {{Du}, Pu and {Wang}, Jian-Min},
        title = "{The Radius-Luminosity Relationship Depends on Optical Spectra in Active Galactic Nuclei}",
      journal = {\apj},
         year = 2019,
        month = nov,
       volume = {886},
       number = {1},
          eid = {42},
        pages = {42},
          doi = {10.3847/1538-4357/ab4908},
archivePrefix = {arXiv},
       eprint = {1909.06735},
 primaryClass = {astro-ph.GA},
       adsurl = {https://ui.adsabs.harvard.edu/abs/2019ApJ...886...42D}
}

@ARTICLE{Elvis1994,
       author = {{Elvis}, Martin and {Wilkes}, Belinda J. and {McDowell}, Jonathan C. and {Green}, Richard F. and {Bechtold}, Jill and {Willner}, S.~P. and {Oey}, M.~S. and {Polomski}, Elisha and {Cutri}, Roc},
        title = "{Atlas of Quasar Energy Distributions}",
      journal = {\apjs},
         year = 1994,
        month = nov,
       volume = {95},
        pages = {1},
          doi = {10.1086/192093},
       adsurl = {https://ui.adsabs.harvard.edu/abs/1994ApJS...95....1E}
}

@ARTICLE{Fitzpatrick1999,
       author = {{Fitzpatrick}, Edward L.},
        title = "{Correcting for the Effects of Interstellar Extinction}",
      journal = {\pasp},
         year = 1999,
        month = jan,
       volume = {111},
       number = {755},
        pages = {63-75},
          doi = {10.1086/316293},
archivePrefix = {arXiv},
       eprint = {astro-ph/9809387},
 primaryClass = {astro-ph},
       adsurl = {https://ui.adsabs.harvard.edu/abs/1999PASP..111...63F}
}

@ARTICLE{Fioc1997,
       author = {{Fioc}, M. and {Rocca-Volmerange}, B.},
        title = "{PEGASE: a UV to NIR spectral evolution model of galaxies. Application to the calibration of bright galaxy counts.}",
      journal = {\aap},
         year = 1997,
        month = oct,
       volume = {326},
        pages = {950-962},
          doi = {10.48550/arXiv.astro-ph/9707017},
archivePrefix = {arXiv},
       eprint = {astro-ph/9707017},
 primaryClass = {astro-ph},
       adsurl = {https://ui.adsabs.harvard.edu/abs/1997A&A...326..950F}
}

@ARTICLE{Fukugita1996,
       author = {{Fukugita}, M. and {Ichikawa}, T. and {Gunn}, J.~E. and {Doi}, M. and {Shimasaku}, K. and {Schneider}, D.~P.},
        title = "{The Sloan Digital Sky Survey Photometric System}",
      journal = {\aj},
         year = 1996,
        month = apr,
       volume = {111},
        pages = {1748},
          doi = {10.1086/117915},
       adsurl = {https://ui.adsabs.harvard.edu/abs/1996AJ....111.1748F}
}

@ARTICLE{Ge2016,
       author = {{Ge}, Xue and {Bian}, Wei-Hao and {Jiang}, Xiao-Lei and {Liu}, Wen-Shuai and {Wang}, Xiao-Feng},
        title = "{The underlying driver for the C IV Baldwin effect in QSOs with 0 < z < 5}",
      journal = {\mnras},
         year = 2016,
        month = oct,
       volume = {462},
       number = {1},
        pages = {966-976},
          doi = {10.1093/mnras/stw1605},
archivePrefix = {arXiv},
       eprint = {1608.02172},
 primaryClass = {astro-ph.GA},
       adsurl = {https://ui.adsabs.harvard.edu/abs/2016MNRAS.462..966G}
}

@ARTICLE{Hagen2023,
       author = {{Hagen}, Scott and {Done}, Chris},
        title = "{Estimating black hole spin from AGN SED fitting: the impact of general-relativistic ray tracing}",
      journal = {\mnras},
         year = 2023,
        month = nov,
       volume = {525},
       number = {3},
        pages = {3455-3467},
          doi = {10.1093/mnras/stad2499},
archivePrefix = {arXiv},
       eprint = {2304.01253},
 primaryClass = {astro-ph.HE},
       adsurl = {https://ui.adsabs.harvard.edu/abs/2023MNRAS.525.3455H}
}

@ARTICLE{Ho2008,
       author = {{Ho}, L.~C.},
        title = "{Nuclear activity in nearby galaxies.}",
      journal = {\araa},
         year = 2008,
        month = sep,
       volume = {46},
        pages = {475-539},
          doi = {10.1146/annurev.astro.45.051806.110546},
archivePrefix = {arXiv},
       eprint = {0803.2268},
 primaryClass = {astro-ph},
       adsurl = {https://ui.adsabs.harvard.edu/abs/2008ARA&A..46..475H}
}

@ARTICLE{Jin2012,
       author = {{Jin}, Chichuan and {Ward}, Martin and {Done}, Chris and {Gelbord}, Jonathan},
        title = "{A combined optical and X-ray study of unobscured type 1 active galactic nuclei - I. Optical spectra and spectral energy distribution modelling}",
      journal = {\mnras},
         year = 2012,
        month = mar,
       volume = {420},
       number = {3},
        pages = {1825-1847},
          doi = {10.1111/j.1365-2966.2011.19805.x},
archivePrefix = {arXiv},
       eprint = {1109.2069},
 primaryClass = {astro-ph.HE},
       adsurl = {https://ui.adsabs.harvard.edu/abs/2012MNRAS.420.1825J}
}

@ARTICLE{KH2013,
       author = {{Kormendy}, John and {Ho}, Luis C.},
        title = "{Coevolution (Or Not) of Supermassive Black Holes and Host Galaxies}",
      journal = {\araa},
         year = 2013,
        month = aug,
       volume = {51},
       number = {1},
        pages = {511-653},
          doi = {10.1146/annurev-astro-082708-101811},
archivePrefix = {arXiv},
       eprint = {1304.7762},
 primaryClass = {astro-ph.CO},
       adsurl = {https://ui.adsabs.harvard.edu/abs/2013ARA&A..51..511K}
}

@ARTICLE{Kaspi2000,
       author = {{Kaspi}, Shai and {Smith}, Paul S. and {Netzer}, Hagai and {Maoz}, Dan and {Jannuzi}, Buell T. and {Giveon}, Uriel},
        title = "{Reverberation Measurements for 17 Quasars and the Size-Mass-Luminosity Relations in Active Galactic Nuclei}",
      journal = {\apj},
         year = 2000,
        month = apr,
       volume = {533},
       number = {2},
        pages = {631-649},
          doi = {10.1086/308704},
archivePrefix = {arXiv},
       eprint = {astro-ph/9911476},
 primaryClass = {astro-ph},
       adsurl = {https://ui.adsabs.harvard.edu/abs/2000ApJ...533..631K}
}

@ARTICLE{Krawczy2013,
       author = {{Krawczyk}, Coleman M. and {Richards}, Gordon T. and {Mehta}, Sajjan S. and {Vogeley}, Michael S. and {Gallagher}, S.~C. and {Leighly}, Karen M. and {Ross}, Nicholas P. and {Schneider}, Donald P.},
        title = "{Mean Spectral Energy Distributions and Bolometric Corrections for Luminous Quasars}",
      journal = {\apjs},
         year = 2013,
        month = may,
       volume = {206},
       number = {1},
          eid = {4},
        pages = {4},
          doi = {10.1088/0067-0049/206/1/4},
archivePrefix = {arXiv},
       eprint = {1304.5573},
 primaryClass = {astro-ph.CO},
       adsurl = {https://ui.adsabs.harvard.edu/abs/2013ApJS..206....4K}
}

@ARTICLE{Kubota2018,
       author = {{Kubota}, Aya and {Done}, Chris},
        title = "{A physical model of the broad-band continuum of AGN and its implications for the UV/X relation and optical variability}",
      journal = {\mnras},
         year = 2018,
        month = oct,
       volume = {480},
       number = {1},
        pages = {1247-1262},
          doi = {10.1093/mnras/sty1890},
archivePrefix = {arXiv},
       eprint = {1804.00171},
 primaryClass = {astro-ph.HE},
       adsurl = {https://ui.adsabs.harvard.edu/abs/2018MNRAS.480.1247K}
}

@ARTICLE{Gaia2018,
       author = {{Gaia Collaboration} and {Brown}, A.~G.~A. and {Vallenari}, A. and {Prusti}, T. and {de Bruijne}, J.~H.~J. and {Babusiaux}, C. and {Bailer-Jones}, C.~A.~L. and {Biermann}, M. and {Evans}, D.~W. and {Eyer}, L. and {Jansen}, F. and {Jordi}, C. and {Klioner}, S.~A. and {Lammers}, U. and {Lindegren}, L. and {Luri}, X. and {Mignard}, F. and {Panem}, C. and {Pourbaix}, D. and {Randich}, S. and {Sartoretti}, P. and {Siddiqui}, H.~I. and {Soubiran}, C. and {van Leeuwen}, F. and {Walton}, N.~A. and {Arenou}, F. and {Bastian}, U. and {Cropper}, M. and {Drimmel}, R. and {Katz}, D. and {Lattanzi}, M.~G. and {Bakker}, J. and {Cacciari}, C. and {Casta{\~n}eda}, J. and {Chaoul}, L. and {Cheek}, N. and {De Angeli}, F. and {Fabricius}, C. and {Guerra}, R. and {Holl}, B. and {Masana}, E. and {Messineo}, R. and {Mowlavi}, N. and {Nienartowicz}, K. and {Panuzzo}, P. and {Portell}, J. and {Riello}, M. and {Seabroke}, G.~M. and {Tanga}, P. and {Th{\'e}venin}, F. and {Gracia-Abril}, G. and {Comoretto}, G. and {Garcia-Reinaldos}, M. and {Teyssier}, D. and {Altmann}, M. and {Andrae}, R. and {Audard}, M. and {Bellas-Velidis}, I. and {Benson}, K. and {Berthier}, J. and {Blomme}, R. and {Burgess}, P. and {Busso}, G. and {Carry}, B. and {Cellino}, A. and {Clementini}, G. and {Clotet}, M. and {Creevey}, O. and {Davidson}, M. and {De Ridder}, J. and {Delchambre}, L. and {Dell'Oro}, A. and {Ducourant}, C. and {Fern{\'a}ndez-Hern{\'a}ndez}, J. and {Fouesneau}, M. and {Fr{\'e}mat}, Y. and {Galluccio}, L. and {Garc{\'\i}a-Torres}, M. and {Gonz{\'a}lez-N{\'u}{\~n}ez}, J. and {Gonz{\'a}lez-Vidal}, J.~J. and {Gosset}, E. and {Guy}, L.~P. and {Halbwachs}, J.-L. and {Hambly}, N.~C. and {Harrison}, D.~L. and {Hern{\'a}ndez}, J. and {Hestroffer}, D. and {Hodgkin}, S.~T. and {Hutton}, A. and {Jasniewicz}, G. and {Jean-Antoine-Piccolo}, A. and {Jordan}, S. and {Korn}, A.~J. and {Krone-Martins}, A. and {Lanzafame}, A.~C. and {Lebzelter}, T. and {L{\"o}ffler}, W. and {Manteiga}, M. and {Marrese}, P.~M. and {Mart{\'\i}n-Fleitas}, J.~M. and {Moitinho}, A. and {Mora}, A. and {Muinonen}, K. and {Osinde}, J. and {Pancino}, E. and {Pauwels}, T. and {Petit}, J.-M. and {Recio-Blanco}, A. and {Richards}, P.~J. and {Rimoldini}, L. and {Robin}, A.~C. and {Sarro}, L.~M. and {Siopis}, C. and {Smith}, M. and {Sozzetti}, A. and {S{\"u}veges}, M. and {Torra}, J. and {van Reeven}, W. and {Abbas}, U. and {Abreu Aramburu}, A. and {Accart}, S. and {Aerts}, C. and {Altavilla}, G. and {{\'A}lvarez}, M.~A. and {Alvarez}, R. and {Alves}, J. and {Anderson}, R.~I. and {Andrei}, A.~H. and {Anglada Varela}, E. and {Antiche}, E. and {Antoja}, T. and {Arcay}, B. and {Astraatmadja}, T.~L. and {Bach}, N. and {Baker}, S.~G. and {Balaguer-N{\'u}{\~n}ez}, L. and {Balm}, P. and {Barache}, C. and {Barata}, C. and {Barbato}, D. and {Barblan}, F. and {Barklem}, P.~S. and {Barrado}, D. and {Barros}, M. and {Barstow}, M.~A. and {Bartholom{\'e} Mu{\~n}oz}, S. and {Bassilana}, J.-L. and {Becciani}, U. and {Bellazzini}, M. and {Berihuete}, A. and {Bertone}, S. and {Bianchi}, L. and {Bienaym{\'e}}, O. and {Blanco-Cuaresma}, S. and {Boch}, T. and {Boeche}, C. and {Bombrun}, A. and {Borrachero}, R. and {Bossini}, D. and {Bouquillon}, S. and {Bourda}, G. and {Bragaglia}, A. and {Bramante}, L. and {Breddels}, M.~A. and {Bressan}, A. and {Brouillet}, N. and {Br{\"u}semeister}, T. and {Brugaletta}, E. and {Bucciarelli}, B. and {Burlacu}, A. and {Busonero}, D. and {Butkevich}, A.~G. and {Buzzi}, R. and {Caffau}, E. and {Cancelliere}, R. and {Cannizzaro}, G. and {Cantat-Gaudin}, T. and {Carballo}, R. and {Carlucci}, T. and {Carrasco}, J.~M. and {Casamiquela}, L. and {Castellani}, M. and {Castro-Ginard}, A. and {Charlot}, P. and {Chemin}, L. and {Chiavassa}, A. and {Cocozza}, G. and {Costigan}, G. and {Cowell}, S. and {Crifo}, F. and {Crosta}, M. and {Crowley}, C. and {Cuypers}, J. and {Dafonte}, C. and {Damerdji}, Y. and {Dapergolas}, A. and {David}, P. and {David}, M. and {de Laverny}, P. and {De Luise}, F.},
        title = "{Gaia Data Release 2. Summary of the contents and survey properties}",
      journal = {\aap},
         year = 2018,
        month = aug,
       volume = {616},
          eid = {A1},
        pages = {A1},
          doi = {10.1051/0004-6361/201833051},
archivePrefix = {arXiv},
       eprint = {1804.09365},
 primaryClass = {astro-ph.GA},
       adsurl = {https://ui.adsabs.harvard.edu/abs/2018A&A...616A...1G}
}

@ARTICLE{George2000,
       author = {{George}, I.~M. and {Turner}, T.~J. and {Yaqoob}, T. and {Netzer}, H. and {Laor}, A. and {Mushotzky}, R.~F. and {Nandra}, K. and {Takahashi}, T.},
        title = "{X-Ray Observations of Optically Selected, Radio-quiet Quasars. I. The ASCA Results}",
      journal = {\apj},
         year = 2000,
        month = mar,
       volume = {531},
       number = {1},
        pages = {52-80},
          doi = {10.1086/308461},
archivePrefix = {arXiv},
       eprint = {astro-ph/9910218},
 primaryClass = {astro-ph},
       adsurl = {https://ui.adsabs.harvard.edu/abs/2000ApJ...531...52G}
}

@ARTICLE{Glikman2022,
       author = {{Glikman}, E. and {Lacy}, M. and {LaMassa}, S. and {Bradley}, C. and {Djorgovski}, S.~G. and {Urrutia}, T. and {Gates}, E.~L. and {Graham}, M.~J. and {Urry}, M. and {Yoon}, I.},
        title = "{The WISE-2MASS Survey: Red Quasars Into the Radio Quiet Regime}",
      journal = {\apj},
         year = 2022,
        month = aug,
       volume = {934},
       number = {2},
          eid = {119},
        pages = {119},
          doi = {10.3847/1538-4357/ac6bee},
archivePrefix = {arXiv},
       eprint = {2204.13745},
 primaryClass = {astro-ph.GA},
       adsurl = {https://ui.adsabs.harvard.edu/abs/2022ApJ...934..119G}
}

@ARTICLE{Liu2021,
       author = {{Liu}, Hezhen and {Luo}, B. and {Brandt}, W.~N. and {Brotherton}, Michael S. and {Gallagher}, S.~C. and {Ni}, Q. and {Shemmer}, Ohad and {Timlin}, III, J.~D.},
        title = "{On the Observational Difference between the Accretion Disk-Corona Connections among Super- and Sub-Eddington Accreting Active Galactic Nuclei}",
      journal = {\apj},
         year = 2021,
        month = apr,
       volume = {910},
       number = {2},
          eid = {103},
        pages = {103},
          doi = {10.3847/1538-4357/abe37f},
archivePrefix = {arXiv},
       eprint = {2102.02832},
 primaryClass = {astro-ph.GA},
       adsurl = {https://ui.adsabs.harvard.edu/abs/2021ApJ...910..103L}
}

@ARTICLE{Liu2022,
       author = {{Liu}, Yan-Sheng and {Bian}, Wei-Hao},
        title = "{The Relation between the Optical Fe II Emission and the Dimensionless Accretion Rate for Active Galactic Nuclei}",
      journal = {\apj},
         year = 2022,
        month = oct,
       volume = {937},
       number = {2},
          eid = {82},
        pages = {82},
          doi = {10.3847/1538-4357/ac8b84},
archivePrefix = {arXiv},
       eprint = {2210.00684},
 primaryClass = {astro-ph.GA},
       adsurl = {https://ui.adsabs.harvard.edu/abs/2022ApJ...937...82L}
}

@ARTICLE{Lyke2020,
       author = {{Lyke}, Brad W. and {Higley}, Alexandra N. and {McLane}, J.~N. and {Schurhammer}, Danielle P. and {Myers}, Adam D. and {Ross}, Ashley J. and {Dawson}, Kyle and {Chabanier}, Sol{\`e}ne and {Martini}, Paul and {Busca}, Nicol{\'a}s G. and {Mas des Bourboux}, H{\'e}lion du and {Salvato}, Mara and {Streblyanska}, Alina and {Zarrouk}, Pauline and {Burtin}, Etienne and {Anderson}, Scott F. and {Bautista}, Julian and {Bizyaev}, Dmitry and {Brandt}, W.~N. and {Brinkmann}, Jonathan and {Brownstein}, Joel R. and {Comparat}, Johan and {Green}, Paul and {de la Macorra}, Axel and {Mu{\~n}oz Guti{\'e}rrez}, Andrea and {Hou}, Jiamin and {Newman}, Jeffrey A. and {Palanque-Delabrouille}, Nathalie and {P{\^a}ris}, Isabelle and {Percival}, Will J. and {Petitjean}, Patrick and {Rich}, James and {Rossi}, Graziano and {Schneider}, Donald P. and {Smith}, Alexander and {Vivek}, M. and {Weaver}, Benjamin Alan},
        title = "{The Sloan Digital Sky Survey Quasar Catalog: Sixteenth Data Release}",
      journal = {\apjs},
         year = 2020,
        month = sep,
       volume = {250},
       number = {1},
          eid = {8},
        pages = {8},
          doi = {10.3847/1538-4365/aba623},
archivePrefix = {arXiv},
       eprint = {2007.09001},
 primaryClass = {astro-ph.GA},
       adsurl = {https://ui.adsabs.harvard.edu/abs/2020ApJS..250....8L}
}

@ARTICLE{Marconi2004,
       author = {{Marconi}, A. and {Risaliti}, G. and {Gilli}, R. and {Hunt}, L.~K. and {Maiolino}, R. and {Salvati}, M.},
        title = "{Local supermassive black holes, relics of active galactic nuclei and the X-ray background}",
      journal = {\mnras},
         year = 2004,
        month = jun,
       volume = {351},
       number = {1},
        pages = {169-185},
          doi = {10.1111/j.1365-2966.2004.07765.x},
archivePrefix = {arXiv},
       eprint = {astro-ph/0311619},
 primaryClass = {astro-ph},
       adsurl = {https://ui.adsabs.harvard.edu/abs/2004MNRAS.351..169M}
}

@ARTICLE{Meiksin2006,
       author = {{Meiksin}, Avery},
        title = "{Colour corrections for high-redshift objects due to intergalactic attenuation}",
      journal = {\mnras},
         year = 2006,
        month = jan,
       volume = {365},
       number = {3},
        pages = {807-812},
          doi = {10.1111/j.1365-2966.2005.09756.x},
archivePrefix = {arXiv},
       eprint = {astro-ph/0512435},
 primaryClass = {astro-ph},
       adsurl = {https://ui.adsabs.harvard.edu/abs/2006MNRAS.365..807M}
}

@ARTICLE{Martin2005,
       author = {{Martin}, D. Christopher and {Fanson}, James and {Schiminovich}, David and {Morrissey}, Patrick and {Friedman}, Peter G. and {Barlow}, Tom A. and {Conrow}, Tim and {Grange}, Robert and {Jelinsky}, Patrick N. and {Milliard}, Bruno and {Siegmund}, Oswald H.~W. and {Bianchi}, Luciana and {Byun}, Yong-Ik and {Donas}, Jose and {Forster}, Karl and {Heckman}, Timothy M. and {Lee}, Young-Wook and {Madore}, Barry F. and {Malina}, Roger F. and {Neff}, Susan G. and {Rich}, R. Michael and {Small}, Todd and {Surber}, Frank and {Szalay}, Alex S. and {Welsh}, Barry and {Wyder}, Ted K.},
        title = "{The Galaxy Evolution Explorer: A Space Ultraviolet Survey Mission}",
      journal = {\apjl},
         year = 2005,
        month = jan,
       volume = {619},
       number = {1},
        pages = {L1-L6},
          doi = {10.1086/426387},
archivePrefix = {arXiv},
       eprint = {astro-ph/0411302},
 primaryClass = {astro-ph},
       adsurl = {https://ui.adsabs.harvard.edu/abs/2005ApJ...619L...1M}
}

@ARTICLE{Lawrence2007,
       author = {{Lawrence}, A. and {Warren}, S.~J. and {Almaini}, O. and {Edge}, A.~C. and {Hambly}, N.~C. and {Jameson}, R.~F. and {Lucas}, P. and {Casali}, M. and {Adamson}, A. and {Dye}, S. and {Emerson}, J.~P. and {Foucaud}, S. and {Hewett}, P. and {Hirst}, P. and {Hodgkin}, S.~T. and {Irwin}, M.~J. and {Lodieu}, N. and {McMahon}, R.~G. and {Simpson}, C. and {Smail}, I. and {Mortlock}, D. and {Folger}, M.},
        title = "{The UKIRT Infrared Deep Sky Survey (UKIDSS)}",
      journal = {\mnras},
         year = 2007,
        month = aug,
       volume = {379},
       number = {4},
        pages = {1599-1617},
          doi = {10.1111/j.1365-2966.2007.12040.x},
archivePrefix = {arXiv},
       eprint = {astro-ph/0604426},
 primaryClass = {astro-ph},
       adsurl = {https://ui.adsabs.harvard.edu/abs/2007MNRAS.379.1599L}
}

@ARTICLE{Molina2023,
       author = {{Molina}, Juan and {Shangguan}, Jinyi and {Wang}, Ran and {Ho}, Luis C. and {Bauer}, Franz E. and {Treister}, Ezequiel},
        title = "{Lack of Correlations between Cold Molecular Gas and AGN Properties in Type 1 AGNs at z {\ensuremath{\lesssim}} 0.5}",
      journal = {\apj},
         year = 2023,
        month = jun,
       volume = {950},
       number = {1},
          eid = {60},
        pages = {60},
          doi = {10.3847/1538-4357/acc9b4},
archivePrefix = {arXiv},
       eprint = {2304.01017},
 primaryClass = {astro-ph.GA},
       adsurl = {https://ui.adsabs.harvard.edu/abs/2023ApJ...950...60M}
}

@BOOK{Netzer2013,
       author = {{Netzer}, Hagai},
        title = "{The Physics and Evolution of Active Galactic Nuclei}",
         year = 2013,
       adsurl = {https://ui.adsabs.harvard.edu/abs/2013peag.book.....N}
}

@ARTICLE{Oke1968,
       author = {{Oke}, J.~B. and {Sandage}, Allan},
        title = "{Energy Distributions, K Corrections, and the Stebbins-Whitford Effect for Giant Elliptical Galaxies}",
      journal = {\apj},
         year = 1968,
        month = oct,
       volume = {154},
        pages = {21},
          doi = {10.1086/149737},
       adsurl = {https://ui.adsabs.harvard.edu/abs/1968ApJ...154...21O}
}

@ARTICLE{Peterson2004,
       author = {{Peterson}, B.~M. and {Ferrarese}, L. and {Gilbert}, K.~M. and {Kaspi}, S. and {Malkan}, M.~A. and {Maoz}, D. and {Merritt}, D. and {Netzer}, H. and {Onken}, C.~A. and {Pogge}, R.~W. and {Vestergaard}, M. and {Wandel}, A.},
        title = "{Central Masses and Broad-Line Region Sizes of Active Galactic Nuclei. II. A Homogeneous Analysis of a Large Reverberation-Mapping Database}",
      journal = {\apj},
         year = 2004,
        month = oct,
       volume = {613},
       number = {2},
        pages = {682-699},
          doi = {10.1086/423269},
archivePrefix = {arXiv},
       eprint = {astro-ph/0407299},
 primaryClass = {astro-ph},
       adsurl = {https://ui.adsabs.harvard.edu/abs/2004ApJ...613..682P}
}

@ARTICLE{Richards2006a,
       author = {{Richards}, Gordon T. and {Lacy}, Mark and {Storrie-Lombardi}, Lisa J. and {Hall}, Patrick B. and {Gallagher}, S.~C. and {Hines}, Dean C. and {Fan}, Xiaohui and {Papovich}, Casey and {Vanden Berk}, Daniel E. and {Trammell}, George B. and {Schneider}, Donald P. and {Vestergaard}, Marianne and {York}, Donald G. and {Jester}, Sebastian and {Anderson}, Scott F. and {Budav{\'a}ri}, Tam{\'a}s and {Szalay}, Alexander S.},
        title = "{Spectral Energy Distributions and Multiwavelength Selection of Type 1 Quasars}",
      journal = {\apjs},
         year = 2006,
        month = oct,
       volume = {166},
       number = {2},
        pages = {470-497},
          doi = {10.1086/506525},
archivePrefix = {arXiv},
       eprint = {astro-ph/0601558},
 primaryClass = {astro-ph},
       adsurl = {https://ui.adsabs.harvard.edu/abs/2006ApJS..166..470R}
}

@ARTICLE{Richards2006b,
       author = {{Richards}, Gordon T. and {Strauss}, Michael A. and {Fan}, Xiaohui and {Hall}, Patrick B. and {Jester}, Sebastian and {Schneider}, Donald P. and {Vanden Berk}, Daniel E. and {Stoughton}, Chris and {Anderson}, Scott F. and {Brunner}, Robert J. and {Gray}, Jim and {Gunn}, James E. and {Ivezi{\'c}}, {\v{Z}}eljko and {Kirkland}, Margaret K. and {Knapp}, G.~R. and {Loveday}, Jon and {Meiksin}, Avery and {Pope}, Adrian and {Szalay}, Alexander S. and {Thakar}, Anirudda R. and {Yanny}, Brian and {York}, Donald G. and {Barentine}, J.~C. and {Brewington}, Howard J. and {Brinkmann}, J. and {Fukugita}, Masataka and {Harvanek}, Michael and {Kent}, Stephen M. and {Kleinman}, S.~J. and {Krzesi{\'n}ski}, Jurek and {Long}, Daniel C. and {Lupton}, Robert H. and {Nash}, Thomas and {Neilsen}, Jr., Eric H. and {Nitta}, Atsuko and {Schlegel}, David J. and {Snedden}, Stephanie A.},
        title = "{The Sloan Digital Sky Survey Quasar Survey: Quasar Luminosity Function from Data Release 3}",
      journal = {\aj},
         year = 2006,
        month = jun,
       volume = {131},
       number = {6},
        pages = {2766-2787},
          doi = {10.1086/503559},
archivePrefix = {arXiv},
       eprint = {astro-ph/0601434},
 primaryClass = {astro-ph},
       adsurl = {https://ui.adsabs.harvard.edu/abs/2006AJ....131.2766R}
}

@ARTICLE{Rivera2021,
       author = {{Calistro Rivera}, G. and {Alexander}, D.~M. and {Rosario}, D.~J. and {Harrison}, C.~M. and {Stalevski}, M. and {Rakshit}, S. and {Fawcett}, V.~A. and {Morabito}, L.~K. and {Klindt}, L. and {Best}, P.~N. and {Bonato}, M. and {Bowler}, R.~A.~A. and {Costa}, T. and {Kondapally}, R.},
        title = "{The multiwavelength properties of red QSOs: Evidence for dusty winds as the origin of QSO reddening}",
      journal = {\aap},
         year = 2021,
        month = may,
       volume = {649},
          eid = {A102},
        pages = {A102},
          doi = {10.1051/0004-6361/202040214},
archivePrefix = {arXiv},
       eprint = {2103.02610},
 primaryClass = {astro-ph.GA},
       adsurl = {https://ui.adsabs.harvard.edu/abs/2021A&A...649A.102C}
}

@ARTICLE{Rosen2016,
       author = {{Rosen}, S.~R. and {Webb}, N.~A. and {Watson}, M.~G. and {Ballet}, J. and {Barret}, D. and {Braito}, V. and {Carrera}, F.~J. and {Ceballos}, M.~T. and {Coriat}, M. and {Della Ceca}, R. and {Denkinson}, G. and {Esquej}, P. and {Farrell}, S.~A. and {Freyberg}, M. and {Gris{\'e}}, F. and {Guillout}, P. and {Heil}, L. and {Koliopanos}, F. and {Law-Green}, D. and {Lamer}, G. and {Lin}, D. and {Martino}, R. and {Michel}, L. and {Motch}, C. and {Nebot Gomez-Moran}, A. and {Page}, C.~G. and {Page}, K. and {Page}, M. and {Pakull}, M.~W. and {Pye}, J. and {Read}, A. and {Rodriguez}, P. and {Sakano}, M. and {Saxton}, R. and {Schwope}, A. and {Scott}, A.~E. and {Sturm}, R. and {Traulsen}, I. and {Yershov}, V. and {Zolotukhin}, I.},
        title = "{The XMM-Newton serendipitous survey. VII. The third XMM-Newton serendipitous source catalogue}",
      journal = {\aap},
         year = 2016,
        month = may,
       volume = {590},
          eid = {A1},
        pages = {A1},
          doi = {10.1051/0004-6361/201526416},
archivePrefix = {arXiv},
       eprint = {1504.07051},
 primaryClass = {astro-ph.HE},
       adsurl = {https://ui.adsabs.harvard.edu/abs/2016A&A...590A...1R}
}

@ARTICLE{Siebenmorgen2015,
       author = {{Siebenmorgen}, Ralf and {Heymann}, Frank and {Efstathiou}, Andreas},
        title = "{Self-consistent two-phase AGN torus models{\ensuremath{\star}}. SED library for observers}",
      journal = {\aap},
         year = 2015,
        month = nov,
       volume = {583},
          eid = {A120},
        pages = {A120},
          doi = {10.1051/0004-6361/201526034},
archivePrefix = {arXiv},
       eprint = {1508.04343},
 primaryClass = {astro-ph.GA},
       adsurl = {https://ui.adsabs.harvard.edu/abs/2015A&A...583A.120S}
}

@ARTICLE{Schlegel1998,
       author = {{Schlegel}, David J. and {Finkbeiner}, Douglas P. and {Davis}, Marc},
        title = "{Maps of Dust Infrared Emission for Use in Estimation of Reddening and Cosmic Microwave Background Radiation Foregrounds}",
      journal = {\apj},
         year = 1998,
        month = jun,
       volume = {500},
       number = {2},
        pages = {525-553},
          doi = {10.1086/305772},
archivePrefix = {arXiv},
       eprint = {astro-ph/9710327},
 primaryClass = {astro-ph},
       adsurl = {https://ui.adsabs.harvard.edu/abs/1998ApJ...500..525S}
}

@ARTICLE{SS73,
       author = {{Shakura}, N.~I. and {Sunyaev}, R.~A.},
        title = "{Black holes in binary systems. Observational appearance.}",
      journal = {\aap},
         year = 1973,
        month = jan,
       volume = {24},
        pages = {337-355},
       adsurl = {https://ui.adsabs.harvard.edu/abs/1973A&A....24..337S}
}

@ARTICLE{Shang2011,
       author = {{Shang}, Zhaohui and {Brotherton}, Michael S. and {Wills}, Beverley J. and {Wills}, D. and {Cales}, Sabrina L. and {Dale}, Daniel A. and {Green}, Richard F. and {Runnoe}, Jessie C. and {Nemmen}, Rodrigo S. and {Gallagher}, Sarah C. and {Ganguly}, Rajib and {Hines}, Dean C. and {Kelly}, Benjamin J. and {Kriss}, Gerard A. and {Li}, Jun and {Tang}, Baitian and {Xie}, Yanxia},
        title = "{The Next Generation Atlas of Quasar Spectral Energy Distributions from Radio to X-Rays}",
      journal = {\apjs},
         year = 2011,
        month = sep,
       volume = {196},
       number = {1},
          eid = {2},
        pages = {2},
          doi = {10.1088/0067-0049/196/1/2},
archivePrefix = {arXiv},
       eprint = {1107.1855},
 primaryClass = {astro-ph.CO},
       adsurl = {https://ui.adsabs.harvard.edu/abs/2011ApJS..196....2S}
}

@ARTICLE{Shen2011,
       author = {{Shen}, Yue and {Richards}, Gordon T. and {Strauss}, Michael A. and {Hall}, Patrick B. and {Schneider}, Donald P. and {Snedden}, Stephanie and {Bizyaev}, Dmitry and {Brewington}, Howard and {Malanushenko}, Viktor and {Malanushenko}, Elena and {Oravetz}, Dan and {Pan}, Kaike and {Simmons}, Audrey},
        title = "{A Catalog of Quasar Properties from Sloan Digital Sky Survey Data Release 7}",
      journal = {\apjs},
         year = 2011,
        month = jun,
       volume = {194},
       number = {2},
          eid = {45},
        pages = {45},
          doi = {10.1088/0067-0049/194/2/45},
archivePrefix = {arXiv},
       eprint = {1006.5178},
 primaryClass = {astro-ph.CO},
       adsurl = {https://ui.adsabs.harvard.edu/abs/2011ApJS..194...45S}
}

@ARTICLE{Shen2014,
       author = {{Shen}, Yue and {Ho}, Luis C.},
        title = "{The diversity of quasars unified by accretion and orientation}",
      journal = {\nat},
         year = 2014,
        month = sep,
       volume = {513},
       number = {7517},
        pages = {210-213},
          doi = {10.1038/nature13712},
archivePrefix = {arXiv},
       eprint = {1409.2887},
 primaryClass = {astro-ph.GA},
       adsurl = {https://ui.adsabs.harvard.edu/abs/2014Natur.513..210S}
}

@ARTICLE{Shi2014,
       author = {{Shi}, Yong and {Rieke}, G.~H. and {Ogle}, P.~M. and {Su}, K.~Y.~L. and {Balog}, Z.},
        title = "{Infrared Spectra and Photometry Of Complete Samples of Palomar-Green and Two Micron All Sky Survey Quasars}",
      journal = {\apjs},
         year = 2014,
        month = oct,
       volume = {214},
       number = {2},
          eid = {23},
        pages = {23},
          doi = {10.1088/0067-0049/214/2/23},
archivePrefix = {arXiv},
       eprint = {1408.5909},
 primaryClass = {astro-ph.GA},
       adsurl = {https://ui.adsabs.harvard.edu/abs/2014ApJS..214...23S}
}

@ARTICLE{Skrutskie2006,
       author = {{Skrutskie}, M.~F. and {Cutri}, R.~M. and {Stiening}, R. and {Weinberg}, M.~D. and {Schneider}, S. and {Carpenter}, J.~M. and {Beichman}, C. and {Capps}, R. and {Chester}, T. and {Elias}, J. and {Huchra}, J. and {Liebert}, J. and {Lonsdale}, C. and {Monet}, D.~G. and {Price}, S. and {Seitzer}, P. and {Jarrett}, T. and {Kirkpatrick}, J.~D. and {Gizis}, J.~E. and {Howard}, E. and {Evans}, T. and {Fowler}, J. and {Fullmer}, L. and {Hurt}, R. and {Light}, R. and {Kopan}, E.~L. and {Marsh}, K.~A. and {McCallon}, H.~L. and {Tam}, R. and {Van Dyk}, S. and {Wheelock}, S.},
        title = "{The Two Micron All Sky Survey (2MASS)}",
      journal = {\aj},
         year = 2006,
        month = feb,
       volume = {131},
       number = {2},
        pages = {1163-1183},
          doi = {10.1086/498708},
       adsurl = {https://ui.adsabs.harvard.edu/abs/2006AJ....131.1163S}
}

@ARTICLE{Steffen2006,
       author = {{Steffen}, A.~T. and {Strateva}, I. and {Brandt}, W.~N. and {Alexander}, D.~M. and {Koekemoer}, A.~M. and {Lehmer}, B.~D. and {Schneider}, D.~P. and {Vignali}, C.},
        title = "{The X-Ray-to-Optical Properties of Optically Selected Active Galaxies over Wide Luminosity and Redshift Ranges}",
      journal = {\aj},
         year = 2006,
        month = jun,
       volume = {131},
       number = {6},
        pages = {2826-2842},
          doi = {10.1086/503627},
archivePrefix = {arXiv},
       eprint = {astro-ph/0602407},
 primaryClass = {astro-ph},
       adsurl = {https://ui.adsabs.harvard.edu/abs/2006AJ....131.2826S}
}

@ARTICLE{Stolc2023,
       author = {{{\v{S}}tolc}, Marcel and {Zaja{\v{c}}ek}, Michal and {Czerny}, Bo{\.z}ena and {Karas}, Vladim{\'\i}r},
        title = "{Spectral energy distribution profiles from AGN accretion disc in multigap set-up}",
      journal = {\mnras},
         year = 2023,
        month = jun,
       volume = {522},
       number = {2},
        pages = {2869-2884},
          doi = {10.1093/mnras/stad1127},
archivePrefix = {arXiv},
       eprint = {2304.03015},
 primaryClass = {astro-ph.HE},
       adsurl = {https://ui.adsabs.harvard.edu/abs/2023MNRAS.522.2869S}
}

@ARTICLE{Sulentic2000,
       author = {{Sulentic}, J.~W. and {Marziani}, P. and {Dultzin-Hacyan}, D.},
        title = "{Phenomenology of Broad Emission Lines in Active Galactic Nuclei}",
      journal = {\araa},
         year = 2000,
        month = jan,
       volume = {38},
        pages = {521-571},
          doi = {10.1146/annurev.astro.38.1.521},
       adsurl = {https://ui.adsabs.harvard.edu/abs/2000ARA&A..38..521S}
}

@ARTICLE{Yu2020a,
       author = {{Yu}, Li-Ming and {Zhao}, Bi-Xuan and {Bian}, Wei-Hao and {Wang}, Chan and {Ge}, Xue},
        title = "{An extended size-luminosity relation for the reverberation-mapped AGNs: the role of the accretion rate}",
      journal = {\mnras},
         year = 2020,
        month = feb,
       volume = {491},
       number = {4},
        pages = {5881-5896},
          doi = {10.1093/mnras/stz3387},
archivePrefix = {arXiv},
       eprint = {1911.10802},
 primaryClass = {astro-ph.GA},
       adsurl = {https://ui.adsabs.harvard.edu/abs/2020MNRAS.491.5881Y}
}

@ARTICLE{Vanden2001,
       author = {{Vanden Berk}, Daniel E. and {Richards}, Gordon T. and {Bauer}, Amanda and {Strauss}, Michael A. and {Schneider}, Donald P. and {Heckman}, Timothy M. and {York}, Donald G. and {Hall}, Patrick B. and {Fan}, Xiaohui and {Knapp}, G.~R. and {Anderson}, Scott F. and {Annis}, James and {Bahcall}, Neta A. and {Bernardi}, Mariangela and {Briggs}, John W. and {Brinkmann}, J. and {Brunner}, Robert and {Burles}, Scott and {Carey}, Larry and {Castander}, Francisco J. and {Connolly}, A.~J. and {Crocker}, J.~H. and {Csabai}, Istv{\'a}n and {Doi}, Mamoru and {Finkbeiner}, Douglas and {Friedman}, Scott and {Frieman}, Joshua A. and {Fukugita}, Masataka and {Gunn}, James E. and {Hennessy}, G.~S. and {Ivezi{\'c}}, {\v{Z}}eljko and {Kent}, Stephen and {Kunszt}, Peter Z. and {Lamb}, D.~Q. and {Leger}, R. French and {Long}, Daniel C. and {Loveday}, Jon and {Lupton}, Robert H. and {Meiksin}, Avery and {Merelli}, Aronne and {Munn}, Jeffrey A. and {Newberg}, Heidi Jo and {Newcomb}, Matt and {Nichol}, R.~C. and {Owen}, Russell and {Pier}, Jeffrey R. and {Pope}, Adrian and {Rockosi}, Constance M. and {Schlegel}, David J. and {Siegmund}, Walter A. and {Smee}, Stephen and {Snir}, Yehuda and {Stoughton}, Chris and {Stubbs}, Christopher and {SubbaRao}, Mark and {Szalay}, Alexander S. and {Szokoly}, Gyula P. and {Tremonti}, Christy and {Uomoto}, Alan and {Waddell}, Patrick and {Yanny}, Brian and {Zheng}, Wei},
        title = "{Composite Quasar Spectra from the Sloan Digital Sky Survey}",
      journal = {\aj},
         year = 2001,
        month = aug,
       volume = {122},
       number = {2},
        pages = {549-564},
          doi = {10.1086/321167},
archivePrefix = {arXiv},
       eprint = {astro-ph/0105231},
 primaryClass = {astro-ph},
       adsurl = {https://ui.adsabs.harvard.edu/abs/2001AJ....122..549V}
}

@ARTICLE{Wang2012,
       author = {{Wang}, Huiyuan and {Zhou}, Hongyan and {Yuan}, Weimin and {Wang}, Tinggui},
        title = "{Metallicity and Quasar Outflows}",
      journal = {\apjl},
         year = 2012,
        month = jun,
       volume = {751},
       number = {2},
          eid = {L23},
        pages = {L23},
          doi = {10.1088/2041-8205/751/2/L23},
archivePrefix = {arXiv},
       eprint = {1204.5234},
 primaryClass = {astro-ph.CO},
       adsurl = {https://ui.adsabs.harvard.edu/abs/2012ApJ...751L..23W}
}

@ARTICLE{Wang2014,
       author = {{Wang}, Jian-Min and {Du}, Pu and {Li}, Yan-Rong and {Ho}, Luis C. and {Hu}, Chen and {Bai}, Jin-Ming},
        title = "{A New Approach to Constrain Black Hole Spins in Active Galaxies Using Optical Reverberation Mapping}",
      journal = {\apjl},
         year = 2014,
        month = sep,
       volume = {792},
       number = {1},
          eid = {L13},
        pages = {L13},
          doi = {10.1088/2041-8205/792/1/L13},
archivePrefix = {arXiv},
       eprint = {1408.2341},
 primaryClass = {astro-ph.HE},
       adsurl = {https://ui.adsabs.harvard.edu/abs/2014ApJ...792L..13W}
}

@ARTICLE{Wright2010,
       author = {{Wright}, Edward L. and {Eisenhardt}, Peter R.~M. and {Mainzer}, Amy K. and {Ressler}, Michael E. and {Cutri}, Roc M. and {Jarrett}, Thomas and {Kirkpatrick}, J. Davy and {Padgett}, Deborah and {McMillan}, Robert S. and {Skrutskie}, Michael and {Stanford}, S.~A. and {Cohen}, Martin and {Walker}, Russell G. and {Mather}, John C. and {Leisawitz}, David and {Gautier}, III, Thomas N. and {McLean}, Ian and {Benford}, Dominic and {Lonsdale}, Carol J. and {Blain}, Andrew and {Mendez}, Bryan and {Irace}, William R. and {Duval}, Valerie and {Liu}, Fengchuan and {Royer}, Don and {Heinrichsen}, Ingolf and {Howard}, Joan and {Shannon}, Mark and {Kendall}, Martha and {Walsh}, Amy L. and {Larsen}, Mark and {Cardon}, Joel G. and {Schick}, Scott and {Schwalm}, Mark and {Abid}, Mohamed and {Fabinsky}, Beth and {Naes}, Larry and {Tsai}, Chao-Wei},
        title = "{The Wide-field Infrared Survey Explorer (WISE): Mission Description and Initial On-orbit Performance}",
      journal = {\aj},
         year = 2010,
        month = dec,
       volume = {140},
       number = {6},
        pages = {1868-1881},
          doi = {10.1088/0004-6256/140/6/1868},
archivePrefix = {arXiv},
       eprint = {1008.0031},
 primaryClass = {astro-ph.IM},
       adsurl = {https://ui.adsabs.harvard.edu/abs/2010AJ....140.1868W}
}

@ARTICLE{Wyder2005,
       author = {{Wyder}, Ted K. and {Treyer}, Marie A. and {Milliard}, Bruno and {Schiminovich}, David and {Arnouts}, St{\'e}phane and {Budav{\'a}ri}, Tam{\'a}s and {Barlow}, Tom A. and {Bianchi}, Luciana and {Byun}, Yong-Ik and {Donas}, Jos{\'e} and {Forster}, Karl and {Friedman}, Peter G. and {Heckman}, Timothy M. and {Jelinsky}, Patrick N. and {Lee}, Young-Wook and {Madore}, Barry F. and {Malina}, Roger F. and {Martin}, D. Christopher and {Morrissey}, Patrick and {Neff}, Susan G. and {Rich}, R. Michael and {Siegmund}, Oswald H.~W. and {Small}, Todd and {Szalay}, Alex S. and {Welsh}, Barry Y.},
        title = "{The Ultraviolet Galaxy Luminosity Function in the Local Universe from GALEX Data}",
      journal = {\apjl},
         year = 2005,
        month = jan,
       volume = {619},
       number = {1},
        pages = {L15-L18},
          doi = {10.1086/424735},
archivePrefix = {arXiv},
       eprint = {astro-ph/0411364},
 primaryClass = {astro-ph},
       adsurl = {https://ui.adsabs.harvard.edu/abs/2005ApJ...619L..15W}
}

@ARTICLE{Zdziarski1995,
       author = {{Zdziarski}, Andrzej A. and {Johnson}, W. Neil and {Done}, Chris and {Smith}, David and {McNaron-Brown}, Kellie},
        title = "{The Average X-Ray/Gamma-Ray Spectra of Seyfert Galaxies from GINGA and OSSE and the Origin of the Cosmic X-Ray Background}",
      journal = {\apjl},
         year = 1995,
        month = jan,
       volume = {438},
        pages = {L63},
          doi = {10.1086/187716},
       adsurl = {https://ui.adsabs.harvard.edu/abs/1995ApJ...438L..63Z}
}

@ARTICLE{Zhang2014,
       author = {{Zhang}, Shaohua and {Wang}, Huiyuan and {Wang}, Tinggui and {Xing}, Feijun and {Zhang}, Kai and {Zhou}, Hongyan and {Jiang}, Peng},
        title = "{Outflow and Hot Dust Emission in Broad Absorption Line Quasars}",
      journal = {\apj},
         year = 2014,
        month = may,
       volume = {786},
       number = {1},
          eid = {42},
        pages = {42},
          doi = {10.1088/0004-637X/786/1/42},
archivePrefix = {arXiv},
       eprint = {1403.3166},
 primaryClass = {astro-ph.GA},
       adsurl = {https://ui.adsabs.harvard.edu/abs/2014ApJ...786...42Z}
}

@ARTICLE{Panda2018,
       author = {{Panda}, Swayamtrupta and {Czerny}, Bo{\.z}ena and {Adhikari}, Tek P. and {Hryniewicz}, Krzysztof and {Wildy}, Conor and {Kuraszkiewicz}, Joanna and {{\'S}niegowska}, Marzena},
        title = "{Modeling of the Quasar Main Sequence in the Optical Plane}",
      journal = {\apj},
         year = 2018,
        month = oct,
       volume = {866},
       number = {2},
          eid = {115},
        pages = {115},
          doi = {10.3847/1538-4357/aae209},
archivePrefix = {arXiv},
       eprint = {1806.08571},
 primaryClass = {astro-ph.HE},
       adsurl = {https://ui.adsabs.harvard.edu/abs/2018ApJ...866..115P}
}

@ARTICLE{Zhuang2018,
       author = {{Zhuang}, Ming-Yang and {Ho}, Luis C. and {Shangguan}, Jinyi},
        title = "{The Infrared Emission and Opening Angle of the Torus in Quasars}",
      journal = {\apj},
         year = 2018,
        month = aug,
       volume = {862},
       number = {2},
          eid = {118},
        pages = {118},
          doi = {10.3847/1538-4357/aacc2d},
archivePrefix = {arXiv},
       eprint = {1806.03783},
 primaryClass = {astro-ph.GA},
       adsurl = {https://ui.adsabs.harvard.edu/abs/2018ApJ...862..118Z}
}

@ARTICLE{Wu2022,
       author = {{Wu}, Qiaoya and {Shen}, Yue},
        title = "{A Catalog of Quasar Properties from Sloan Digital Sky Survey Data Release 16}",
      journal = {\apjs},
         year = 2022,
        month = dec,
       volume = {263},
       number = {2},
          eid = {42},
        pages = {42},
          doi = {10.3847/1538-4365/ac9ead},
archivePrefix = {arXiv},
       eprint = {2209.03987},
 primaryClass = {astro-ph.GA},
       adsurl = {https://ui.adsabs.harvard.edu/abs/2022ApJS..263...42W}
}

@ARTICLE{Urry1995,
       author = {{Urry}, C. Megan and {Padovani}, Paolo},
        title = "{Unified Schemes for Radio-Loud Active Galactic Nuclei}",
      journal = {\pasp},
         year = 1995,
        month = sep,
       volume = {107},
        pages = {803},
          doi = {10.1086/133630},
archivePrefix = {arXiv},
       eprint = {astro-ph/9506063},
 primaryClass = {astro-ph},
       adsurl = {https://ui.adsabs.harvard.edu/abs/1995PASP..107..803U}
}

@ARTICLE{Netzer2015,
       author = {{Netzer}, Hagai},
        title = "{Revisiting the Unified Model of Active Galactic Nuclei}",
      journal = {\araa},
         year = 2015,
        month = aug,
       volume = {53},
        pages = {365-408},
          doi = {10.1146/annurev-astro-082214-122302},
archivePrefix = {arXiv},
       eprint = {1505.00811},
 primaryClass = {astro-ph.GA},
       adsurl = {https://ui.adsabs.harvard.edu/abs/2015ARA&A..53..365N}
}

@ARTICLE{Padovani2017,
       author = {{Padovani}, P. and {Alexander}, D.~M. and {Assef}, R.~J. and {De Marco}, B. and {Giommi}, P. and {Hickox}, R.~C. and {Richards}, G.~T. and {Smol{\v{c}}i{\'c}}, V. and {Hatziminaoglou}, E. and {Mainieri}, V. and {Salvato}, M.},
        title = "{Active galactic nuclei: what's in a name?}",
      journal = {\aapr},
         year = 2017,
        month = aug,
       volume = {25},
       number = {1},
          eid = {2},
        pages = {2},
          doi = {10.1007/s00159-017-0102-9},
archivePrefix = {arXiv},
       eprint = {1707.07134},
 primaryClass = {astro-ph.GA},
       adsurl = {https://ui.adsabs.harvard.edu/abs/2017A&ARv..25....2P}
}

@ARTICLE{Shields1978,
       author = {{Shields}, G.~A.},
        title = "{Thermal continuum from accretion disks in quasars}",
      journal = {\nat},
         year = 1978,
        month = apr,
       volume = {272},
       number = {5655},
        pages = {706-708},
          doi = {10.1038/272706a0},
       adsurl = {https://ui.adsabs.harvard.edu/abs/1978Natur.272..706S}
}

@ARTICLE{Czerny1987,
       author = {{Czerny}, Bozena and {Elvis}, Martin},
        title = "{Constraints on Quasar Accretion Disks from the Optical/Ultraviolet/Soft X-Ray Big Bump}",
      journal = {\apj},
         year = 1987,
        month = oct,
       volume = {321},
        pages = {305},
          doi = {10.1086/165630},
       adsurl = {https://ui.adsabs.harvard.edu/abs/1987ApJ...321..305C}
}

@ARTICLE{ Laor1990,
       author = {{Laor}, Ari},
        title = "{Line Profiles from a Disk around a Rotating Black Hole}",
      journal = {\apj},
         year = 1991,
        month = jul,
       volume = {376},
        pages = {90},
          doi = {10.1086/170257},
       adsurl = {https://ui.adsabs.harvard.edu/abs/1991ApJ...376...90L}
}

@INPROCEEDINGS{Blaes2004,
       author = {{Blaes}, O.},
        title = "{Accretion Disks Based on Real Physics}",
    booktitle = {AGN Physics with the Sloan Digital Sky Survey},
         year = 2004,
       editor = {{Richards}, Gordon T. and {Hall}, Patrick B.},
       series = {Astronomical Society of the Pacific Conference Series},
       volume = {311},
        month = jun,
        pages = {121},
       adsurl = {https://ui.adsabs.harvard.edu/abs/2004ASPC..311..121B}
}

@ARTICLE{Telfer2002,
       author = {{Telfer}, Randal C. and {Zheng}, Wei and {Kriss}, Gerard A. and {Davidsen}, Arthur F.},
        title = "{The Rest-Frame Extreme-Ultraviolet Spectral Properties of Quasi-stellar Objects}",
      journal = {\apj},
         year = 2002,
        month = feb,
       volume = {565},
       number = {2},
        pages = {773-785},
          doi = {10.1086/324689},
archivePrefix = {arXiv},
       eprint = {astro-ph/0109531},
 primaryClass = {astro-ph},
       adsurl = {https://ui.adsabs.harvard.edu/abs/2002ApJ...565..773T}
}

@ARTICLE{Shang2005,
       author = {{Shang}, Zhaohui and {Brotherton}, Michael S. and {Green}, Richard F. and {Kriss}, Gerard A. and {Scott}, Jennifer and {Quijano}, Jessica Kim and {Blaes}, Omer and {Hubeny}, Ivan and {Hutchings}, John and {Kaiser}, Mary Elizabeth and {Koratkar}, Anuradha and {Oegerle}, William and {Zheng}, Wei},
        title = "{Quasars and the Big Blue Bump}",
      journal = {\apj},
         year = 2005,
        month = jan,
       volume = {619},
       number = {1},
        pages = {41-59},
          doi = {10.1086/426134},
archivePrefix = {arXiv},
       eprint = {astro-ph/0409697},
 primaryClass = {astro-ph},
       adsurl = {https://ui.adsabs.harvard.edu/abs/2005ApJ...619...41S}
}

@ARTICLE{Davis2007,
       author = {{Davis}, Shane W. and {Blaes}, Omer M. and {Hubeny}, Ivan and {Turner}, Neal J.},
        title = "{Relativistic Accretion Disk Models of High-State Black Hole X-Ray Binary Spectra}",
      journal = {\apj},
         year = 2005,
        month = mar,
       volume = {621},
       number = {1},
        pages = {372-387},
          doi = {10.1086/427278},
archivePrefix = {arXiv},
       eprint = {astro-ph/0408590},
 primaryClass = {astro-ph},
       adsurl = {https://ui.adsabs.harvard.edu/abs/2005ApJ...621..372D}
}

@ARTICLE{Davis2011,
       author = {{Laor}, Ari and {Davis}, Shane W.},
        title = "{Cold accretion discs and lineless quasars}",
      journal = {\mnras},
         year = 2011,
        month = oct,
       volume = {417},
       number = {1},
        pages = {681-688},
          doi = {10.1111/j.1365-2966.2011.19310.x},
archivePrefix = {arXiv},
       eprint = {1106.4969},
 primaryClass = {astro-ph.CO},
       adsurl = {https://ui.adsabs.harvard.edu/abs/2011MNRAS.417..681L}
}

@INPROCEEDINGS{Runnoe2012a,
       author = {{Runnoe}, Jessie C. and {Brotherton}, M. and {Shang}, Z.},
        title = "{Updating Standard Quasar Bolometric Luminosity Corrections}",
    booktitle = {American Astronomical Society Meeting Abstracts \#219},
         year = 2012,
       series = {American Astronomical Society Meeting Abstracts},
       volume = {219},
        month = jan,
          eid = {243.19},
        pages = {243.19},
       adsurl = {https://ui.adsabs.harvard.edu/abs/2012AAS...21924319R}
}

@ARTICLE{Runnoe2012b,
       author = {{Runnoe}, Jessie C. and {Brotherton}, Michael S. and {Shang}, Zhaohui},
        title = "{Updating quasar bolometric luminosity corrections - II. Infrared bolometric corrections}",
      journal = {\mnras},
         year = 2012,
        month = nov,
       volume = {426},
       number = {4},
        pages = {2677-2688},
          doi = {10.1111/j.1365-2966.2012.21644.x},
archivePrefix = {arXiv},
       eprint = {1207.2124},
 primaryClass = {astro-ph.CO},
       adsurl = {https://ui.adsabs.harvard.edu/abs/2012MNRAS.426.2677R}
}

@ARTICLE{Pennell2017,
       author = {{Pennell}, Alison and {Runnoe}, Jessie C. and {Brotherton}, M.~S.},
        title = "{Updating quasar bolometric luminosity corrections - III. [O III] bolometric corrections}",
      journal = {\mnras},
         year = 2017,
        month = jun,
       volume = {468},
       number = {2},
        pages = {1433-1441},
          doi = {10.1093/mnras/stx556},
archivePrefix = {arXiv},
       eprint = {1703.03431},
 primaryClass = {astro-ph.GA},
       adsurl = {https://ui.adsabs.harvard.edu/abs/2017MNRAS.468.1433P}
}

@ARTICLE{Netzer2019,
       author = {{Netzer}, Hagai},
        title = "{Bolometric correction factors for active galactic nuclei}",
      journal = {\mnras},
         year = 2019,
        month = oct,
       volume = {488},
       number = {4},
        pages = {5185-5191},
          doi = {10.1093/mnras/stz2016},
archivePrefix = {arXiv},
       eprint = {1907.09534},
 primaryClass = {astro-ph.GA},
       adsurl = {https://ui.adsabs.harvard.edu/abs/2019MNRAS.488.5185N}
}

@ARTICLE{Gaskell1985,
       author = {{Gaskell}, C.~M.},
        title = "{Observational evidence for the radiative acceleration of broad-line clouds in Seyfert 1 galaxies and quasars.}",
      journal = {\apj},
         year = 1985,
        month = apr,
       volume = {291},
        pages = {112-116},
          doi = {10.1086/163045},
       adsurl = {https://ui.adsabs.harvard.edu/abs/1985ApJ...291..112G}
}

@ARTICLE{Sulentic2000b,
       author = {{Sulentic}, J.~W. and {Zwitter}, T. and {Marziani}, P. and {Dultzin-Hacyan}, D.},
        title = "{Eigenvector 1: An Optimal Correlation Space for Active Galactic Nuclei}",
      journal = {\apjl},
         year = 2000,
        month = jun,
       volume = {536},
       number = {1},
        pages = {L5-L9},
          doi = {10.1086/312717},
archivePrefix = {arXiv},
       eprint = {astro-ph/0005177},
 primaryClass = {astro-ph},
       adsurl = {https://ui.adsabs.harvard.edu/abs/2000ApJ...536L...5S}
}

@ARTICLE{Sulentic2007a,
       author = {{Sulentic}, Jack W. and {Bachev}, Rumen and {Marziani}, Paola and {Negrete}, C. Alenka and {Dultzin}, Deborah},
        title = "{C IV {\ensuremath{\lambda}}1549 as an Eigenvector 1 Parameter for Active Galactic Nuclei}",
      journal = {\apj},
         year = 2007,
        month = sep,
       volume = {666},
       number = {2},
        pages = {757-777},
          doi = {10.1086/519916},
archivePrefix = {arXiv},
       eprint = {0705.1895},
 primaryClass = {astro-ph},
       adsurl = {https://ui.adsabs.harvard.edu/abs/2007ApJ...666..757S}
}

@INPROCEEDINGS{Sulentic2007b,
       author = {{Sulentic}, J.~W. and {Dultzin-Hacyan}, D. and {Marziani}, P.},
        title = "{Eigenvector 1: Towards AGN Spectroscopic Unification}",
    booktitle = {Revista Mexicana de Astronomia y Astrofisica Conference Series},
         year = 2007,
       editor = {{Kurtz}, Stanley},
       series = {Revista Mexicana de Astronomia y Astrofisica Conference Series},
       volume = {28},
        month = jun,
        pages = {83-88},
       adsurl = {https://ui.adsabs.harvard.edu/abs/2007RMxAC..28...83S}
}

@ARTICLE{Kuraszkiewicz2009,
       author = {{Kuraszkiewicz}, Joanna and {Wilkes}, Belinda J. and {Schmidt}, Gary and {Smith}, Paul S. and {Cutri}, Roc and {Czerny}, Bo{\.z}ena},
        title = "{Principal Component Analysis of the Spectral Energy Distribution and Emission Line Properties of Red 2MASS Active Galactic Nuclei}",
      journal = {\apj},
         year = 2009,
        month = feb,
       volume = {692},
       number = {2},
        pages = {1180-1189},
          doi = {10.1088/0004-637X/692/2/1180},
archivePrefix = {arXiv},
       eprint = {0810.5714},
 primaryClass = {astro-ph},
       adsurl = {https://ui.adsabs.harvard.edu/abs/2009ApJ...692.1180K}
}

@ARTICLE{Zamfir2010,
       author = {{Zamfir}, S. and {Sulentic}, J.~W. and {Marziani}, P. and {Dultzin}, D.},
        title = "{Detailed characterization of H{\ensuremath{\beta}} emission line profile in low-z SDSS quasars}",
      journal = {\mnras},
         year = 2010,
        month = apr,
       volume = {403},
       number = {4},
        pages = {1759-1786},
          doi = {10.1111/j.1365-2966.2009.16236.x},
archivePrefix = {arXiv},
       eprint = {0912.4306},
 primaryClass = {astro-ph.CO},
       adsurl = {https://ui.adsabs.harvard.edu/abs/2010MNRAS.403.1759Z}
}

@ARTICLE{Marziani2015,
       author = {{Sulentic}, Jack and {Marziani}, Paola},
        title = "{Quasars in the 4D Eigenvector 1 Context: a stroll down memory lane}",
      journal = {Frontiers in Astronomy and Space Sciences},
         year = 2015,
        month = oct,
       volume = {2},
          eid = {6},
        pages = {6},
          doi = {10.3389/fspas.2015.00006},
archivePrefix = {arXiv},
       eprint = {1506.01276},
 primaryClass = {astro-ph.GA},
       adsurl = {https://ui.adsabs.harvard.edu/abs/2015FrASS...2....6S}
}

@ARTICLE{Panda2019,
       author = {{Panda}, Swayamtrupta and {Marziani}, Paola and {Czerny}, Bo{\.z}ena},
        title = "{The Quasar Main Sequence Explained by the Combination of Eddington Ratio, Metallicity, and Orientation}",
      journal = {\apj},
         year = 2019,
        month = sep,
       volume = {882},
       number = {2},
          eid = {79},
        pages = {79},
          doi = {10.3847/1538-4357/ab3292},
archivePrefix = {arXiv},
       eprint = {1905.01729},
 primaryClass = {astro-ph.HE},
       adsurl = {https://ui.adsabs.harvard.edu/abs/2019ApJ...882...79P}
}

@ARTICLE{Bahcall1972,
       author = {{Bahcall}, John N. and {Kozlovsky}, Ben-Zion and {Salpeter}, E.~E.},
        title = "{On the Time Dependence of Emission-Line Strengths from a Photoionized Nebula}",
      journal = {\apj},
         year = 1972,
        month = feb,
       volume = {171},
        pages = {467},
          doi = {10.1086/151300},
       adsurl = {https://ui.adsabs.harvard.edu/abs/1972ApJ...171..467B}
}

@ARTICLE{Blandford1982,
       author = {{Blandford}, R.~D. and {McKee}, C.~F.},
        title = "{Reverberation mapping of the emission line regions of Seyfert galaxies and quasars.}",
      journal = {\apj},
         year = 1982,
        month = apr,
       volume = {255},
        pages = {419-439},
          doi = {10.1086/159843},
       adsurl = {https://ui.adsabs.harvard.edu/abs/1982ApJ...255..419B}
}

@ARTICLE{Peterson1993,
       author = {{Peterson}, Bradley M.},
        title = "{Reverberation Mapping of Active Galactic Nuclei}",
      journal = {\pasp},
         year = 1993,
        month = mar,
       volume = {105},
        pages = {247},
          doi = {10.1086/133140},
       adsurl = {https://ui.adsabs.harvard.edu/abs/1993PASP..105..247P}
}

@ARTICLE{Peterson2014,
       author = {{Peterson}, Bradley M.},
        title = "{Measuring the Masses of Supermassive Black Holes}",
      journal = {\ssr},
         year = 2014,
        month = sep,
       volume = {183},
       number = {1-4},
        pages = {253-275},
          doi = {10.1007/s11214-013-9987-4},
       adsurl = {https://ui.adsabs.harvard.edu/abs/2014SSRv..183..253P}
}

@ARTICLE{Vestergaard2002,
       author = {{Vestergaard}, M.},
        title = "{Determining Central Black Hole Masses in Distant Active Galaxies}",
      journal = {\apj},
         year = 2002,
        month = jun,
       volume = {571},
       number = {2},
        pages = {733-752},
          doi = {10.1086/340045},
archivePrefix = {arXiv},
       eprint = {astro-ph/0204106},
 primaryClass = {astro-ph},
       adsurl = {https://ui.adsabs.harvard.edu/abs/2002ApJ...571..733V}
}

@ARTICLE{Greene2005,
       author = {{Greene}, Jenny E. and {Ho}, Luis C.},
        title = "{Estimating Black Hole Masses in Active Galaxies Using the H{\ensuremath{\alpha}} Emission Line}",
      journal = {\apj},
         year = 2005,
        month = sep,
       volume = {630},
       number = {1},
        pages = {122-129},
          doi = {10.1086/431897},
archivePrefix = {arXiv},
       eprint = {astro-ph/0508335},
 primaryClass = {astro-ph},
       adsurl = {https://ui.adsabs.harvard.edu/abs/2005ApJ...630..122G}
}

@ARTICLE{Vestergaard2006,
       author = {{Vestergaard}, Marianne and {Peterson}, Bradley M.},
        title = "{Determining Central Black Hole Masses in Distant Active Galaxies and Quasars. II. Improved Optical and UV Scaling Relationships}",
      journal = {\apj},
         year = 2006,
        month = apr,
       volume = {641},
       number = {2},
        pages = {689-709},
          doi = {10.1086/500572},
archivePrefix = {arXiv},
       eprint = {astro-ph/0601303},
 primaryClass = {astro-ph},
       adsurl = {https://ui.adsabs.harvard.edu/abs/2006ApJ...641..689V}
}

@ARTICLE{Bentz2009,
       author = {{Bentz}, Misty C. and {Peterson}, Bradley M. and {Netzer}, Hagai and {Pogge}, Richard W. and {Vestergaard}, Marianne},
        title = "{The Radius-Luminosity Relationship for Active Galactic Nuclei: The Effect of Host-Galaxy Starlight on Luminosity Measurements. II. The Full Sample of Reverberation-Mapped AGNs}",
      journal = {\apj},
         year = 2009,
        month = may,
       volume = {697},
       number = {1},
        pages = {160-181},
          doi = {10.1088/0004-637X/697/1/160},
archivePrefix = {arXiv},
       eprint = {0812.2283},
 primaryClass = {astro-ph},
       adsurl = {https://ui.adsabs.harvard.edu/abs/2009ApJ...697..160B}
}

\begin{appendix}
\counterwithin{figure}{section}  
\renewcommand{\thefigure}{\Alph{section}.\arabic{figure}} 

\section{Spectral Energy Distribution Examples for Individual Sources}\label{app_A}
\begin{figure*}
\centering
\includegraphics[angle=0,width=5.5in]{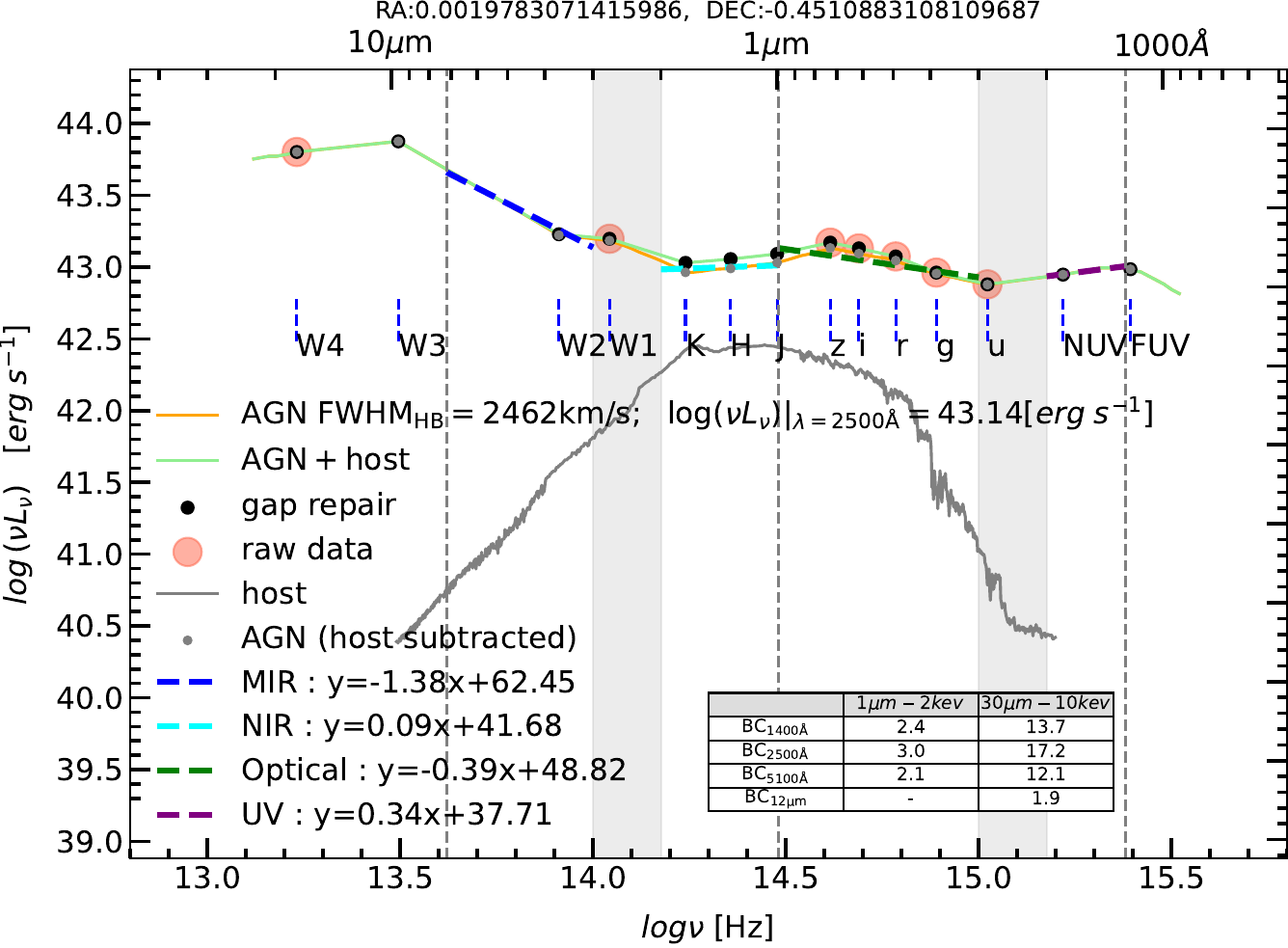}
\includegraphics[angle=0,width=5.5in]{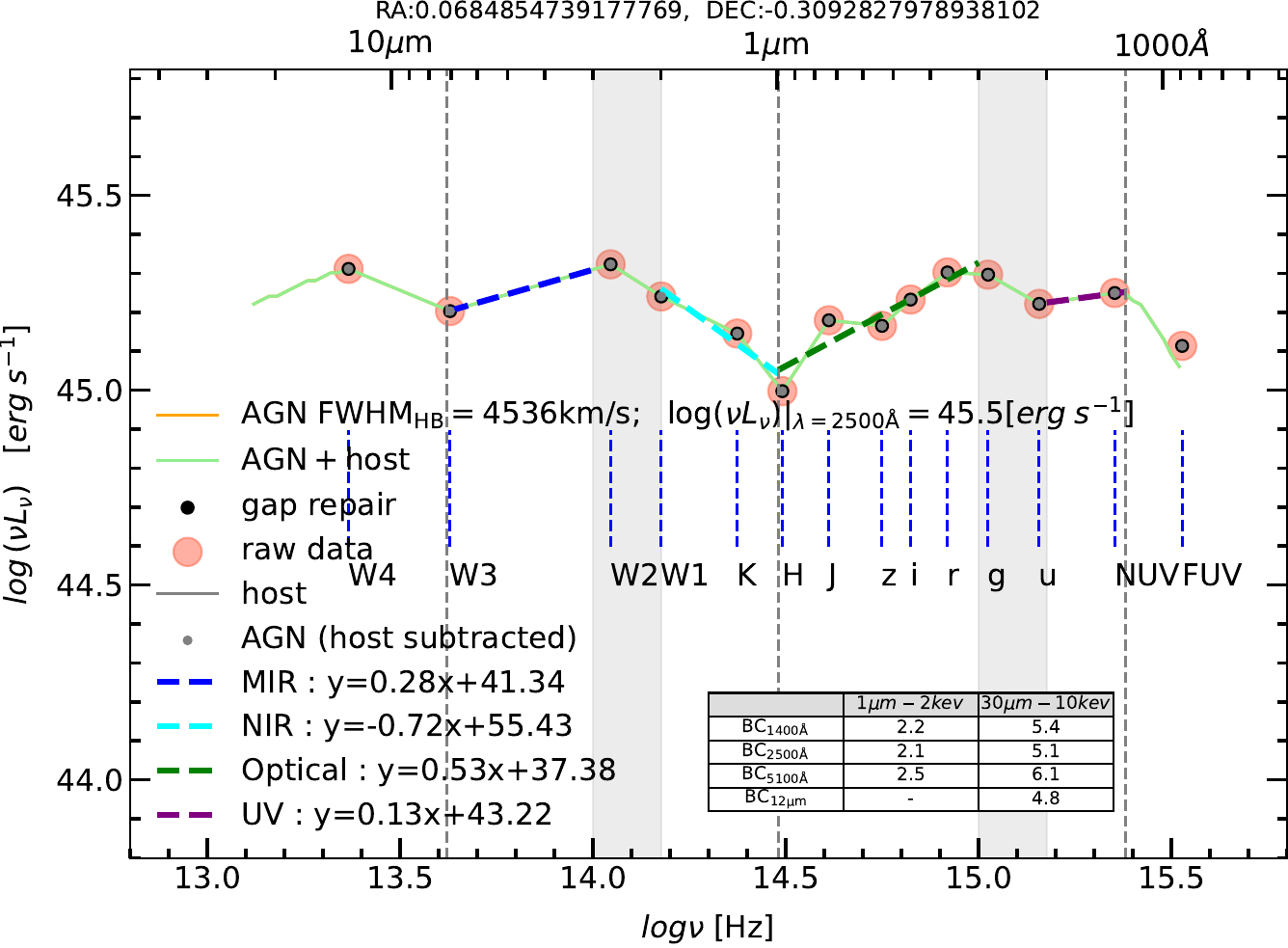}
\caption{SED examples for two individual sources: with host-galaxy correction (top) and without (bottom). Red points mark the original photometric data in the observed filters. Gap-repaired photometry is shown in black; host-corrected photometry is in gray. Green crosses indicate interpolated values at 14 rest-frame effective frequencies. The gray line represents the host-galaxy template, and the yellow cross marks the AGN luminosity after host correction. Blue dashed lines indicate filter positions. Colored lines show spectral slopes: MIR (blue), NIR (cyan), optical (green), and UV (purple). The inset table gives the BCs.}
\label{fig_a1}
\end{figure*}

Figure \ref{fig_a1} shows SED examples for individual source. The top panel illustrates the source corrected  for host-galaxy contamination and bottom panel shows the sources without a host-galaxy correction.
    
\section{Spectral Energy Distributio Slopes versus \rfe, \leddR, \mdot, and $\rm FWHM_{H\beta}$ for Individual Sources}\label{app_D}

\begin{figure*}
\centering
\includegraphics[angle=0,width=2.5in]{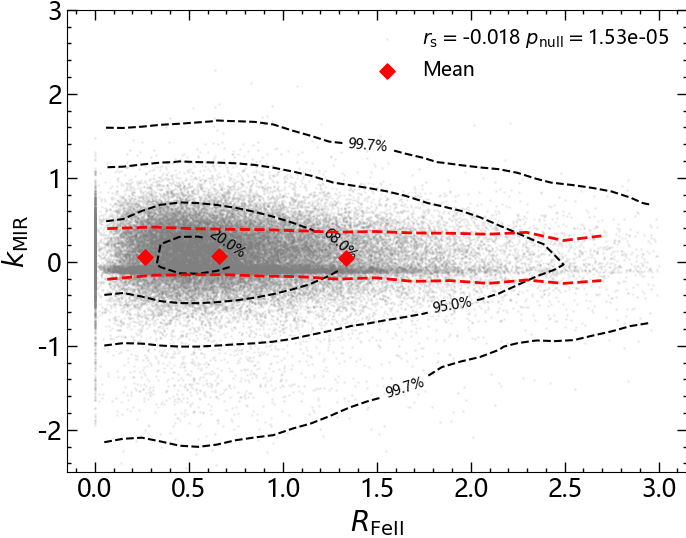}
\includegraphics[angle=0,width=2.5in]{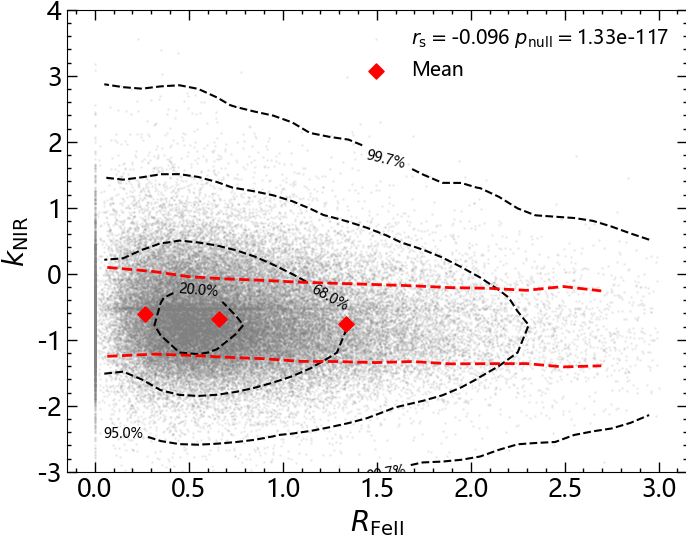}
\includegraphics[angle=0,width=2.5in]{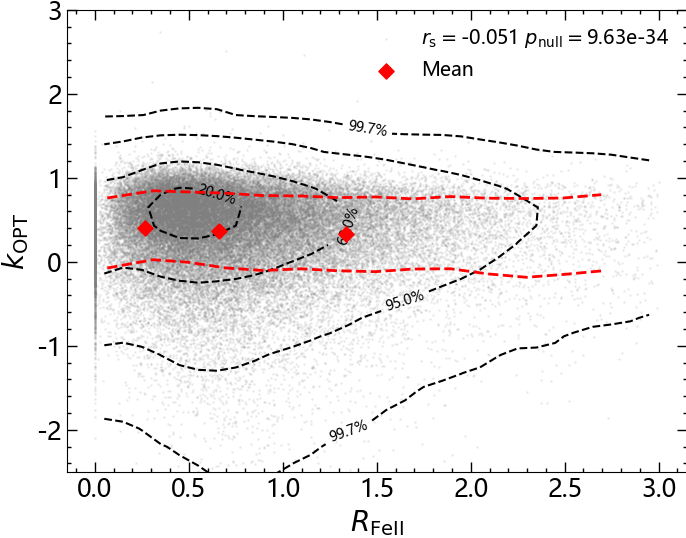}
\includegraphics[angle=0,width=2.5in]{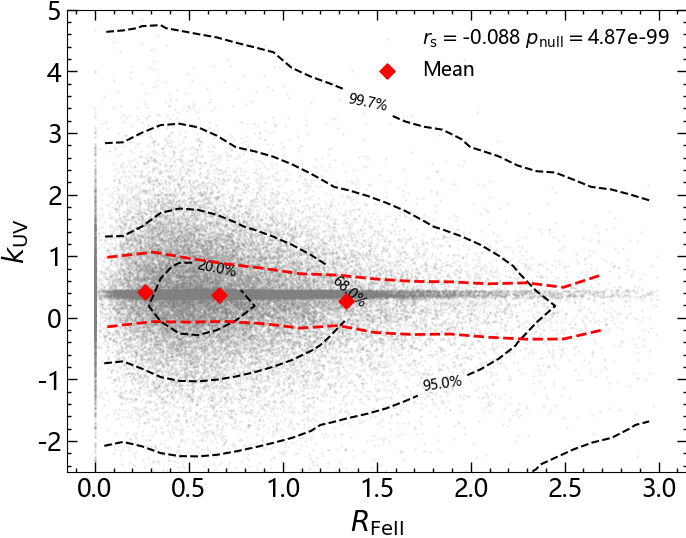}
\caption{SED slopes $k_{\rm MIR}, k_{\rm NIR}, k_{\rm OPT}$, and $k_{\rm UV}$ (left to right, top to bottom) as functions of  \rfe. Black solid curves indicate density contours; red diamonds mark bin means from equal-count binning; red dashed lines show the $1 \sigma$ confidence intervals.}
\label{D_fig1}
\end{figure*}

\begin{figure*}
\centering
\includegraphics[angle=0,width=2.5in]{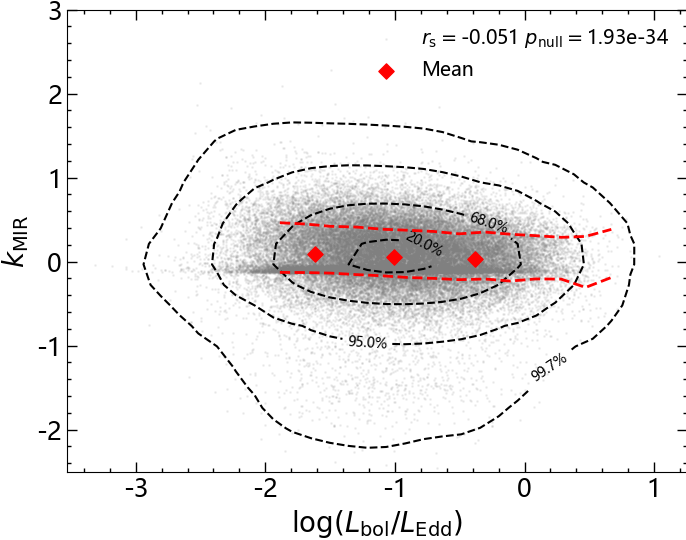}
\includegraphics[angle=0,width=2.5in]{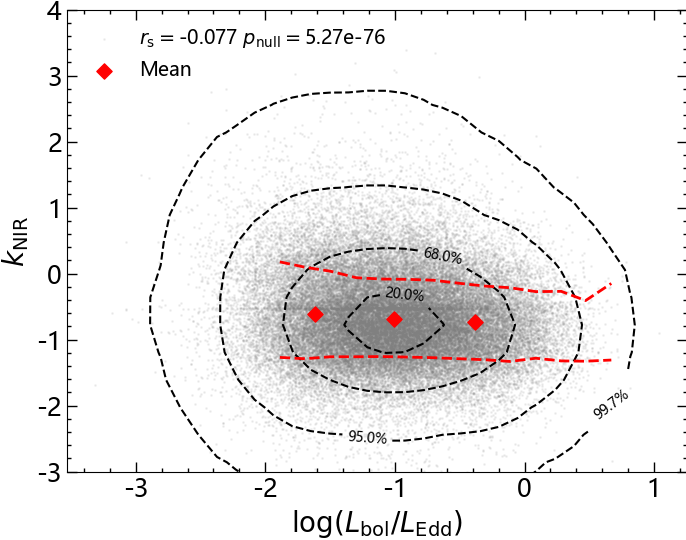}
\includegraphics[angle=0,width=2.5in]{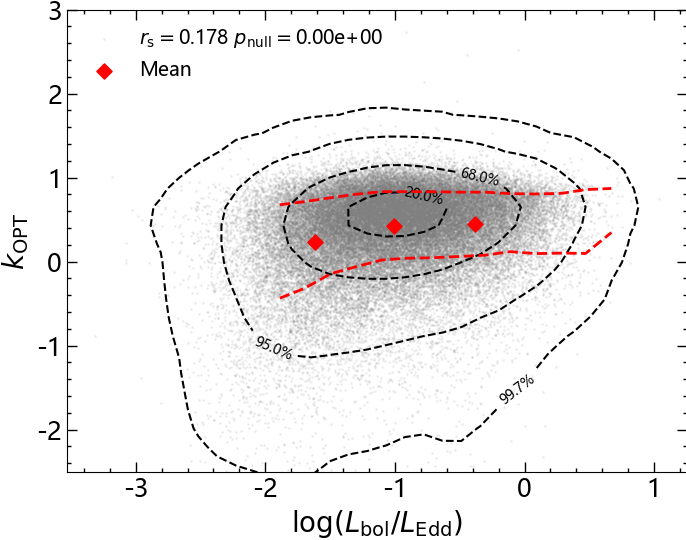}
\includegraphics[angle=0,width=2.5in]{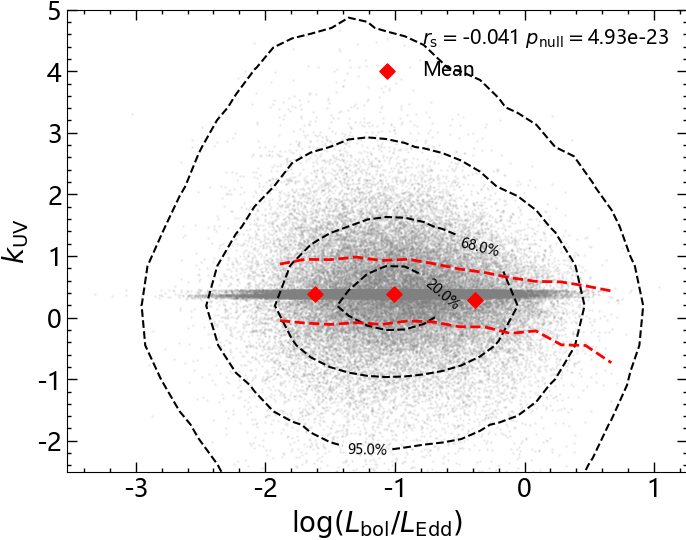}
\caption{Four SED slopes vs. \leddR\ (same layout and symbols as Figure \ref{D_fig1}).}
\label{D_fig2}
\end{figure*}

\begin{figure*}
\centering
\includegraphics[angle=0,width=2.5in]{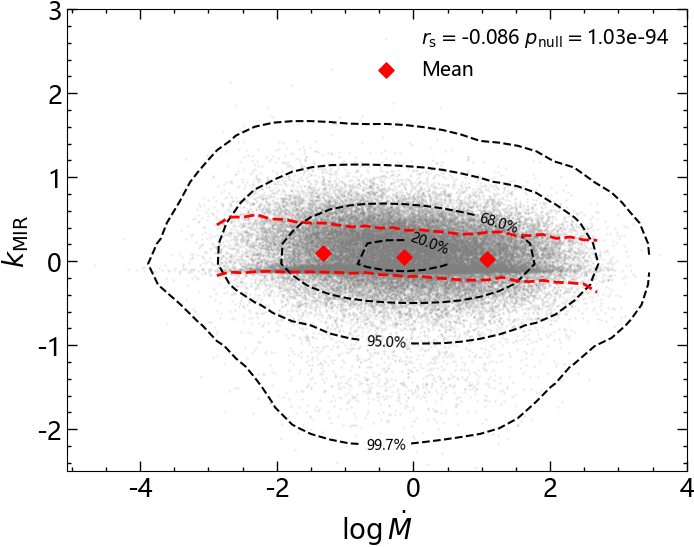}
\includegraphics[angle=0,width=2.5in]{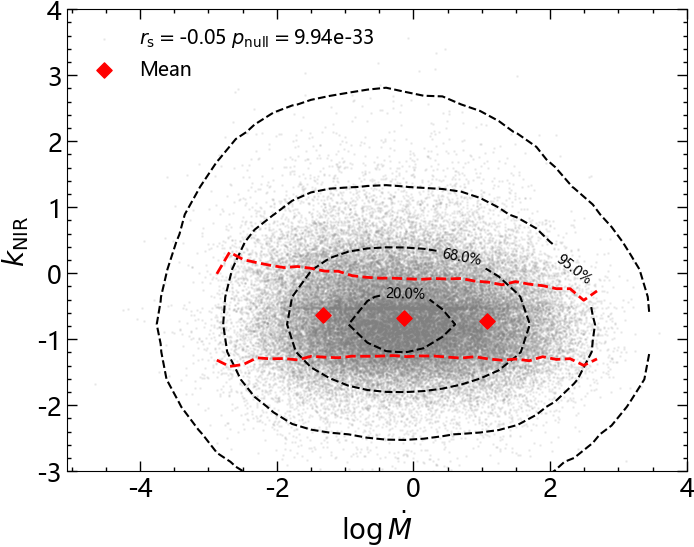}
\includegraphics[angle=0,width=2.5in]{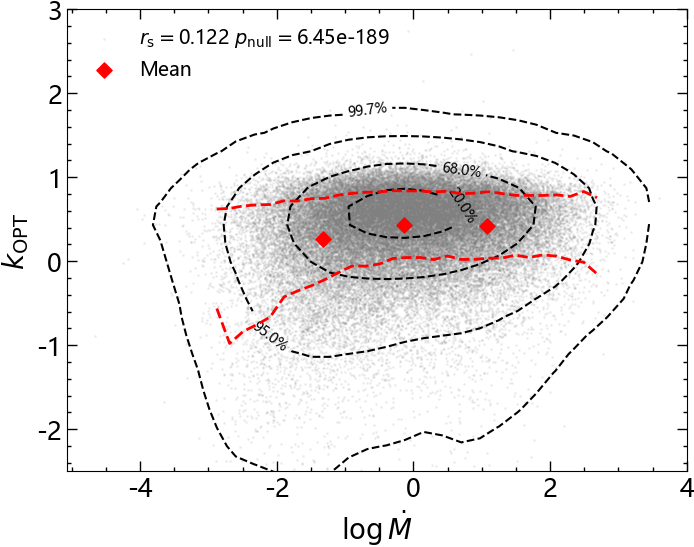}
\includegraphics[angle=0,width=2.5in]{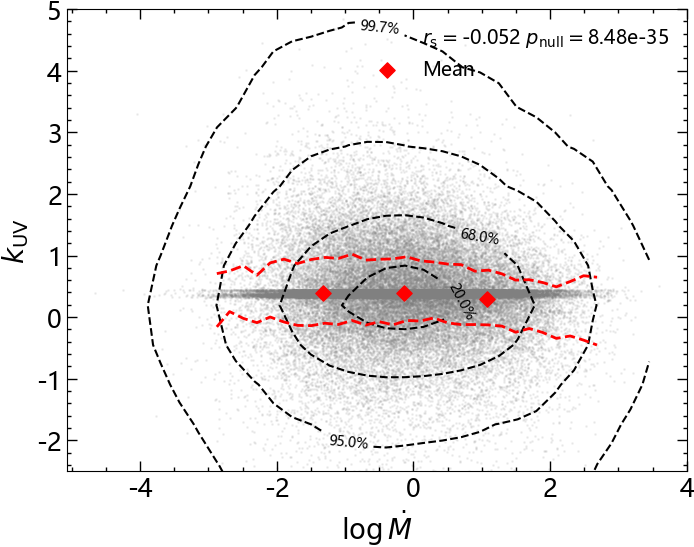}
\caption{Four SED slopes vs. \mdot\ (same layout and symbols as Figure \ref{D_fig1}).}
\label{D_fig3}
\end{figure*}

\begin{figure*}
\centering
\includegraphics[angle=0,width=2.8in]{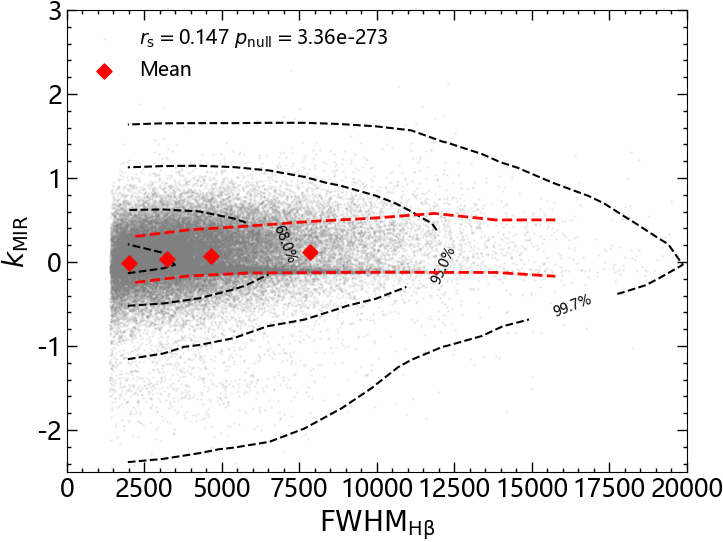}
\includegraphics[angle=0,width=2.8in]{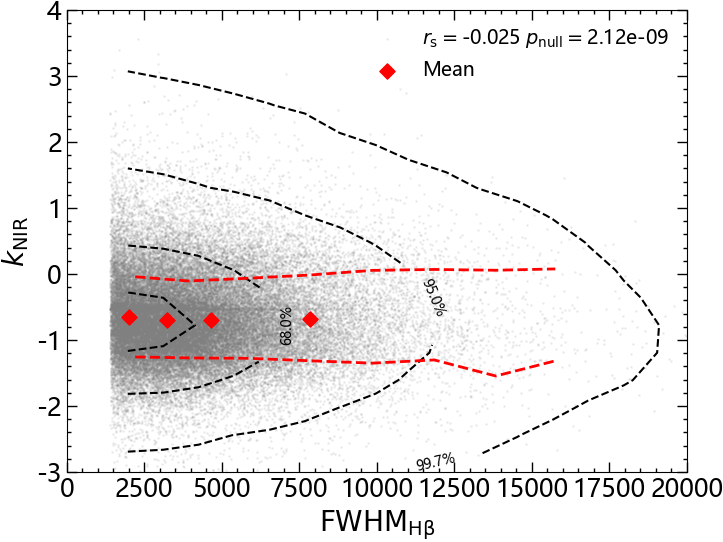}
\includegraphics[angle=0,width=2.8in]{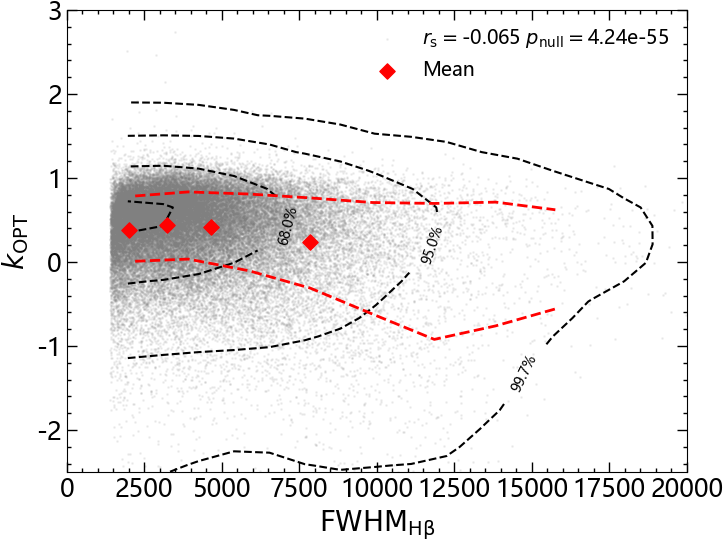}
\includegraphics[angle=0,width=2.8in]{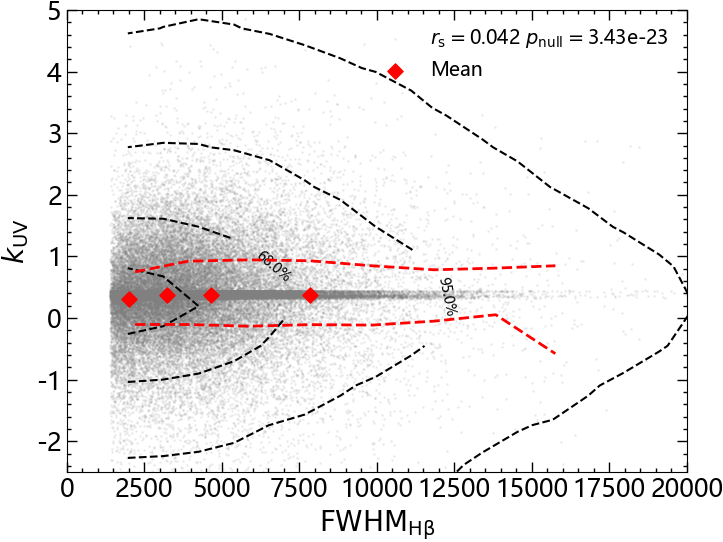}
\caption{Four SED slopes vs. $\rm FWHM_{H\beta}$ (same layout and symbols as Figure \ref{D_fig1}).}
\label{D_fig4}
\end{figure*}

SED slopes versus \rfe, \leddR, \mdot, and $\rm FWHM_{H\beta}$ for individual sources are presented in Figure \ref{D_fig1}, \ref{D_fig2}, \ref{D_fig3}, and Figure \ref{D_fig4} (red diamonds: equal number bin means; dashed lines: $1 \sigma$ boundaries;  $k_{\rm MIR},~ k_{\rm NIR}, ~k_{\rm OPT}$, and $k_{\rm UV} $ corresponding to $k_{1},~ k_{2}, ~k_{3}$, and $k_{4}$ in Figure \ref{fig2}). All relations between  slopes and \rfe, \leddR, \mdot, and $\rm FWHM_{H\beta}$ are significant with $p_{\rm null}< 10^{-6}$ , and theirs $r_s$ values are shown in Table \ref{tab:D}.

The individual-source mean slopes differ slightly from the mean SED results (e.g., UV slopes: 0.43, 0.40, and 0.32 versus 0.41, 0.37, and 0.27 with increasing \rfe\ ; standard deviation $\sim$ 0.7-0.8). For individual sources, those with a larger \rfe\ exhibit redder UV, optical, MIR, and NIR continua, which become increasingly red with increasing \leddR\ (or \mdot) in the MIR, NIR, and UV bands and the optical continuum shows the opposite trend (becoming harder and bluer).  We also find that quasars with larger $\rm FWHM_{\hb}$ show redder optical and NIR continua and bluer UV and MIR continua. They are consistent with the trend derived from the mean SEDs.

Despite the small $r_s$ and numerical discrepancies, the  method consistency and $p_{\rm null}<10^{-6}$ confirm genuine population trends. Intrinsic scatter does not affect mean the SED evolutionary patterns.

\begin{table}[htbp]
\centering
\caption{$r_s$ for Relations between Spectral Energy Distribution Slopes versus \rfe, \leddR, \mdot, and $\rm FWHM_{H\beta}$ for Individual Sources.}
\label{tab:D}
\begin{tabular}{lrrrrrrr}
\hline
$k$ & \rfe\ & \leddR & \mdot & $\rm FWHM_{H\beta}$ \\
\hline
$k_{\rm MIR}$ & $-0.018$  & $-0.051$ & $-0.086$  & $0.147$ \\
$k_{\rm NIR}$ & $-0.096$  & $-0.077$  & $-0.050$  & $-0.025$  \\
$k_{\rm OPT}$ & $-0.051$  & $0.178$ & $0.122$    & $-0.065$  \\
$k_{\rm uv}$ & $-0.088$      & $-0.041$  & $-0.052$ & $0.042$  \\
\hline
\end{tabular}
\end{table}

\section{The Luminosity Dependence of the Mean Spectral Energy Distributions}\label{app_B}
\begin{figure*}
\centering
\includegraphics[angle=0,width=6in]{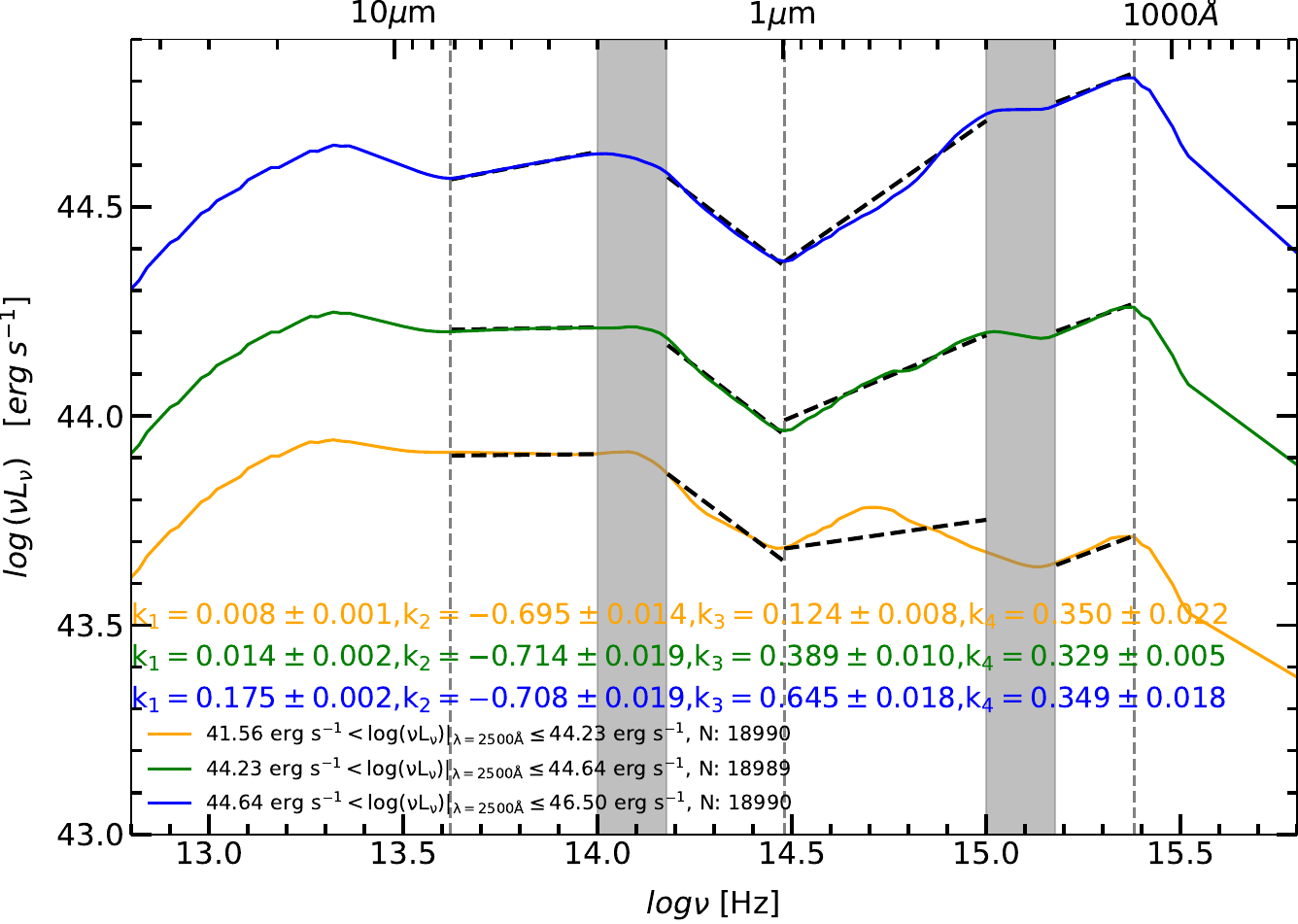}
\caption{Mean SEDs for low-, mid- and high-luminosity quasars. The SED of the bin with the high luminosity shows blue optical, MIR continua. Symbols are the same as Figure \ref{fig3}}.
\label{fig_b1}
\end{figure*}

We have divided the sample into three luminosity bins of nearly equal size, with each bin containing 18,990 quasars. Figure \ref{fig_b1} shows the mean SED of $\log(\nu L_{\nu})$ at 2500 \AA\ for all quasars. The high-luminosity SED has a harder (bluer) MIR and optical spectrum. The low-luminosity SED has a bump around 5000 \AA. The optical spectral slopes become smaller as the luminosity decreases. 

\begin{figure*}
\centering
\includegraphics[angle=0,width=6in]{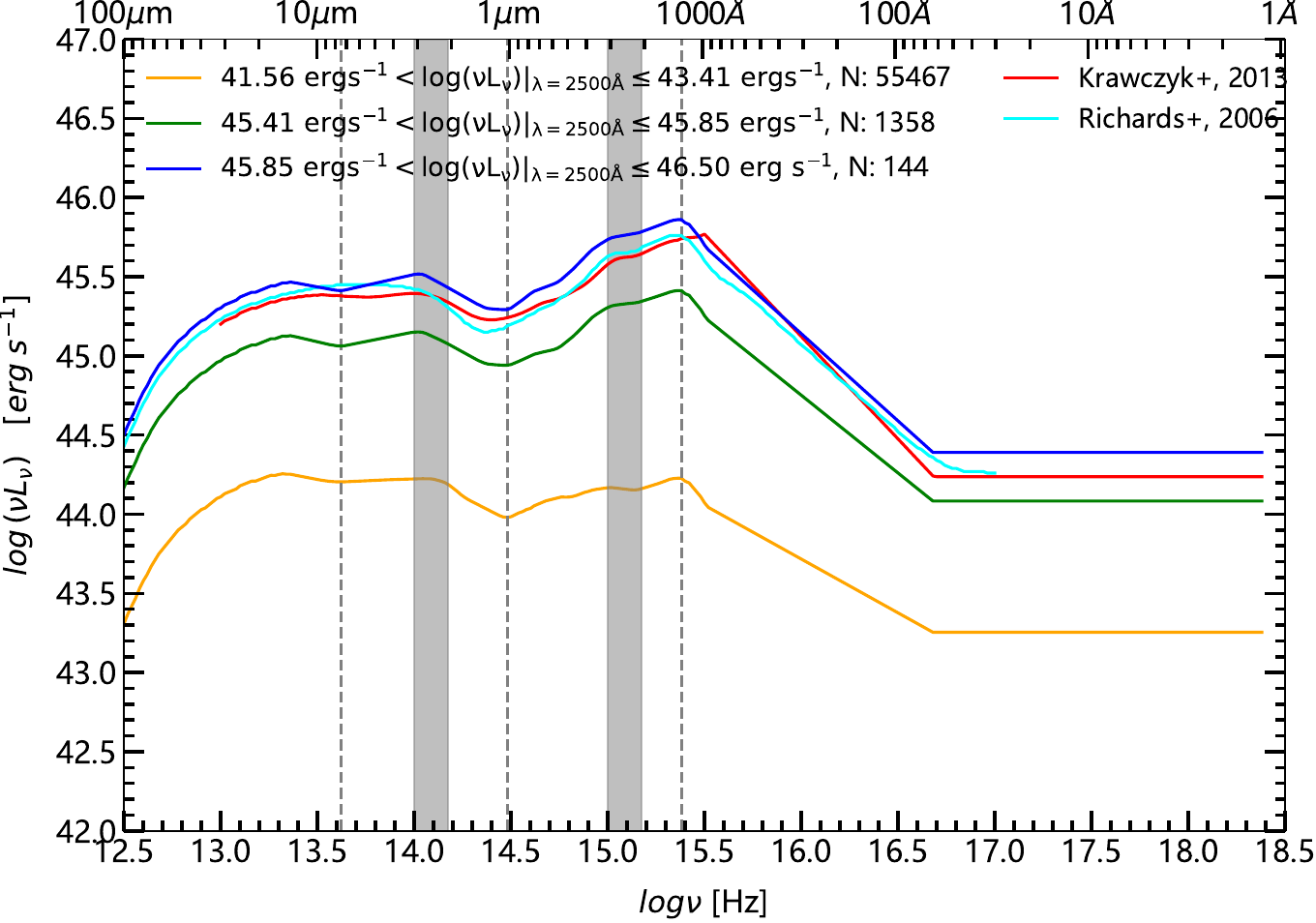}
\includegraphics[angle=0,width=5.8in]{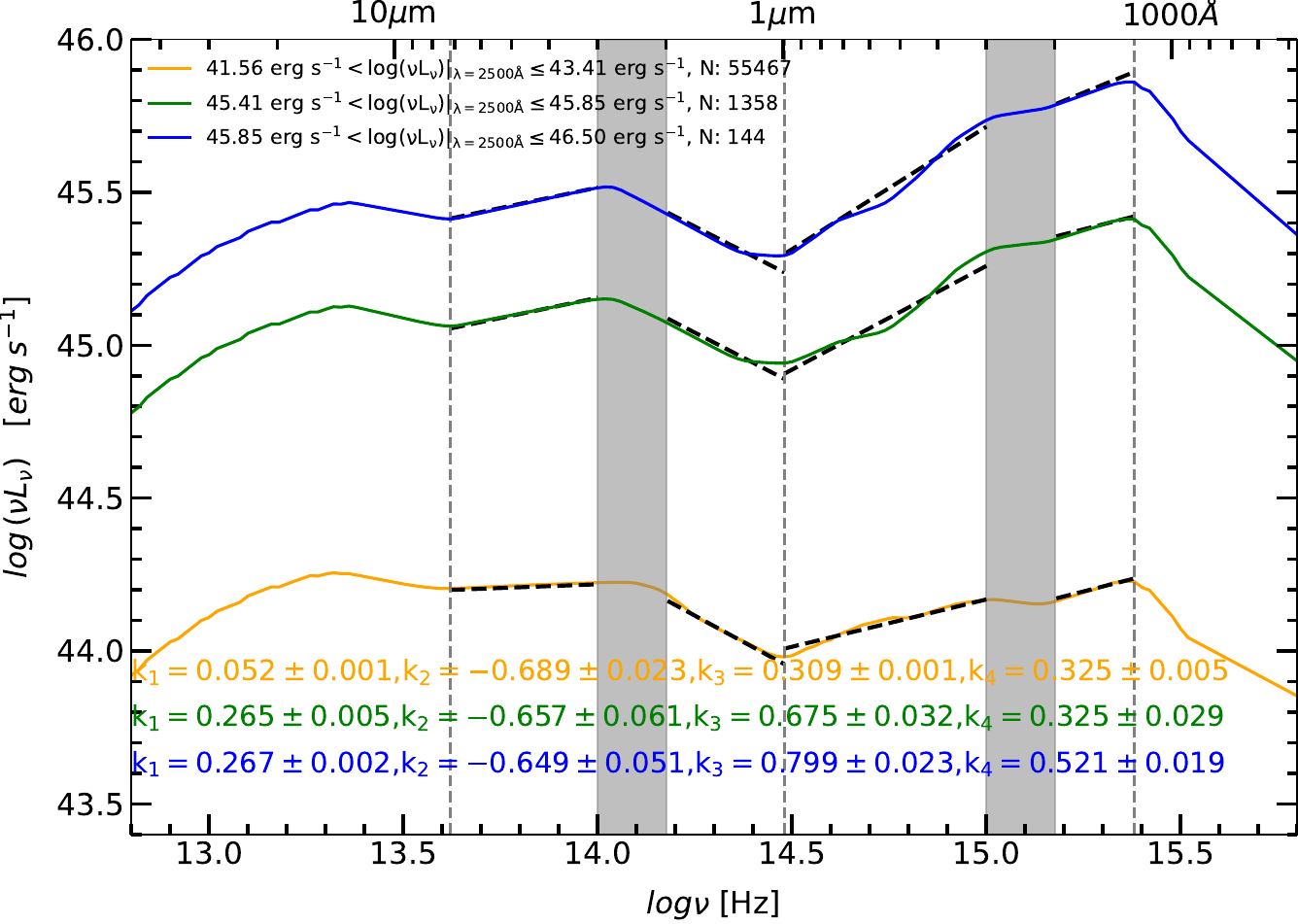}
\caption{The luminosity-dependent mean SEDs with those reported in the literature \citep{Richards2006a,Krawczy2013} in the same luminosity ranges. Top: a comparison of the DR16 SED with the over all mean SEDs of \cite[][cyan]{Richards2006a} and \cite[][red]{Krawczy2013}. Bottom: mean SEDs for low-, mid- and high-luminosity quasars and the corresponding fit results. Symbols are the same as Figure \ref{fig3}.}
\label{fig_b2}
\end{figure*}

The SDSS DR16 quasar catalog (DR16Q) covers broad ranges in luminosity ($44< \rm{log}(L_{\rm bol}/erg\ s^{-1})<48$), and probes lower luminosities than SDSS DR7 quasars \citep{Wu2022}. We compare the luminosity-dependent mean SEDs with those reported in the literature \citep{Richards2006a,Krawczy2013} in the same luminosity ranges. The top panel in Figure \ref{fig_b2} shows a comparison of the DR16 SED with the overall mean SEDs of  \cite[][cyan]{Richards2006a} and \cite[][red]{Krawczy2013}. Our mean SED closely matches that of \cite{Richards2006a} in the UV--optical--NIR range, but differs markedly at MIR to FIR wavelengths. In the IR it is similar to \cite{Krawczy2013}, and at 912 \text{\AA} it again agrees with \cite{Richards2006a}. We use the same range with others only for comparison, but recognize that the number of sources in each bin is highly nonuniform can have a significant impact; indeed can affect the bump visible in our mean SEDs. The detailed differences depend on both the method and the sample used. In the bottom panel in Figure \ref{fig_b2}, with increasing luminosity, the UV slope increases from 0.33 to 0.52, the optical slope increases from 0.31 to 0.80, the NIR slope increases from -0.69 to -0.65, and the MIR slope increases from 0.05 to 0.27.

\section{Relation between $\rm BC_{5100}$ and \lv}\label{app_C} 
BCs are likely to differ from one object to the next because of the accretion rate, the accretion efficiency, and other factors. We investigate how the BCs are dependent on quasar luminosity. The blue line in Figure \ref{fig_c1} shows the luminosity-independent BC that assumes a linear dependence $\mathrm{BC}_{5100} = 53 - \log(L_{5100})$ \citep{Netzer2013}. The median values of $\rm BC_{5100}$ decrease monotonically from $\approx$ 7.5 to $\approx$ 6 as $\log (\lv/\ergs)$ rises from 44.9 to 45.8. The data follow the empirical relation $\mathrm{BC}_{5100} = 53 - \log(L_{5100})$ to within $\pm 1$dex scatter in high-luminosity ranges ($\log (\lv/\ergs)> 44.6$), however in low-luminosity ranges ($\log (\lv/\ergs)< 44.6$), the data scatter far from the empirical relation. 

\begin{figure*}
\centering
\includegraphics[angle=0,width=4in]{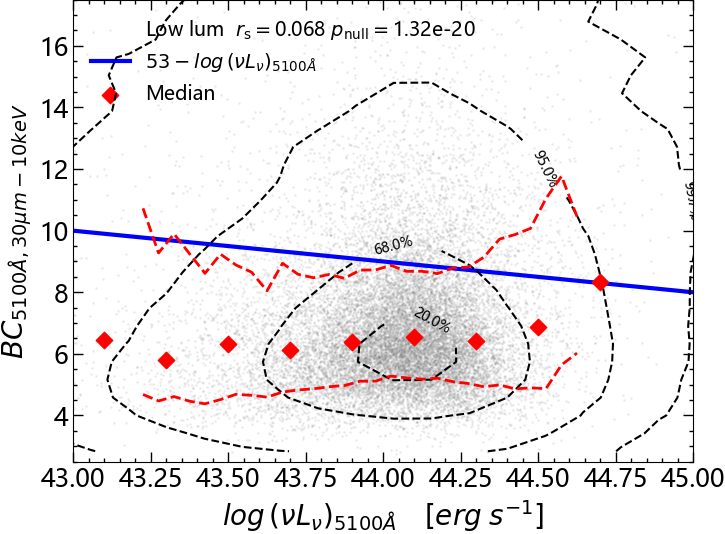}
\includegraphics[angle=0,width=4in]{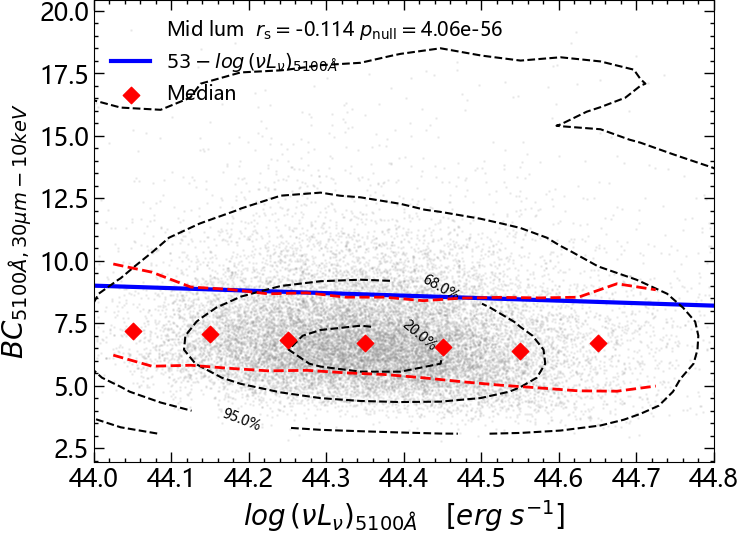}
\includegraphics[angle=0,width=4in]{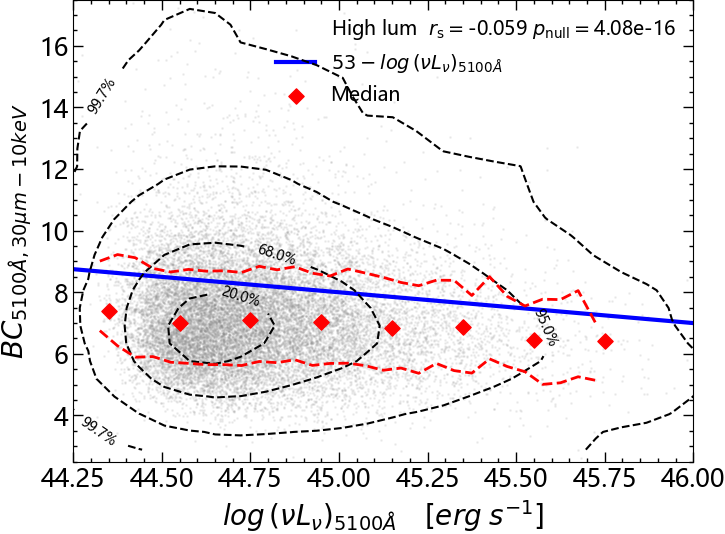}
\caption{Relation between $\rm BC_{5100}$ and \lv\ in different luminosity bins. The blue line shows $\mathrm{BC}_{5100} = 53 - \log(L_{5100})$ \citep{Netzer2013}. Gray points show every source in DR16, black dashed curves are density contours, red diamonds mark the medians, and red dashed curves are $1 \sigma$ uncertainties. From top to bottom: low-, intermediate-, and high-luminosity bins, respectively.}
\label{fig_c2}
\end{figure*}

Owing to the large discrepancies at low luminosities in the current SED, we analyzed the relation between $\rm BC_{5100}$ and \lv\ separately for high-, intermediate-, and low-luminosity subsamples in Figure \ref{fig_c2}. We split the sample into three equally populated bins at 2500 \AA. Almost no correlation between median values of $\rm BC_{5100}$ and \lv\ is found in the lowest-luminosity bin. However, both the intermediate- and high-luminosity bins show an inverse correlation within $1 \sigma$ uncertainties. Low-luminosity sources exert an influence on the relation between $\rm BC_{5100}$ and \lv.

\end{appendix}

\end{document}